\documentclass[]{interact}

\usepackage{epstopdf}
\usepackage{subcaption}

\usepackage[numbers,sort&compress]{natbib}
\bibpunct[, ]{[}{]}{,}{n}{,}{,}

\theoremstyle{plain}

\newtheorem{question}{Open Question}

\theoremstyle{definition}

\theoremstyle{remark}

\newcommand\blfootnote[1]{%
	\begingroup
	\renewcommand\thefootnote{}\footnote{#1}%
	\addtocounter{footnote}{-1}%
	\endgroup
}

\usepackage{tikz}

\usepackage{hyperref}
\hypersetup{
	colorlinks,
	linkcolor={blue!50!black},
	citecolor={blue!50!black},
	urlcolor={blue!50!black}
}

\definecolor{greenhl}{rgb}{0.5,0.7,0.0}
\definecolor{bluegreenur}{rgb}{0.,0.5,1.0}
\definecolor{orangems}{rgb}{1., 0.6, 0.}
\definecolor{likewhite}{rgb}{0.95,0.95,0.95}
\definecolor{someblue}{rgb}{0.0,0.53,0.74} 

\definecolor{greenms}{rgb}{0,0.55,0}

\newcommand{\rev}[1]{\textcolor{black}{#1}}

\graphicspath{{Figures/},{Figures_Weaving/},{Figures_Subdivision/},{Figures_puzzlePhoto/},{Figures_SubdivisionWeaving/}}

\usepackage[normalem]{ulem}

\begin{document}

\articletype{}

\title{\rev{Weaving patterns inspired by the pentagon snub subdivision scheme}}

\author{
\name{Henriette Lipsch\"utz\textsuperscript{a}, Ulrich Reitebuch\textsuperscript{a}, Martin Skrodzki\textsuperscript{b}, and Konrad Polthier\textsuperscript{a}}
\affil{
	\textsuperscript{a}Fachbereich Mathematik und Informatik, Institut f\"ur Mathematik, Freie Universit\"at Berlin, Germany\\
	\textsuperscript{b}Computer Graphics and Visualization, TU Delft, The Netherlands}
}

\maketitle

\begin{abstract}
	Various computer simulations regarding, e.g., the weather or structural mechanics, solve complex problems on a two-dimensional domain.
	They mostly do so by splitting the input domain into a finite set of smaller and simpler elements on which the simulation can be run fast and efficiently.
	This process of splitting can be automatized by using subdivision schemes.
	
	Given the wide range of simulation problems to be tackled, an equally wide range of subdivision schemes is available.
	They create subdivisions that are (mainly) comprised of triangles, quadrilaterals, or hexagons.
	Furthermore, they ensure that (almost) all vertices have the same number of neighboring vertices.
	
	\rev{This paper illustrates a subdivision scheme that splits the input domain into pentagons.
	Repeated application of the scheme gives rise to fractal-like structures.
	Furthermore, the resulting subdivided domain admits to certain weaving patterns.
	These patterns are subsequently generalized to several other subdivision schemes.}
	
	\rev{As a final contribution, we provide paper models illustrating the weaving patterns induced by the pentagonal subdivision scheme.
		Furthermore, we present a jigsaw puzzle illustrating both the subdivision process and the induced weaving pattern.
	These transform the visual and abstract mathematical algorithms into tactile objects that offer exploration possibilities aside from the visual.}
\end{abstract}


\blfootnote{This article is submitted for the JMA Special issue: The Art of Mathematical Illustration.}


\section{Introduction}

For many people all around the world, checking the weather report is part of their daily routine in the morning.
Will it be a sunny day for the planned trip to the beach?
Or will some rain finally water the crops on the field?
While a weather forecast is easily accessible nowadays, its computation is fairly involved and occupies the largest supercomputers in the world.

The underlying mathematical equations that govern the weather cannot be solved explicitly, i.e., it is not possible to write down an analytic solution for them.
A first breakthrough in weather forecast was therefore the development of difference methods that do not solve the entire system of equations, but instead compute an approximate solution.
These solutions are not computed for the entire area of, say, a country, but only for a finite number of points in the country, e.g., at the weather stations where good information is available.
For any place in the region that is not sufficiently close to one of the weather stations, the forecast is computed as a mixture of the solutions from the surrounding weather stations.
Of course, the precision of the computation will depend on the number of points distributed across the country; a single weather station cannot provide a prediction as accurate as a dense network of stations distributed throughout the country.
This was also realized by Lewis Richardson, who first tried (and failed) to predict the weather in 1928~\cite[Ch.~3.8.4]{gustafsson2018scientific}.

Techniques for the solution of such complex problems as weather predictions came up in the 1950's from a completely different scientific branch: structural mechanics.
The equations that govern stresses in mechanical assemblies are also impossible to be solved analytically.
In particular complex geometries, like entire airplanes call for optimizations of, e.g., the weight, while having a high demand for structural integrity.
In order to achieve both, \emph{finite element methods} are used to compute approximations of the arising stresses.
Not the entire hull of the aircraft is considered in the computation, but it is split into a (finite) set of simple elements, such as triangles, quadrilaterals, or other geometric primitives.
Again, like in the case of the weather stations, the computations become more accurate with a growing number of elements~\cite[Ch.~5.6]{gustafsson2018scientific}.

One possibility to generate a fine collection of surface elements for a geometry is the use of \emph{subdivision schemes}.
These start from a coarse input and iteratively divide each geometric primitive into smaller parts until a limit surface is reached.
It is one specific representative from the family of these subdivision schemes that we want to illustrate in this article.
In the following, we will first discuss existing subdivision schemes that create triangles, quadrilaterals, and hexagons.
This discussion motivates the description of a method creating pentagons, which also produces a fractal boundary curve as well as a collection of fractal-like interior curves.

Naturally, subdivision schemes have been illustrated with static figures, dynamic videos or animations, and interactive computer programs.
However, all these are confined to the page of a book or the screen of a computer, tablet, or smartphone, i.e., they relay visual information about the content.
None of them provide a haptic sensation of what it means to have a two-dimensional domain and subdivide it into smaller pieces.
\rev{Starting from the pentagon subdivision scheme, we introduce a coloring that provides a weaving pattern on the subdivided geometries.
The pattern consists of a small number of strands which intertwine in pairs and which weave into a complex pattern.}
\rev{Here, the number of pairs depends on the number of refinement steps executed.
After exploring properties of the coloring and the related weaving patterns for the pentagonal case, we discuss how they can also be applied to other subdivision schemes.
}

\rev{
Finally, this article aims to close the gap between virtual or visual illustrations on the one hand and haptic illustrations on the other hand. 
To do so, we present two paper models that can be manufactured at home.
These paper models transfer the observations made on the weaving patterns into a personal experience by assembling the corresponding subdivisions.
Our final contribution is a two-sided jigsaw puzzle that incorporates both the pentagon subdivision scheme as well as two different views on the weaving patterns.
Thus, the paper models and the jigsaw puzzle provide haptic means to explore both subdivision schemes and the related weaving patterns.
}

\rev{
Throughout the research process for this paper and when creating the illustrations, we encountered several mathematical questions.
While we offer insights to some of them, others remain open and have yet to be answered.
As an invitation to the mathematical community, we pose these open questions prominently throughout the paper.
We hope that this inspires new research, starting from the presented illustrations.
}



\section{Subdivision schemes}
\label{sec:SubdivisionSchemes}

As stated in the introduction, a subdivision scheme works on a set of geometric primitives, such as triangles, quadrilaterals, or hexagons, called a \emph{mesh}.
This input is iteratively split into a number of smaller primitives.
Each of these iterative passes is called a~\emph{refinement step}.
During this process, new vertices, edges, and faces are inserted following the set of rules of the respective subdivision scheme.
In the following, we will denote the respective sets by $V$, $E$, and $F$.
For a vertex~${v\in V}$, we will denote the number of neighbors of~$v$ as the \emph{degree} of~$v$.
\rev{Similarly, for a face~${f\in F}$, we denote the number of its sides as the \emph{degree} of~$f$.}
In the following, we will always assume a planar input mesh.

In general, one distinguishes between \emph{interpolating} and \emph{approximating} subdivision schemes: While the first group keeps the \rev{vertex positions} of the original input mesh intact, the second group does not.
Furthermore, a subdivision scheme is called \emph{primal} if a face~$f$ of the mesh is replaced by several new faces~$f_1,\ldots,f_n$ in the refinement step.
If the vertices of the mesh are replaced by several new vertices, the scheme is called \emph{dual}.
Observe that, depending on the inserted elements, the number of all three components---vertices, edges, and faces---increases. 

\subsection{Triangle meshes}

\begin{figure}
	\includegraphics[width=0.2\textwidth]{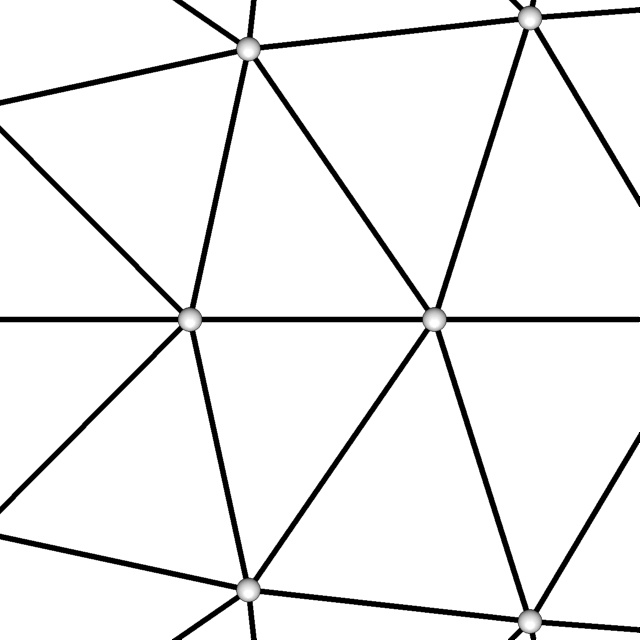}
	\hfill
	\includegraphics[width=0.2\textwidth]{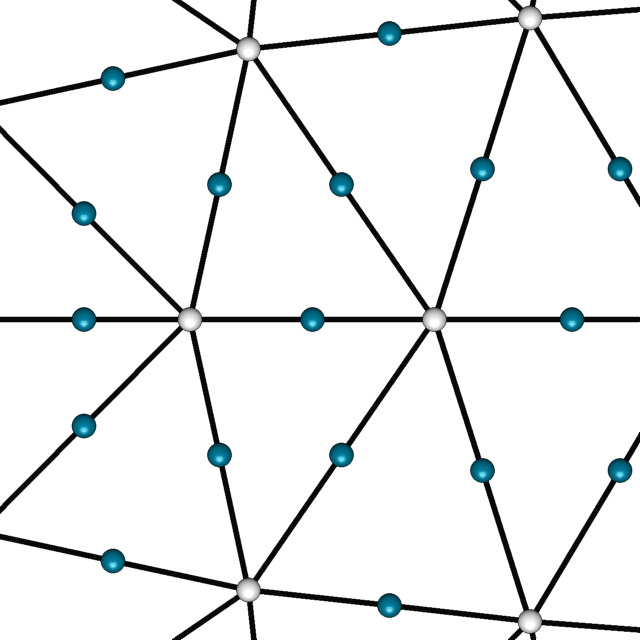}
	\hfill
	\includegraphics[width=0.2\textwidth]{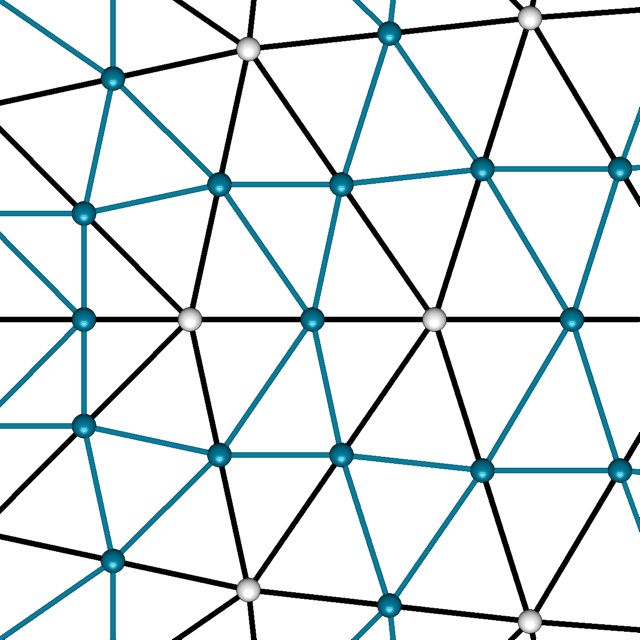}
	\hfill
	\includegraphics[width=0.2\textwidth]{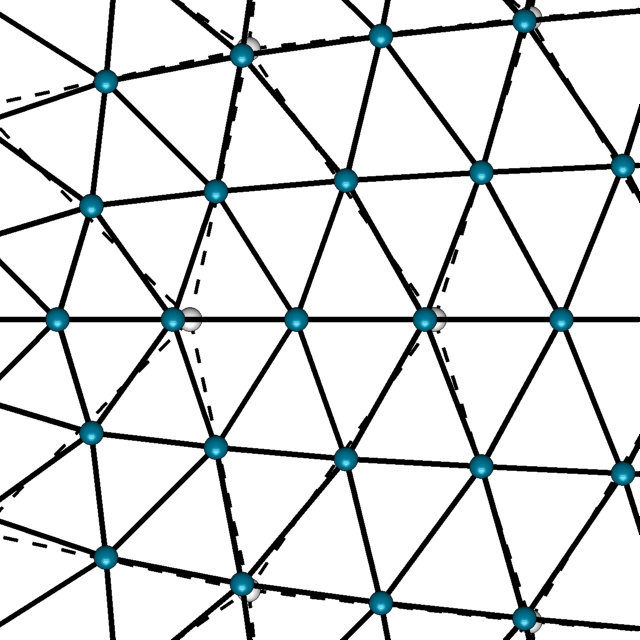}
	\caption{An illustration of the subdivision scheme by Loop~\cite{loop1987smooth}: input mesh, inserting vertices as edge midpoints, connecting the new vertices, and smoothing the mesh to be more regular. 
		White input vertices are moved during smoothing.}
	\label{fig:LoopSubdivision}
\end{figure}

A first example of a subdivision scheme that operates on triangle meshes is the subdivision scheme by Loop.
The general idea is to introduce a new vertex for each edge as a weighted combination of the two triangles that share the edge.
This new vertex is connected to the two old vertices spanning the edge on which it was created as well as to the four other new vertices created as edge midpoints in its neighboring triangles.
Thereby, each newly inserted vertex has degree six.
Finally, in a smoothing operation, the old vertices are moved to a convex combination of their new neighbors, see Figure~\ref{fig:LoopSubdivision}.

\begin{figure}[b]
	\includegraphics[width=0.2\textwidth]{triangulation_Input}
	\hfill
	\includegraphics[width=0.2\textwidth]{triangulation_insertEdgeMidpoints}
	\hfill
	\includegraphics[width=0.2\textwidth]{triangulation_split_1-to-4}
	\hfill
	\includegraphics[width=0.2\textwidth]{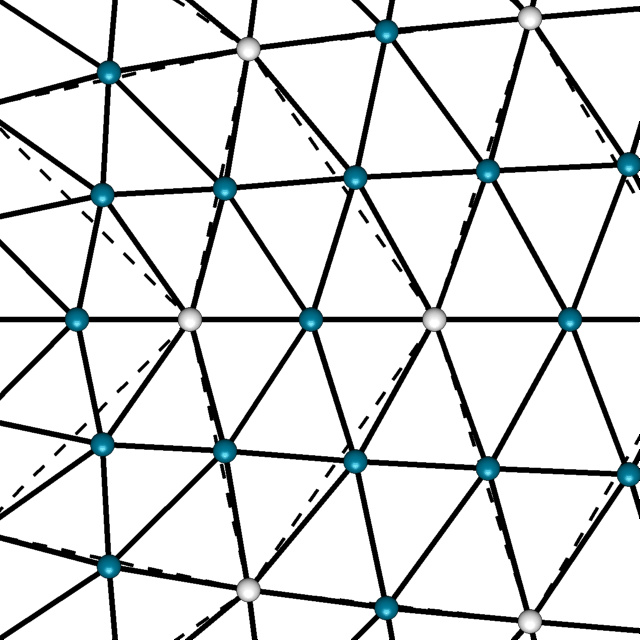}
	\caption{An illustration of the \emph{butterfly} subdivision scheme~\cite{dyn1990butterfly}: input mesh, inserting vertices as edge midpoints, connecting the new vertices, and smoothing the mesh to be more regular.
	White input vertices are retained during smoothing.}
	\label{fig:ButterflySubdivision}
\end{figure}

The \emph{butterfly} subdivision scheme acts based on the same principles~\cite{dyn1990butterfly}.
It differs in the selection of points that contribute to the position of the newly inserted vertices.
See Figure~\ref{fig:ButterflySubdivision} for the result of applying one refinement step of the \emph{butterfly} subdivision scheme to a non-regular triangle mesh.

\begin{figure}
	\includegraphics[width=0.2\textwidth]{triangulation_Input}
	\hfill
	\includegraphics[width=0.2\textwidth]{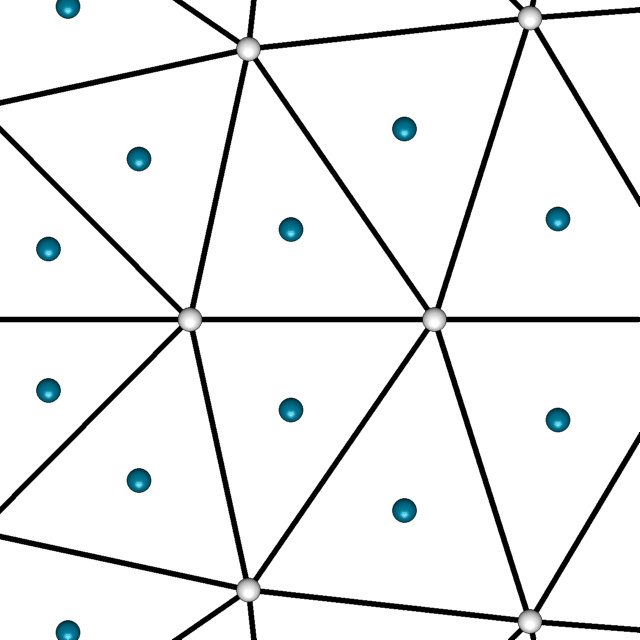}
	\hfill
	\includegraphics[width=0.2\textwidth]{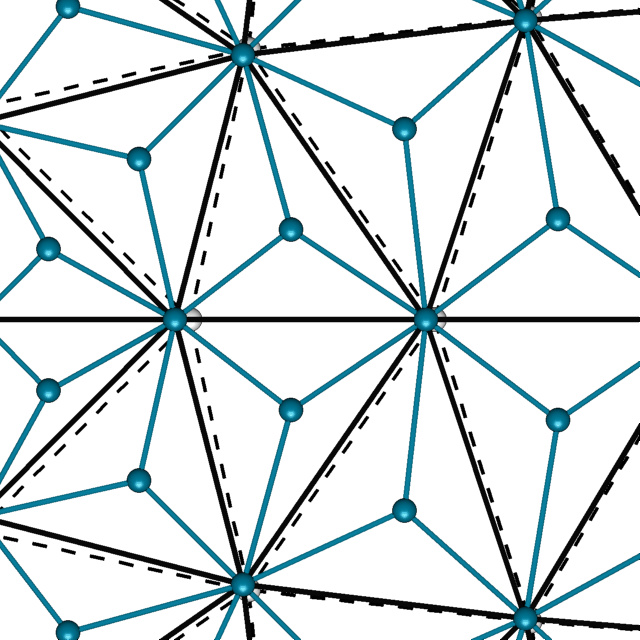}
	\hfill
	\includegraphics[width=0.2\textwidth]{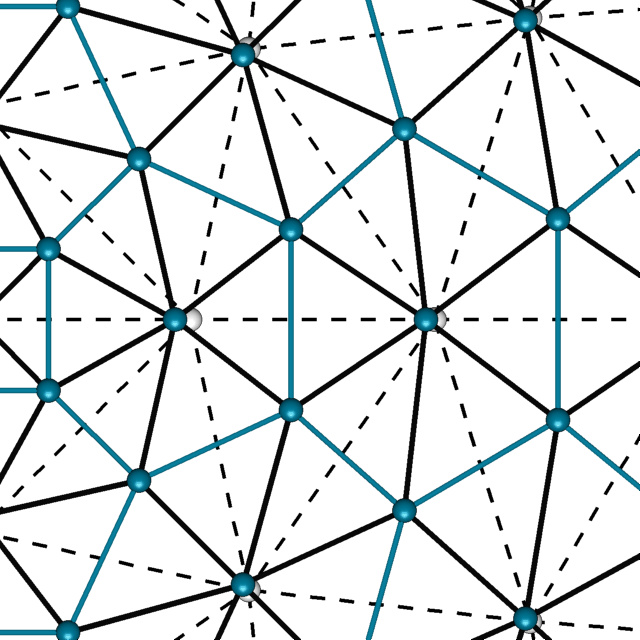}
	\caption{An illustration of the $\sqrt{3}$ subdivision scheme~\cite{kobbelt20003sqrt3}: input mesh, inserting vertices as face barycenters, connecting and smoothing the new vertices, and flipping the original mesh edges.
	White input vertices are moved during smoothing.}
	\label{fig:Sqrt3Subdivision}
\end{figure}

In contrast to these two schemes, consider the $\sqrt{3}$ subdivision scheme~\cite{kobbelt20003sqrt3}.
It operates slightly differently by not creating vertices on the edge midpoints but the triangle barycenters.
These are connected to the three triangle vertices and afterwards, the old edges are flipped, see Figure~\ref{fig:Sqrt3Subdivision}.
Therefore, applying the scheme twice subdivides every input triangle into nine smaller triangles.
Note that all three subdivision schemes listed so far, the \emph{butterfly} and $\sqrt{3}$ subdivision schemes as well as the subdivision scheme by Loop are primal, as they perform face splits.
While the $\sqrt{3}$ subdivision scheme and the subdivision scheme by Loop are approximating, the \emph{butterfly} subdivision scheme is interpolating.

\vspace{0.5cm}

\subsection{Quad and hexagon meshes}

\begin{figure}[b]
	\includegraphics[width=0.2\textwidth]{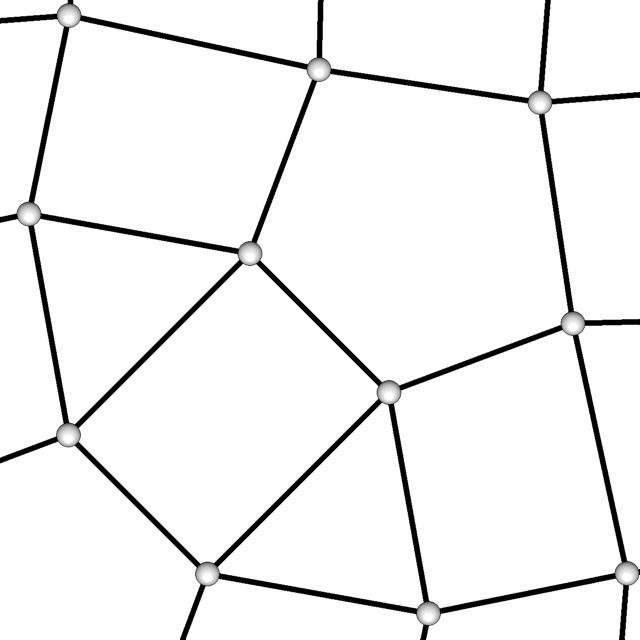}
	\hfill
	\includegraphics[width=0.2\textwidth]{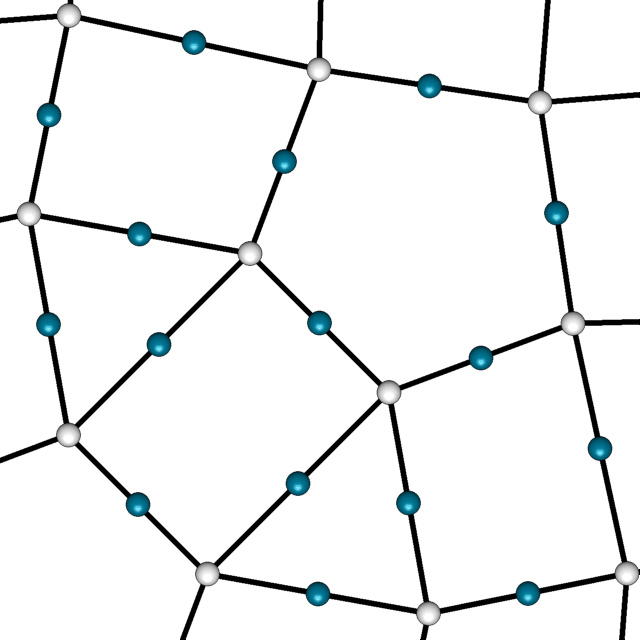}
	\hfill
	\includegraphics[width=0.2\textwidth]{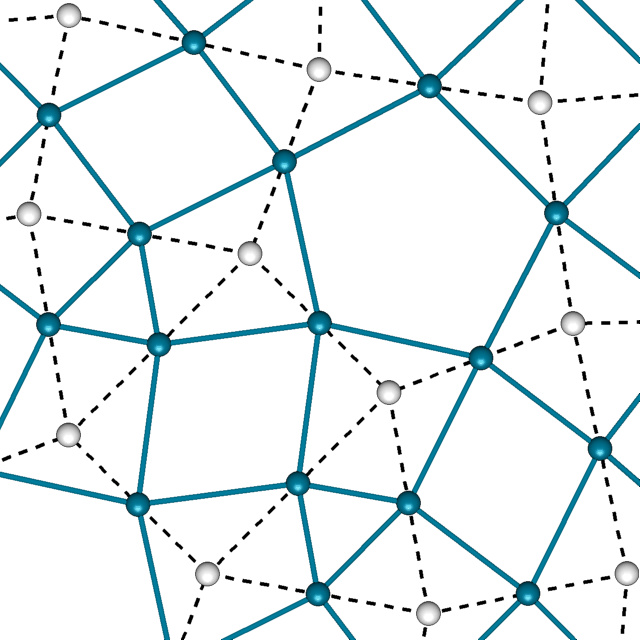}
	\hfill
	\includegraphics[width=0.2\textwidth]{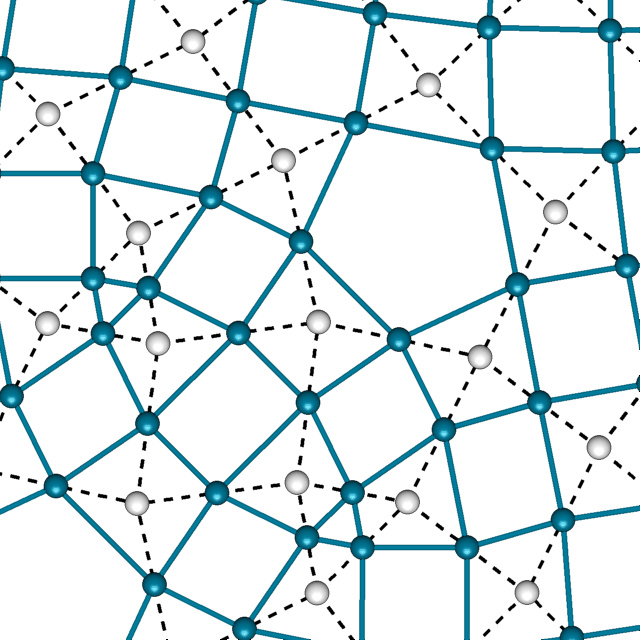}
	\caption{An illustration of the \emph{mid-edge} subdivision scheme~\cite{peters1997simplest}: input mesh, inserting vertices as edge midpoints, and connecting the new vertices. Fourth image shows a second complete refinement step, which illustrates that two steps of the \emph{mid-edge} subdivision scheme are equivalent to one step in the subdivision scheme of Doo-Sabin~\cite{doo1978behaviour}, compare to Figure~\ref{fig:DooSabinSubdivision}.}
	\label{fig:MidEdgeSubdivision}
\end{figure}

Other schemes have been proposed to work on (or create) different mesh types.
The \emph{mid-edge} subdivision scheme was proposed independently by Peters and Reif~\cite{peters1997simplest} as well as by Habib and Warren~\cite{habib1999edge}.
It connects every edge-midpoint to the four midpoints of the edges that share both a vertex and a face with the current edge, see Figure~\ref{fig:MidEdgeSubdivision}.
Thereby, it creates meshes that are quad-\emph{dominant}, i.e., that consist mostly of quadrilateral faces, \rev{at least after applying the scheme a minimum of two times}.
However, vertices and faces of the input mesh, which are of a degree different from four, give rise to faces of the same degree after the refinement step.
For instance, subdivision of a triangle via the \emph{mid-edge} subdivision scheme will at all steps contain a triangular face in the center of the refined mesh.
This scheme is approximating and dual.

The triangle subdivision schemes discussed above do indirectly create a whole different class of meshes.
Namely, their resulting meshes can easily be converted into hexagon-dominant ones.
The procedure to do so simply consists of turning the mesh into its \emph{dual}, i.e., placing a vertex at the center of each face of the mesh, and connect those vertices whose parental faces are connected via an edge.
Thereby, the vertices of the original mesh become faces of the dual, which has as many edges as the degree of the respective vertex.
As the \emph{butterfly} and the $\sqrt{3}$ subdivision schemes and the subdivision scheme by Loop primarily create vertices with degree six\rev{---at least starting from the second refinement step---}, the dual mesh is dominated by hexagons.

It is worth noting at this point that the very first subdivision schemes by Doo-Sabin~\cite{doo1978subdivision,doo1978behaviour} and Catmull-Clark~\cite{catmull1978recursively} are applicable to arbitrary meshes.
The first creates quad-dominant subdivisions, while the second creates meshes that are solely comprised of quadrilaterals.
For a general overview of subdivision schemes see~\cite{sabin2005recent,andersson2010introduction}.

\subsection{Regularity and the absence of pentagonal faces}

The well-known subdivision schemes presented above produce very regular meshes in which all newly inserted vertices have the same degree and all new faces are of a given type (triangles, quadrilaterals, or hexagons).
This regularity does, however, not hold for vertices and faces of the input mesh that are not already of the required type.
\rev{For instance, if a vertex in the input mesh has a degree different from the vertices inserted in the refinement step, it will continue to have this different degree through the whole subdivision procedure or spawn a face of corresponding degree. 
The same is true for faces of the input mesh that have a degree different from the newly inserted faces, they are either retained in the entire refinement step or spawn a vertex of corresponding degree. }
Hence, all irregularities present in the final subdivided mesh are already given with the input mesh. 

\begin{figure}
	\hspace{0.1\textwidth}
	\begin{subfigure}[t]{0.2\textwidth}
		\includegraphics[width=1.\textwidth]{quadDominant_input}
		\caption{Input mesh.}
		\label{fig:InputMeshCatmullClarkDooSabin}
	\end{subfigure}
	\hfill
	\begin{subfigure}[t]{0.2\textwidth}
		\includegraphics[width=1.\textwidth]{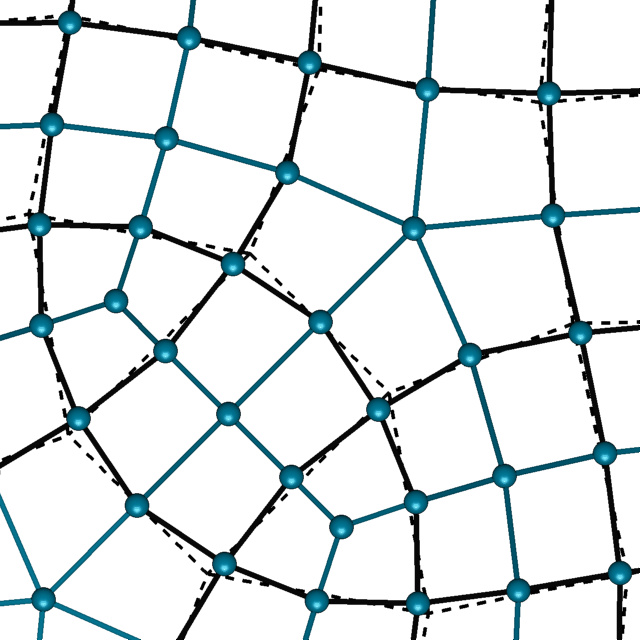}
		\caption{Subdivision by Catmull-Clark~\cite{catmull1978recursively}.}
		\label{fig:CatmullClarkSubdivision}
	\end{subfigure}
	\hfill
	\begin{subfigure}[t]{0.2\textwidth}
		\includegraphics[width=1.\textwidth]{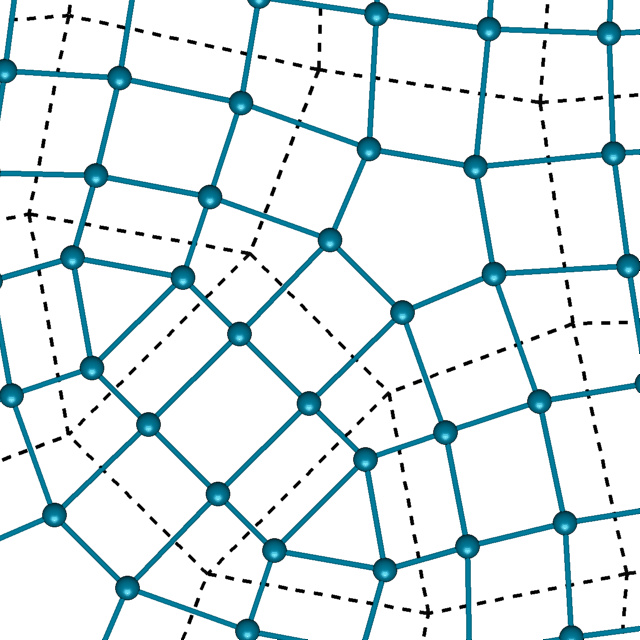}
		\caption{Subdivision by Doo-Sabin~\cite{doo1978behaviour}.}
		\label{fig:DooSabinSubdivision}
	\end{subfigure}
	\hspace{0.1\textwidth}
	\caption{Irregularities of the input mesh retained during the refinement steps. 
		The subdivision scheme of Catmull-Clark~\cite{catmull1978recursively} mainly creates vertices of degree four, but has vertices of different degree wherever the input mesh has a non-quadrilateral face.
		The subdivision scheme of Doo-Sabin~\cite{doo1978behaviour} mainly creates quadrilaterals, but retains non-quadrilateral faces of the input mesh. Also, it is \rev{combinatorially} equivalent to applying the \emph{mid-edge} subdivision scheme twice, confer Figure~\ref{fig:MidEdgeSubdivision}.}
	\label{fig:SingularitiesOfSubdivision}
\end{figure}

This can be observed in Figure~\ref{fig:SingularitiesOfSubdivision} when considering the subdivision schemes of Catmull-Clark~\cite{catmull1978recursively} and Doo-Sabin~\cite{doo1978behaviour} respectively.
The first always creates quadrilateral faces and all vertices inserted after the first refinement step are of degree four.
However, at those places where the input mesh does have a non-quadrilateral face, the scheme creates a vertex with degree different from four.
The second scheme always creates vertices of degree four and all faces inserted after the first refinement step are quadrilaterals.
However, non-quadrilateral faces of the input mesh are retained across all refinement steps.

Note that the subdivision schemes discussed in this section are capable of generating meshes that are dominated or even completely comprised by triangles, quadrilaterals, or hexagons.
In this list, one shape is clearly missing and has not played a role so far: pentagons.
Therefore, in the next section, we will consider a subdivision scheme that creates a pure pentagon mesh from a given input mesh.


\section{The pentagon snub subdivision scheme}
\label{sec:PentagonSnubRefinement}

\rev{
	Given an arbitrary input, the subdivision scheme we investigate in the following produces a mesh consisting solely of pentagons. Vertices newly introduced in a refinement step have either degree three or five, while all vertices originally given in the input keep their degrees throughout all refinement steps applied.
	We originally approached this subdivision scheme from the context of cutting operations executed on Platonic or Archimedean solids.
	Thus, we call it \emph{pentagon snub subdivision scheme} and will explain this naming later in this section.
	The version of the pentagon snub subdivision scheme we will consider here consists of four steps which are listed in the following and will be explained in more detail afterwards.
}

%
%
%

Let~$\mathcal{M}_0$ be a 2-manifold mesh in the plane, consisting of convex faces only and having no self-intersections. 
For later use, we subdivide both the edge set~$E$ and the vertex set~$V$ into two subsets---the outer and inner edges or vertices, respectively. 
An edge is called an \emph{inner edge} if it is incident to exactly two faces, otherwise it is called an \emph{outer edge}. 
A vertex is called an \emph{inner vertex} if it is incident to inner edges only, it is called an \emph{outer vertex} otherwise.

In each refinement step~${t\in\mathbb{N}}$ of \rev{our version of the} subdivision scheme, the following four operations are applied to~$\mathcal{M}_{t}$ in order to obtain the refined mesh~$\mathcal{M}_{t+1}$:
\begin{enumerate}
	\item \textbf{Insertion of vertices and edges}: 
	Each edge~$e$ in~$\mathcal{M}_t$ is replaced by three edges in the form of a stretched ``Z'' (henceforth called a \emph{Z-triplet}). 
	Hence, two vertices are added as well. 
	The first and last vertex of the Z-triplet correspond to the two vertices of~$e$. 
	\item \textbf{Insertion of face barycenters}: 
	To each face~$f$ of~$\mathcal{M}_t$, a new vertex is added at the barycenter of~$f$.
	\item \textbf{Insertion of edges}: 
	Iterate over all edges~$e$ of~$\mathcal{M}_t$.
	If edge~$e$ was an inner edge of~$\mathcal{M}_t$, both new vertices (created during operation~1) are connected to the \rev{respectively closest} barycenter (created during operation~2) of a face incident to~$e$.
	If~$e$ is an outer edge, only one of the newly created vertices is connected to the barycenter of the face incident to~$e$.
	\item \textbf{Smoothing operation}: Simultaneously, the position of each inner vertex is revised using the \rev{barycenters of the faces incident} to it.
\end{enumerate}

\rev{A variant of this subdivision scheme including the first three operations was proposed by Bowers and Stephenson in 1997~\cite{bowers1997regular} and discussed by Akleman et al.~\cite{akleman2004semiregular} while combinatorial aspects are investigated by Yan~\cite[Sec.~3.2]{yan2019pentagonal}.}
Furthermore, a later subsection will relate the first three operations to a notation due to Conway~\cite{conway2016symmetries}, which motivates the naming of the subdivision scheme.
In the following, we will discuss and illustrate these operations in more detail.
Furthermore, we will justify the necessity of smoothing operation.

\subsection{Insertion of vertices and edges}

\begin{figure}
	\centering
	\begin{subfigure}[t]{.23\textwidth}
		\centering
		\includegraphics[width=0.6in]{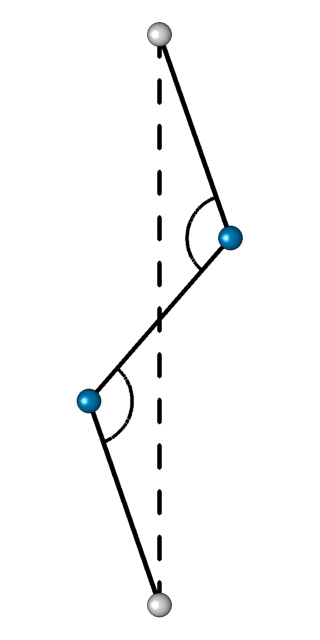}
		\subcaption{Replacing an edge by a Z-triplet.}
		\label{fig:Z-triplet}
	\end{subfigure}
	~
	\begin{subfigure}[t]{0.23\textwidth} 
		\centering
		\includegraphics[width=1.3in]{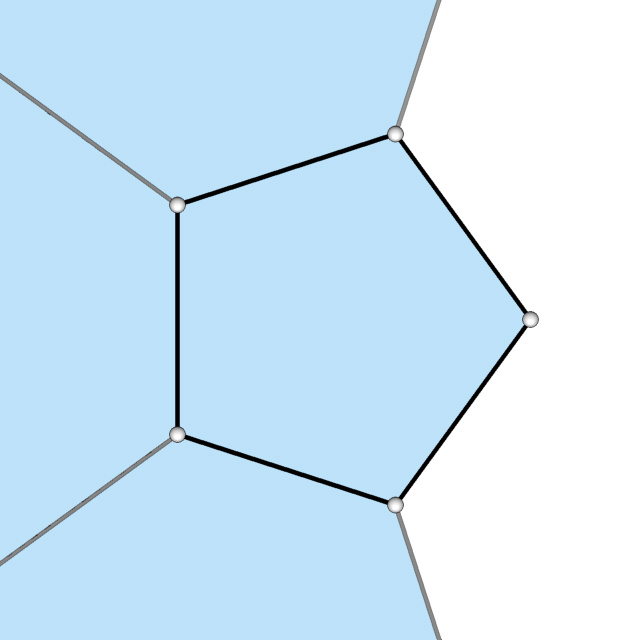}
		\subcaption{Input mesh~$\mathcal{M}_0$, faces in blue.}
       	\label{fig:InputMesh}
	\end{subfigure}
	~ 
	\begin{subfigure}[t]{0.23\textwidth} 
		\centering
		\includegraphics[width=1.3in]{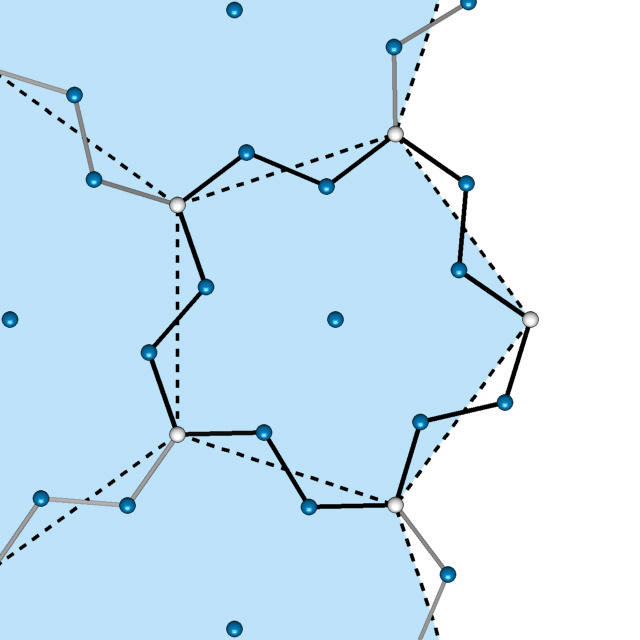}
		\subcaption{Result after operations 1 and 2.}
		\label{fig:Ops1and2}
	\end{subfigure}
	~
	\begin{subfigure}[t]{.23\textwidth}
		\centering
		\includegraphics[width=1.3in]{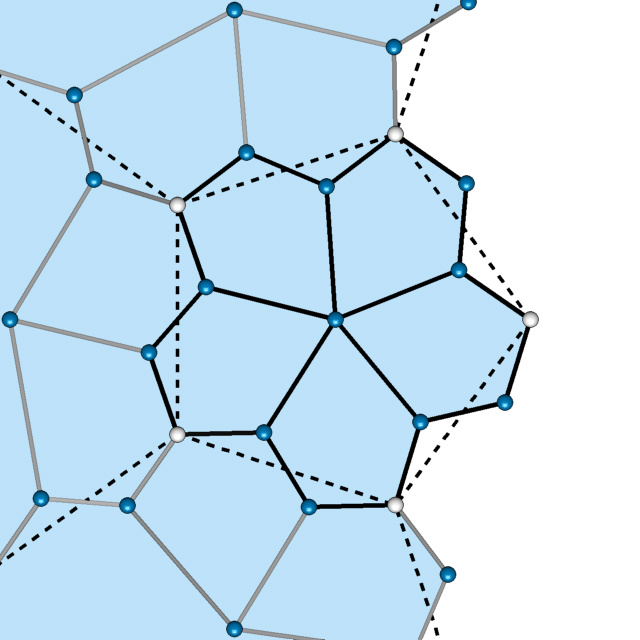}
		\caption{Result after also applying operation 3.}
		\label{fig:Op3}
	\end{subfigure}
	\caption{
		Illustration of a Z-triplet with its two enclosed~$\tfrac{2\pi}{3}$ angles, where the original line that is replaced by the Z-triplet is shown dashed. 
		Also shown is the input mesh~$\mathcal{M}_0$ with its faces colored blue and a boundary to the right, i.e., two inner and three outer vertices are shown. 
		The last two images show operations~1--3 applied in each refinement step: replacing edges by Z-triplets, inserting face barycenters, and connecting new vertices.}
	\label{fig:SnubRefinementSteps}
\end{figure}

During operation~1, every edge is replaced by a Z-triplet in which all edges have equal lengths, see Figure~\ref{fig:Z-triplet}.
The smaller angle enclosed by two edges of the Z-triplet is equal to~$\tfrac{2\pi}{3}$. 
The newly inserted inner vertices have degree three after each refinement step.
\rev{Therefore, this choice of the angle allows for an equal partition of~$2\pi$ among all three angles.
However, in general, the other two angles are not necessarily equal to~$\tfrac{2\pi}{3}$.}
The first and last vertex of each Z-triplet (colored white in Figure~\ref{fig:Z-triplet}) are identical to the vertices of the edge it replaces. 
Hence, the edge length of the Z-triplet depends on the length of the replaced edge, but is completely determined by the angle choice. 
Within operation~2, a new vertex is added at the barycenter of every face of~$\mathcal{M}_t$.
See Figure~\ref{fig:InputMesh} for an example of an input mesh~$\mathcal{M}_0$ and Figure~\ref{fig:Ops1and2} for the result of applying operations~1 and~2 on the input once.

The Z-triplets are inserted always in the same way, i.e., running through the boundary of a face, the newly inserted faces appear to the left and to the right alternately. 
This decision is made in the very beginning for one edge.
Propagating the decision along the mesh fixes the setup for the entire procedure.

During operation~3, we iterate over all edges~$e$ of the mesh~$\mathcal{M}_t$. 
For each edge~$e$, the line running through~$e$ divides the plane into two half-planes.
If~$e$ is an inner edge, each half-plane contains exactly one of the faces incident to~$e$.
Furthermore---by construction---each half-plane also contains exactly one of the newly created vertices from operation~1.
These new vertices are connected to the face midpoint (created during operation~2) lying in the same half-plane. 
If~$e$ is an outer edge, exactly one of the newly created vertices lies in the same half-plane as the face~$e$ is incident to. 
Then, this vertex and this face midpoint are connected.
See Figure~\ref{fig:Op3} for an illustration of the result of applying operation~3.

Note that the scheme creates only pentagonal faces.
After operations~1--3 in the first refinement step, vertices from the Z-triplets have degree three, while a vertex that was created as barycenter of a face~$f$ has degree~$n_f$, where~$n_f$ denotes the number of vertices of the face~$f$.
Like in the other subdivision schemes discussed in Section~\ref{sec:SubdivisionSchemes}, these vertices, stemming from the faces of the original mesh, will always keep their degree and not become regular.
\rev{All newly created outer vertices always have degree two or three while outer vertices of higher degree that are already present in the input mesh retain this degree throughout the subdivision process.}
From the second refinement step onward, all newly inserted vertices have degree three or five.

In case of applying the subdivision scheme to a single pentagon as input~$\mathcal{M}_0$, one gets the following recursion result for the number of vertices, edges, and faces of the mesh~$\mathcal{M}_{t+1}$, after~$t$ refinement steps, in terms of the vertex set~$V_t$, the edge set~$E_t$, and the face set~$F_t$ of~$\mathcal{M}_t$:
\begin{align}
\label{equ:PentagonCombinatorics}
\left|V_{t+1}\right| = \left|V_t\right| + 2\left|E_t\right| + \left|F_t\right|, 
\quad \left|E_{t+1}\right| = 3\left|E_t\right| + 5\left|F_t\right|,
\quad \left|F_{t+1}\right| = 5\left|F_t\right|,
\end{align}
while the following holds for an arbitrary input mesh:
\begin{align}
\label{equ:ArbitraryInputCombinatorics}
\left|V_{t+1}\right| = \left|V_t\right| + 2\left|E_t\right| + \left|F_t\right|, 
\quad \left|E_{t+1}\right| = 3\left|E_t\right|+ \sum_{f\in F_t} n_f, 
\quad \left|F_{t+1}\right| = \sum_{f\in F_t} n_f.
\end{align}
Thus, by iteratively applying operations~1--3, we have created a subdivision scheme that creates pentagon meshes from an arbitrary input.
\rev{So far, the operations described here match the approach of Bowers and Stephenson~\cite{bowers1997regular}, Yan~\cite{yan2019pentagonal}, and Akleman et al.~\cite{akleman2004semiregular} combinatorially.}
However, \rev{embeddings of the resulting meshes, when following the steps as outlined above, generally} lack an important \rev{geometric} property, which we will address in the following.

\subsection{The smoothing operation}

\begin{figure}
	\centering
	\begin{subfigure}[b]{0.3\textwidth}
		\centering
		\includegraphics[width=1.\textwidth]{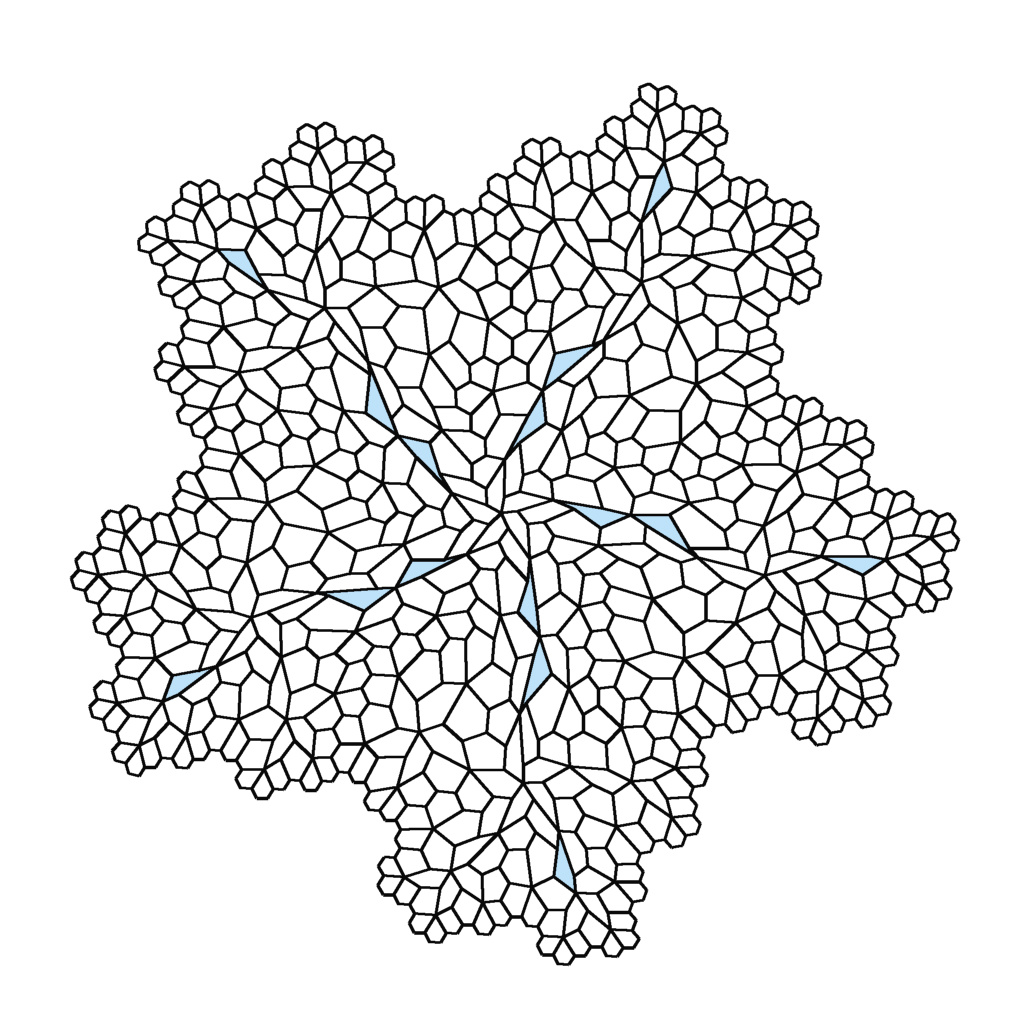}
		\caption{Non-smoothed refinement, applying operations 1--3 four times.}
		\label{fig:NonSmoothRefinement}
	\end{subfigure}
	~
	\begin{subfigure}[b]{0.3\textwidth}
		\centering
		\includegraphics[width=0.9\textwidth]{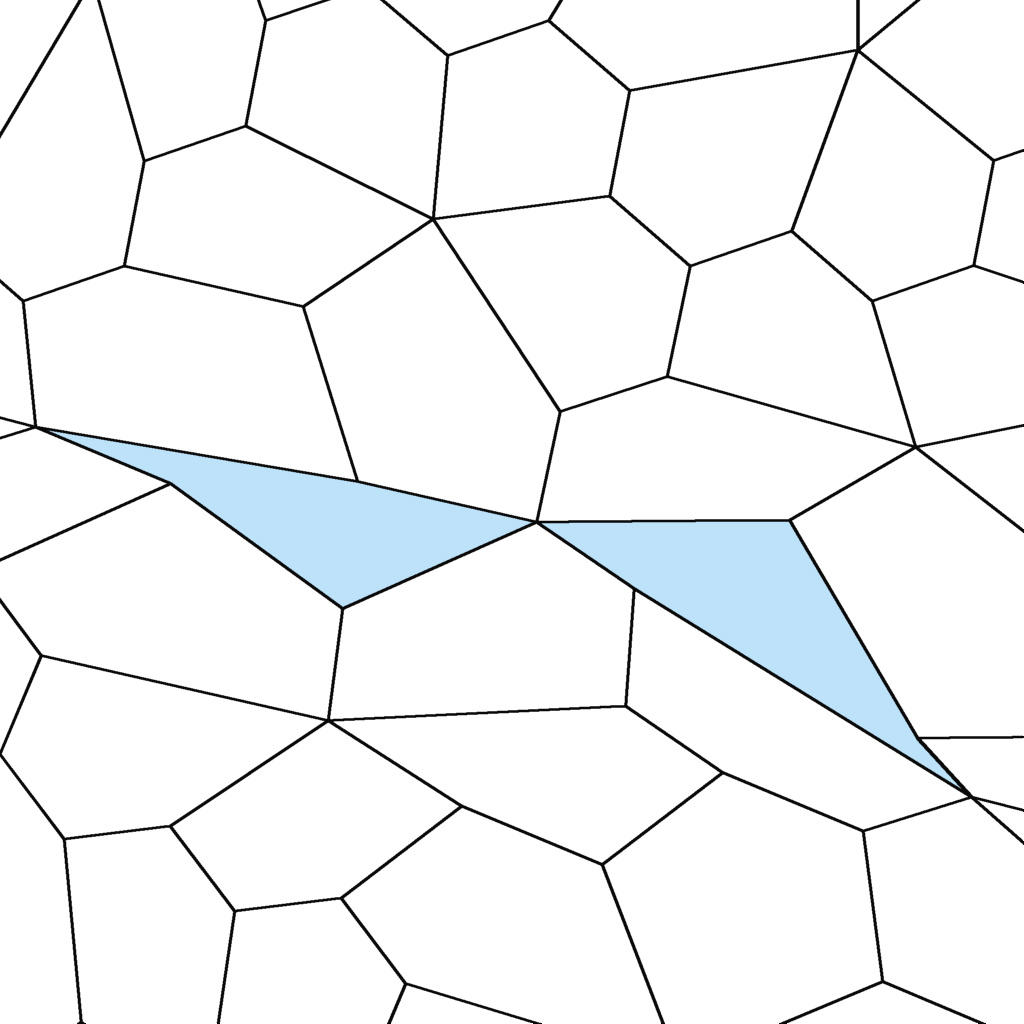}
		\caption{Detail of non-smoothed refinement, which results in non-convex faces.}
		\label{fig:NonSmoothRefinementDetail}
	\end{subfigure}
	~
	\begin{subfigure}[b]{0.3\textwidth}
		\centering
		\includegraphics[width=1.\textwidth]{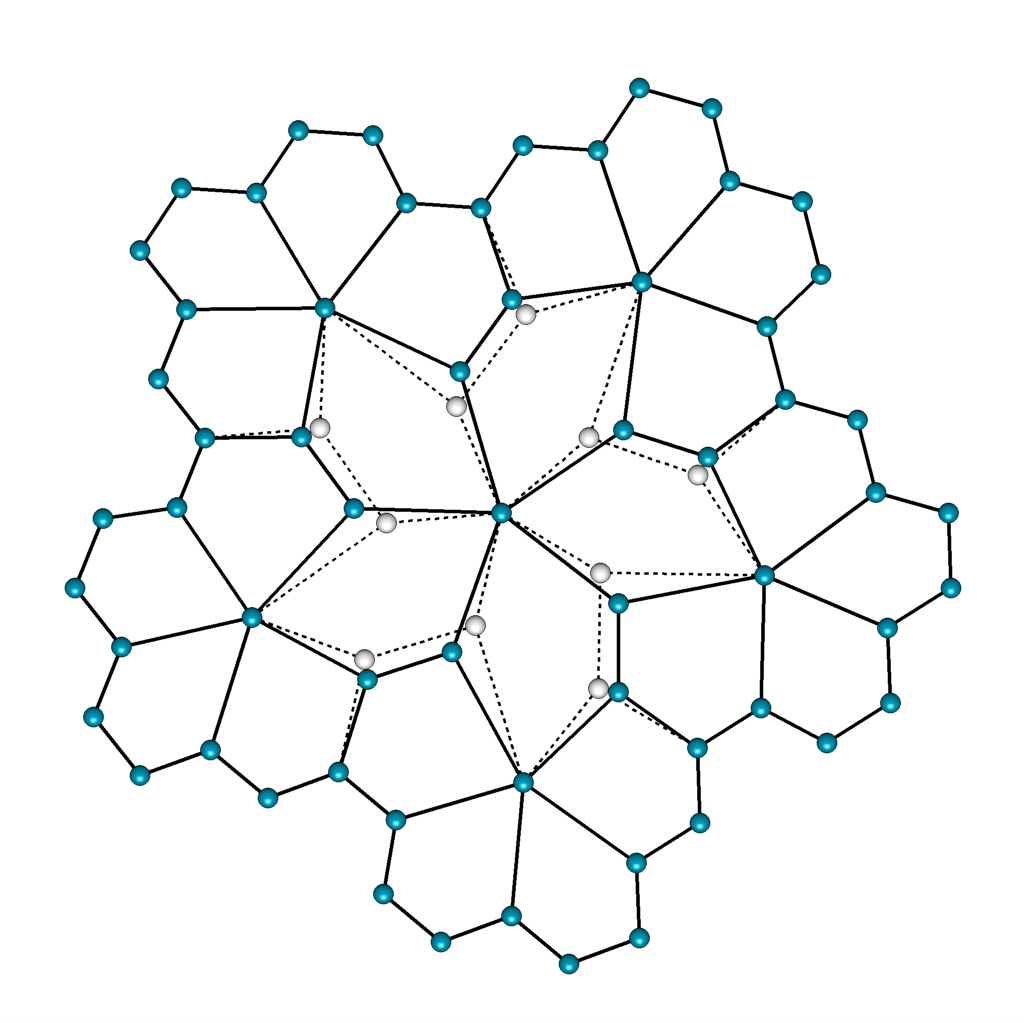}
		\caption{Illustration of the smoothing: white vertices are moved to blue ones.}
		\label{fig:SmoothRefinement}
	\end{subfigure}
	\caption{Simply applying operations 1--3 as given above results in non-convex faces. A smoothing operation prevents this by nudging faces into a more convex position.}
	\label{fig:smoothingOp}
\end{figure}

As it can be seen in the blue highlighted faces in Figure~\ref{fig:NonSmoothRefinement}, after several refinement steps, some of the newly created pentagons contain edges which are---in comparison to the other edges of the same face---relatively long. 
Also, some of the new faces are not convex any more (see Figure~\ref{fig:NonSmoothRefinementDetail}, non-convex faces colored in blue).
\rev{Because of the focus on combinatorial aspects, both the work of Bowers and Stephenson~\cite{bowers1997regular} and that of Yan~\cite{yan2019pentagonal} do not address this.
Akleman et al.~\cite{akleman2004semiregular} do propose a second, different procedure that ensures convexity, which we will discuss after presenting our approach to preserving convexity of the faces.
}

\rev{To overcome the creation of non-convex faces and to avoid intersections of new edges, we introduce a smoothing operation that is applied to the vertices as operation~4 in each refinement step.}
However, this operation is not applied to all vertices.
If a vertex is an outer vertex, its position is maintained, while the inner vertices are possibly moved to a new position. 

The smoothing is performed as follows.
For each face~$f$ of the mesh~$\mathcal{M}_t$, as obtained after operation~3, its barycenter~$b_f$ is computed.
Then, for each inner vertex~$v$, its new position is determined as the average of the face barycenters~$b_f$ of those faces~$f$ to which~$v$ is incident.
After determining the new positions, all vertices are updated at once. 
Since all inner vertices may be moved in the smoothing operation, the scheme is not interpolating.
\rev{See Figure~\ref{fig:SmoothRefinement} for an illustration of the smoothing operation.}

\begin{figure}[b]
	\centering
	\begin{subfigure}[b]{0.23\textwidth} 
		\centering
		\includegraphics[width=1.2in]{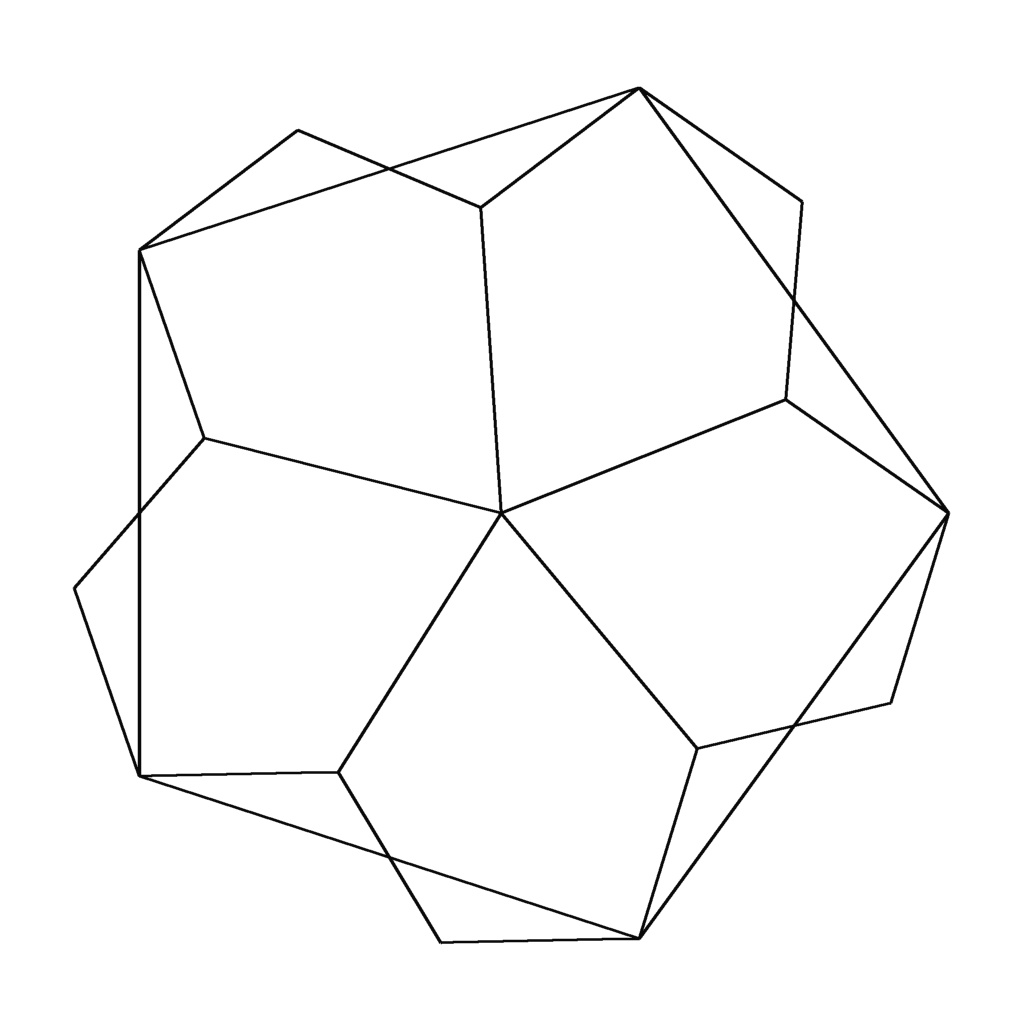}
	\end{subfigure}
	~ 
	\begin{subfigure}[b]{0.23\textwidth} 
		\centering
		\includegraphics[width=1.2in]{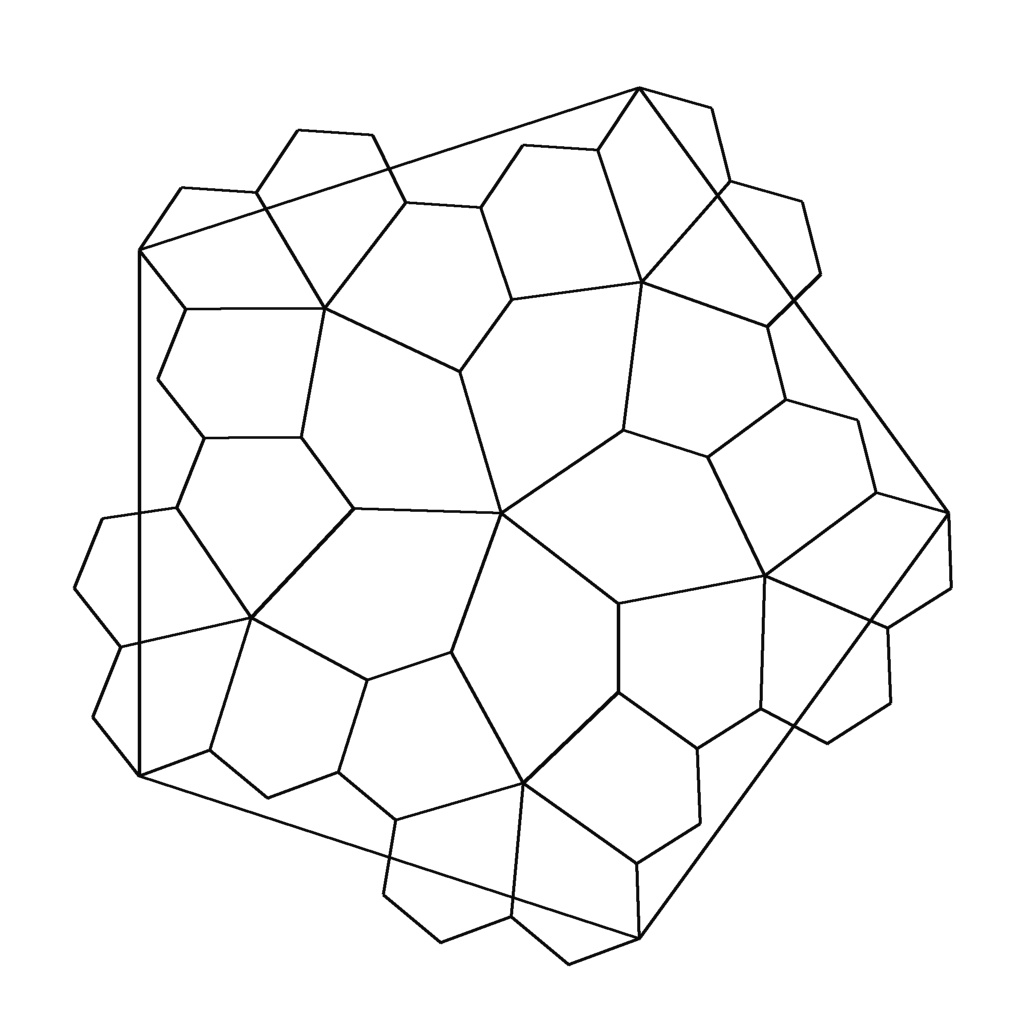}
	\end{subfigure}
	~
	\begin{subfigure}[b]{.23\textwidth}
		\centering
		\includegraphics[width=1.2in]{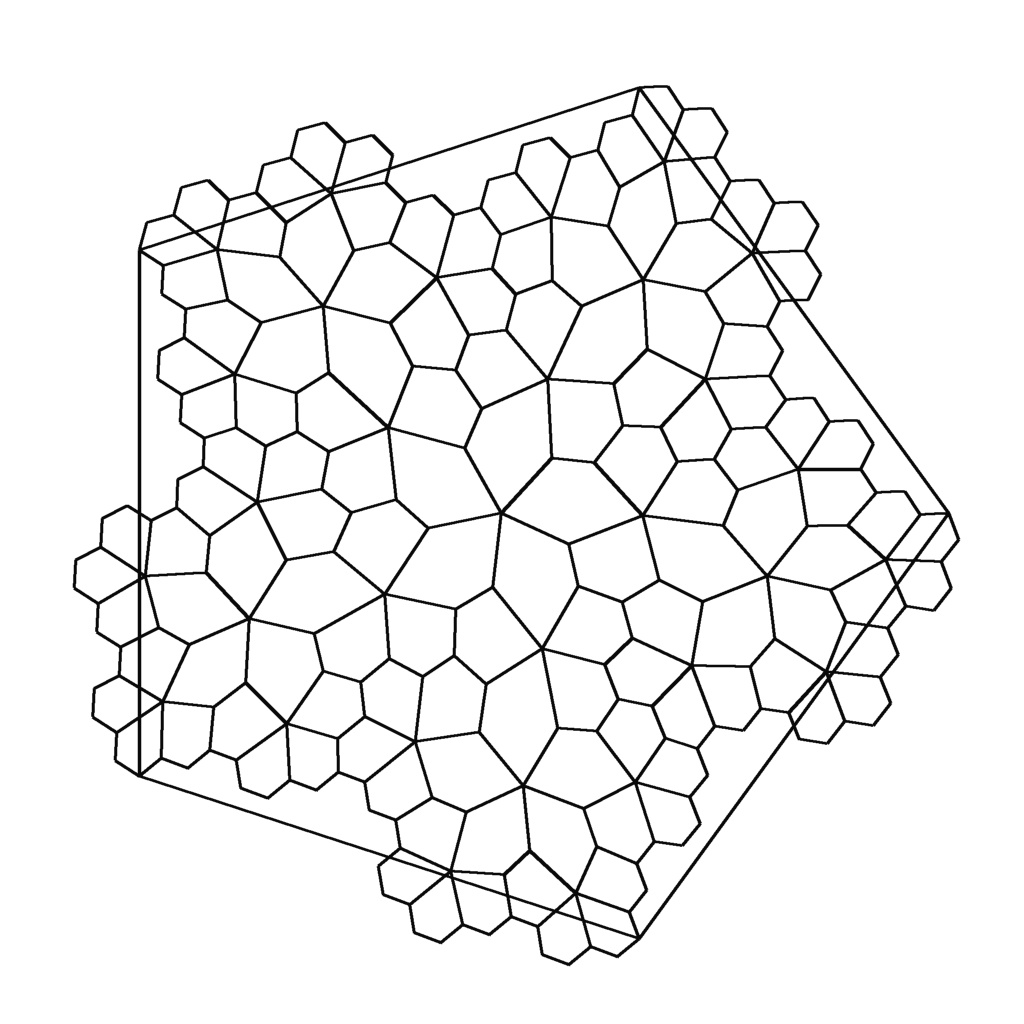}
	\end{subfigure}
	~
	\begin{subfigure}[b]{.23\textwidth}
		\centering
		\includegraphics[width=1.2in]{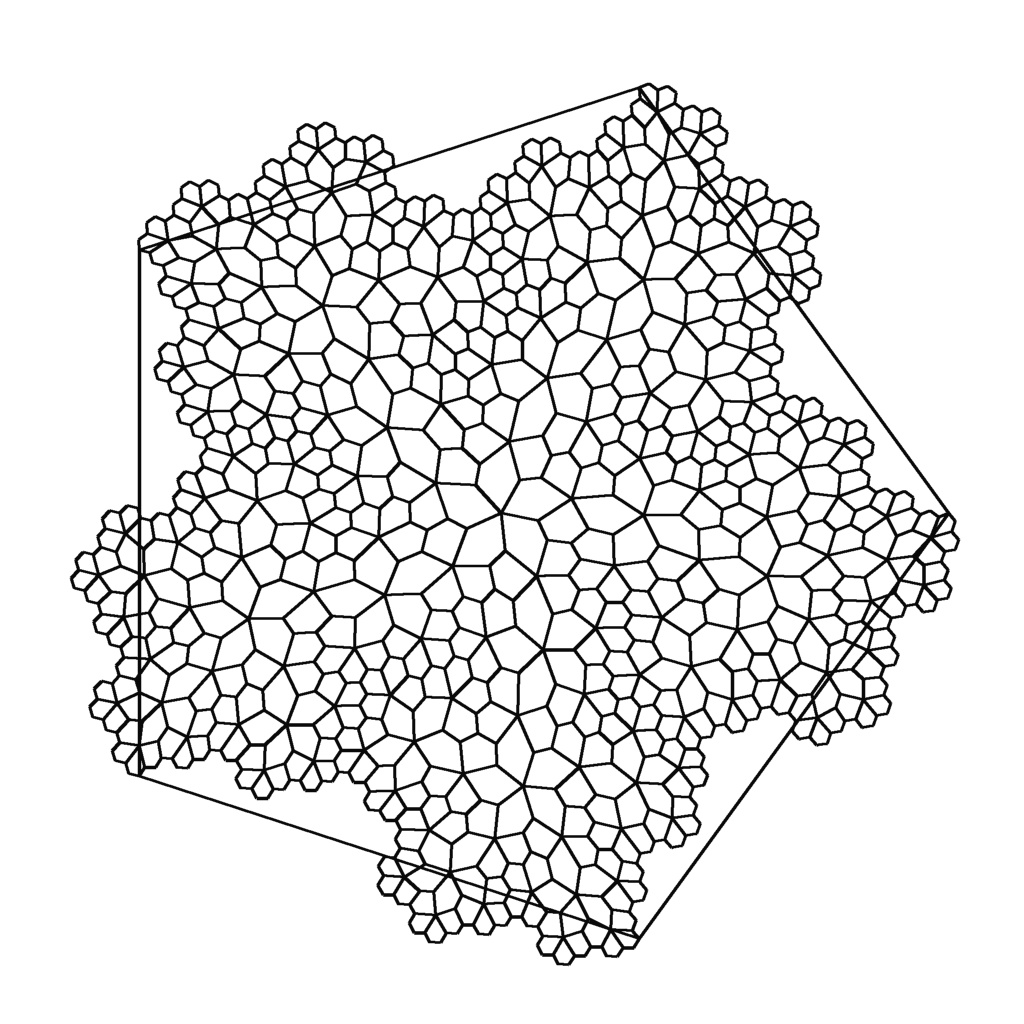}
	\end{subfigure}
	\caption{From left to right: operations~1--4 of pentagon snub refinement with smoothing, applied to an initial pentagon (shown in the background).
	Note the creation of more regular faces, compared to Figure~\ref{fig:NonSmoothRefinement}.}
	\label{fig:FirstSnubRefinementSteps}
\end{figure}

Regular pentagons do not tessellate the plane---therefore, the created pentagons cannot all be regular.
Figure~\ref{fig:FirstSnubRefinementSteps} shows the first four refinement steps of the presented subdivision scheme, applied to a regular pentagon. 
Since the input pentagon does not induce a vertex with degree different from three or five, all inner vertices have degree three or five after every refinement step---some inner pentagons have one vertex with degree five and four with degree three, some have two vertices with degree five and three with degree three.
In 2D, the presented refinement scheme has the following properties: it is primal (faces split into faces), approximates the vertices of the original mesh because of the smoothing operation, and interpolates the outer vertices of the input mesh.

\begin{figure}
	\centering
	\begin{minipage}[b]{.23\textwidth}
		\includegraphics[width=1.4in]{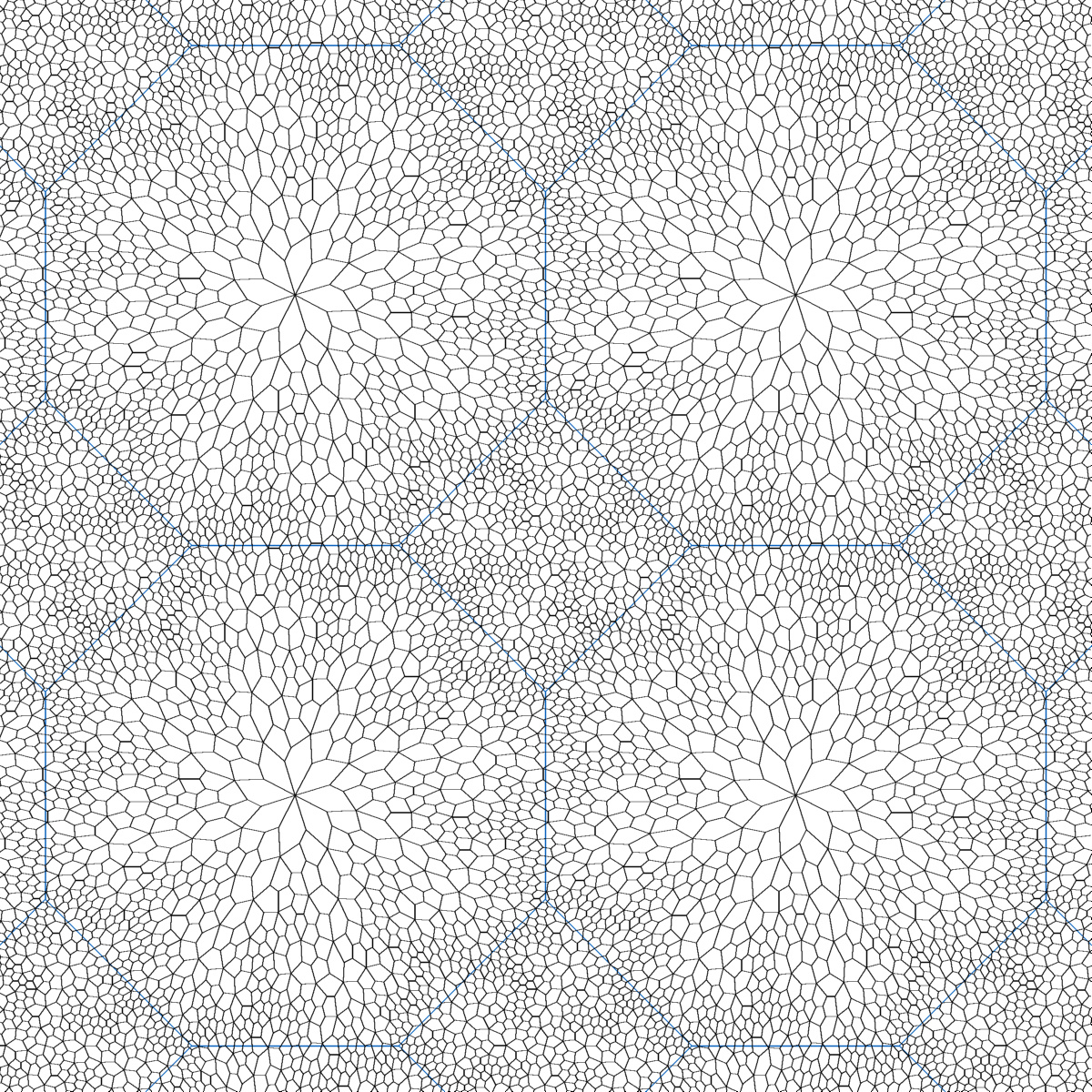}
	\end{minipage}
	~
	\begin{minipage}[b]{.23\textwidth}
		\includegraphics[width=1.4in]{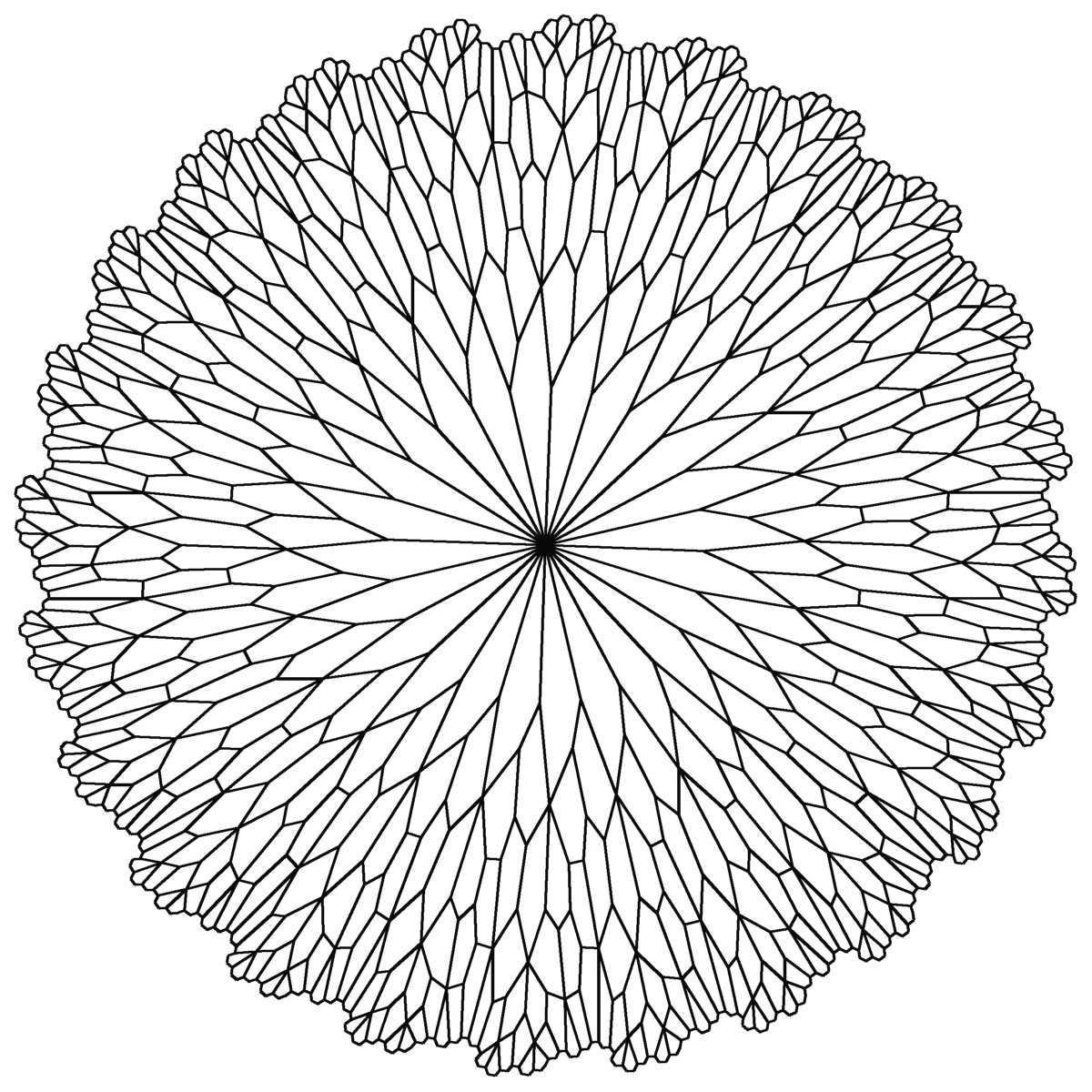}
	\end{minipage}
	~
	\begin{minipage}[b]{.23\textwidth}
		\includegraphics[width=1.4in]{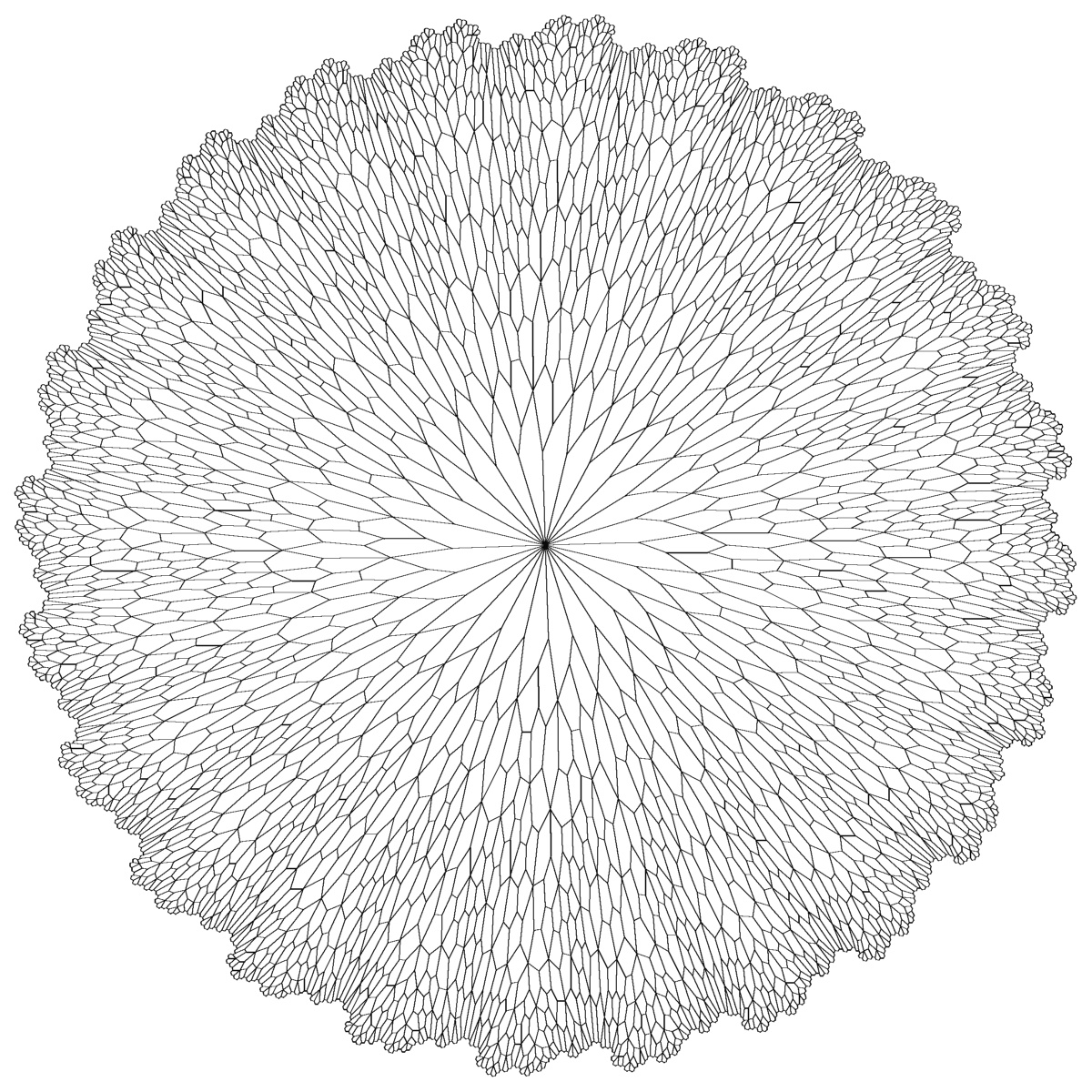}
	\end{minipage}
	~
	\begin{minipage}[b]{.23\textwidth}
		\includegraphics[width=1.4in]{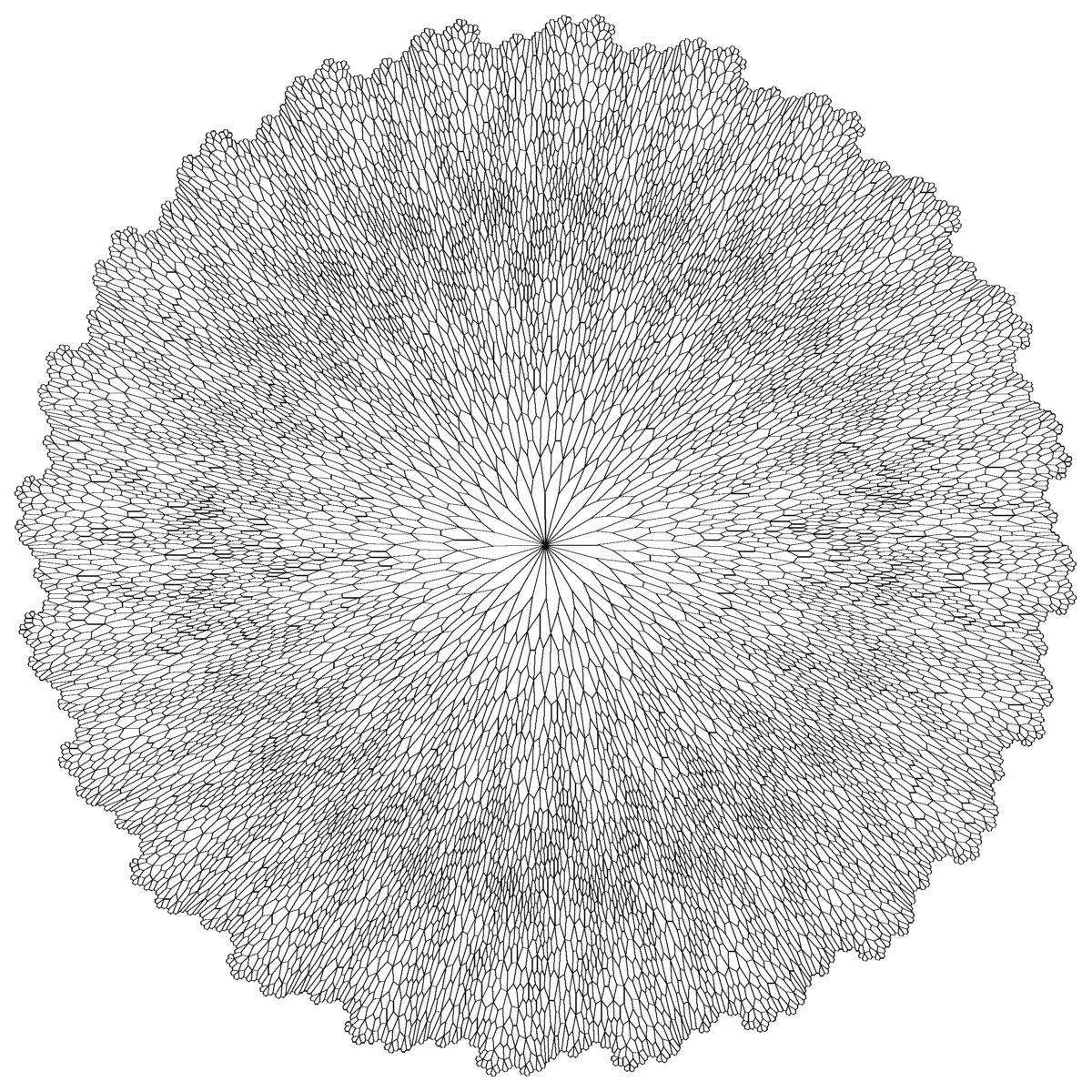}
	\end{minipage}
	\caption{From left to right: four times subdivided 8-4-mesh, three times subdivided regular 24-gon, four times subdivided regular 24-gon, and three times subdivided triangulated 24-gon.}
	\label{fig:snubRefinementNonPentagonal}
\end{figure}

While the proposed smoothing operation cannot guarantee convex faces and the avoidance of edge-intersections, we have not encountered any such behavior in our experiments with reasonably regular input meshes.
As stated in Equation~(\ref{equ:ArbitraryInputCombinatorics}), our subdivision scheme is applicable to arbitrary input meshes, see several examples in Figure~\ref{fig:snubRefinementNonPentagonal}, all of which exhibit convex faces.
\rev{Our work with various examples motivates the following question, which is posed as a problem for further investigation:}

\begin{question}
	\rev{Does the smoothing operation always keep tiles convex?}
\end{question}

\rev{In the above}, we have approached the creation of pentagonal faces from the perspective of subdivision schemes in two dimensions.
\rev{
	As briefly stated above, in the context of searching for a ``regular'' pentagonal tiling of the plane, Bowers and Stephenson introduce a pentagonal subdivision scheme that splits a given pentagon into six smaller ones, see~\cite[Fig.~4]{bowers1997regular}.
	The obtained tiling has a direct relation to circle packings and allows for several generalizations.
	Relaxing one of their parameters yields a result with the same combinatorial structure as the subdivision scheme presented above (compare Figure~\ref{fig:Op3} to \cite[Fig.~13(a)]{bowers1997regular}).
	This combinatorial structure is investigated by Yan~\cite{yan2019pentagonal}.
	For both works~\cite{bowers1997regular,yan2019pentagonal}, it is unclear how to embed the obtained meshes, while our approach creates a corresponding embedding on the fly.
}

\rev{
	Our fourth operation was added in order to obtain as-convex-as-possible faces.
	Note that Akleman et al.\@ present a different approach towards this end, which they denote as ``Pentagon preserving algorithm``~\cite[Sec.~5.2]{akleman2004semiregular}.
	They propose to split all edges at their midpoints and move a copy of these midpoints towards the involved face centers.
	This results in the creation of scaled, smaller faces within each original face of the mesh.
	By construction, these are convex}\footnote{\rev{The smaller copies of each face are convex by congruency while the other parts cannot include a vertex angle larger or equal to~$\pi$ without violating the property that each point of the smaller part has to be an element of the convex original face.}}, \rev{however, they are only pentagonal if their corresponding original face was a pentagon.
	This is in contrast to the scheme presented here, which always creates pentagonal faces from arbitrary input meshes.
}

\rev{Instead of taking a two-dimensional tiling- or subdivision-based approach}, faces can also be created from a three-dimensional perspective when considering operations performed on polyhedra.
In their book, Conway, Burgiel, and Goodman-Strauss present a naming scheme for Archimedean and Catalan solids~\cite[Ch.~21]{conway2016symmetries}.
The \rev{first three operations of the} scheme discussed here correspond \rev{combinatorially} to the process of forming the dual of the snub dodecahedron, written as~\verb|dsD| in their notation, \rev{and also know as \emph{pentagonal hexacontahedron}, one of the \emph{Catalan solids}}.
This \rev{relation to polyhedra} motivates our naming choice for the subdivision scheme and explains the title of this section.
George Hart collects an expansion of the proposed notation on a website~\cite{hart1998conway}.
\rev{Note that these symmetry-preserving operations can be cast into an even more general framework~\cite{brinkmann2017comparing}.}
In the terminology of Hart, the subdivision scheme presented above corresponds to the gyro of the dodecahedron, written as~\verb|gD|.

\rev{
	Thus, the combinatorial aspects of operations 1--3 of the presented subdivision scheme have their roots in literature of both tilings and polyhedra.
	The additional element we added is of geometric nature, by adding the smoothing operation, aiming at embeddings whose faces are always convex.
}
\rev{In the following}, we will keep the two-dimensional perspective and further explore aspects of continued refinement steps.

\section{Fractal-like}

The last image shown in Figure~\ref{fig:FirstSnubRefinementSteps} gives the impression that the presented subdivision scheme does not only create meshes composed of pentagons, but---for a single input pentagon---additionally results in an object resembling a fractal.
\rev{A similar observation has been made by Bowers and Stephenson for their geometric embedding~\cite[Part~III]{bowers1997regular}.}
Recall that a \emph{fractal} is a naturally or artificially created object characterized by several properties: self-similarity, having fractal dimension, and being produced by an iterative procedure~\cite[Ch.~3]{mandelbrot1983fractal}.
Because of the iterative nature of the subdivision scheme discussed in Section~\ref{sec:PentagonSnubRefinement}, attention should be paid to the first two properties. 

In terms of self-similarity, the object resulting from the subdivision scheme fulfills this property in a broad manner---after the first refinement step, all faces are pentagons.
Therefore, from the second refinement step onward, all faces and their replacements are combinatorially self-similar.
However, due to the smoothing operation, this self-similarity does not hold geometrically, i.e., the pentagons cannot be scaled to be pairwise congruent.

\begin{figure}[hb]
	\centering
	\begin{minipage}[c]{.45\textwidth}
		\centering
		\includegraphics[width=2in]{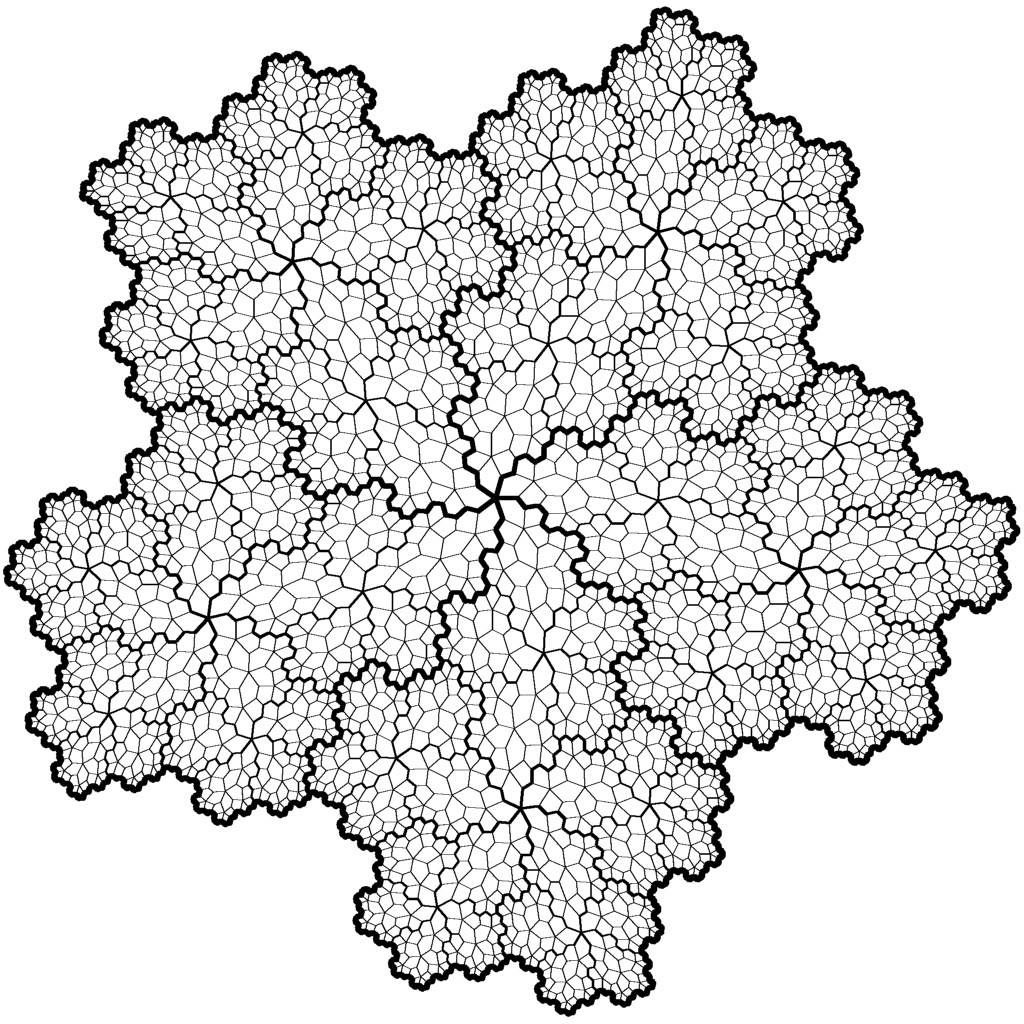}
	\end{minipage}
	~
	\begin{minipage}[c]{.45\textwidth}
		\begin{tikzpicture}[scale=0.5]
			\draw[dashed] (0,0) -- (10,0);
			\draw[thick] (0,0) -- (3.5714,1.23) -- (6.4286,-1.23) -- (10,0);
			\draw[thick,fill=likewhite] (0,0) circle(0.15);
			\draw[thick,fill=likewhite] (10,0) circle(0.15);
			\draw[thick,fill=someblue] (3.5714,1.23) circle(0.15);
			\draw[thick,fill=someblue] (6.4286,-1.23) circle(0.15);
			\draw[thick] (2,0) arc (0:19.1:2cm);
			\node(alpha) at (1.5,0.2){$\alpha$};
			\node(a) at (1.8,1){$F$};
		\end{tikzpicture}
		
		\vspace{3mm}
		
		\begin{tabular}{llll}
			$\bigtriangledown:$ & $\text{rot}(\alpha)$, & $\triangle:$ & $\text{rot}(-\alpha)$,\\
			$+:$ & $ \text{rot}\left( \tfrac{\pi}{3} \right)$, & $-:$ & $\text{rot}\left( -\tfrac{\pi}{3} \right)$,\\
			\multicolumn{4}{c}{$F \rightarrow \bigtriangledown F - F + F \triangle$}
		\end{tabular}
	\end{minipage}
	\caption{Fractal curves contained in the edge set after five refinement steps; corresponding Lindenmayer system, for which~$\alpha$ can be calculated from the known quantities.}
	\label{fig:FractalCurvesAndLindenmayer}
\end{figure}
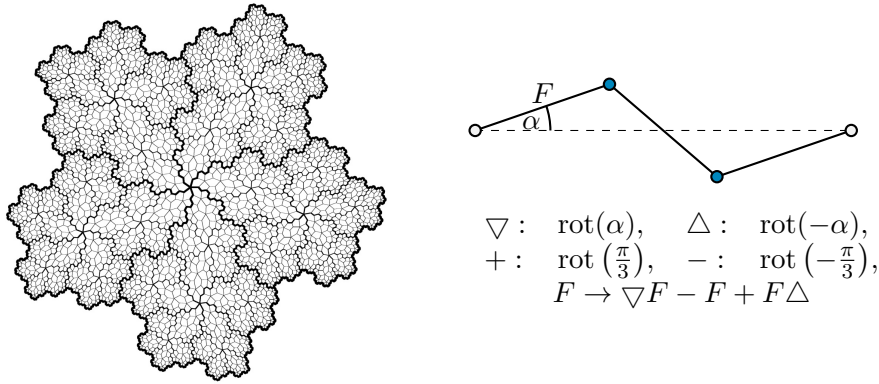

While self-similarity cannot be established, the situation is different when regarding the property of fractal dimension. 
Consider the edges of the subdivided pentagon in Figure~\ref{fig:FirstSnubRefinementSteps}. 
As discussed in Section~\ref{sec:PentagonSnubRefinement}, in each refinement step, every edge is replaced by three edges in the shape of a stretched Z, see Figure~\ref{fig:Z-triplet}. 
As known from the Koch curve, which is defined by replacing an edge 
\begin{tikzpicture}
	\draw (0,0) -- (1,0);
\end{tikzpicture}
by \begin{tikzpicture}
	\draw (0,0) -- (0.33,0) -- (0.5,0.288) -- (0.66,0) -- (1,0);
\end{tikzpicture}
, this procedure of replacing edges can be formulated in terms of a Lindenmayer system, see~\cite[Ch.~1]{prusinkiewicz1989lindenmayer} for theory of these systems and Figure~\ref{fig:FractalCurvesAndLindenmayer} for the concrete system relevant for the presented subdivision scheme.

The three edges constructed in operation~1 of Section~\ref{sec:PentagonSnubRefinement} are constructed with equal length and smaller enclosed angle~$\tfrac{2\pi}{3}$. 
Thus, in each refinement step, the new boundary edges are~$\tfrac{1}{\sqrt{7}}$ times as long as in the previous step.
Furthermore, there are three times as many boundary edges, thus the total length of the boundary grows by the factor~$\tfrac{3}{\sqrt{7}}$ in each step. 
Hence, the limit boundary curve has a fractal dimension of
\vspace{0.2cm}
\begin{align*}
	D = -\tfrac{\log(3)}{\log (1/\sqrt{7})} \approx 1.12915.
\end{align*}
\vspace{0.1cm}

Note that this statement holds for the boundary curve shown in Figure~\ref{fig:FractalCurvesAndLindenmayer} and more generally for the resulting boundary curve of any reasonably regular input mesh.
Inner edges are, once they are created in a refinement step~$t$, iteratively replaced by smaller edges.
This amounts to the generation of~$3^{t'-t}$ edges in a later refinement step~$t'$. 
Visually, these curves are very similar to the curves at the boundary (see Figure~\ref{fig:FractalCurvesAndLindenmayer}), but because of the smoothing operation applied to inner vertices, computing the total length of these curves in a given refinement step~$t$ is not as easy as at the boundary.

\rev{
	In regard to the curves involving inner vertices, we can offer some observations.
	First of all, there are two types of these curves: Those that have at least one fixed point as end-point and those that have two inner vertices at end-points.
	Here, fixed points are vertices that are not moved during a refinement step.
	Such points lie, e.g., on the boundary curve as the vertices of the boundary remain in place after their creation, because the smoothing step does not apply to them.
	Also, the center point of the pentagon is a fixed point because of symmetry.
	Therefore, the lengths of the five curves running from the center point to the boundary are bounded from below by the distance of the center point to the end-point on the boundary.
	This is not generally true for curves running between two inner vertices, as subsequent smoothing steps can move the end-points closer to each other.
	Numerical experiments suggest that curves between inner vertices do grow with repeated executions of refinement steps.
	However, the growth did not happen with a unique factor, nor did the growth factor change monotonically.
	This motivates the following (set of) open question(s):
}

\begin{question}
	\rev{
		Are the inner curves following a fractal pattern?
		If so, what is their fractal dimension?
		How do they, if at all, relate to the boundary fractal?
	}
\end{question}

\begin{figure}
	\centering
	\begin{subfigure}[b]{0.48\textwidth} 
		\centering
		\includegraphics[width=1\textwidth]{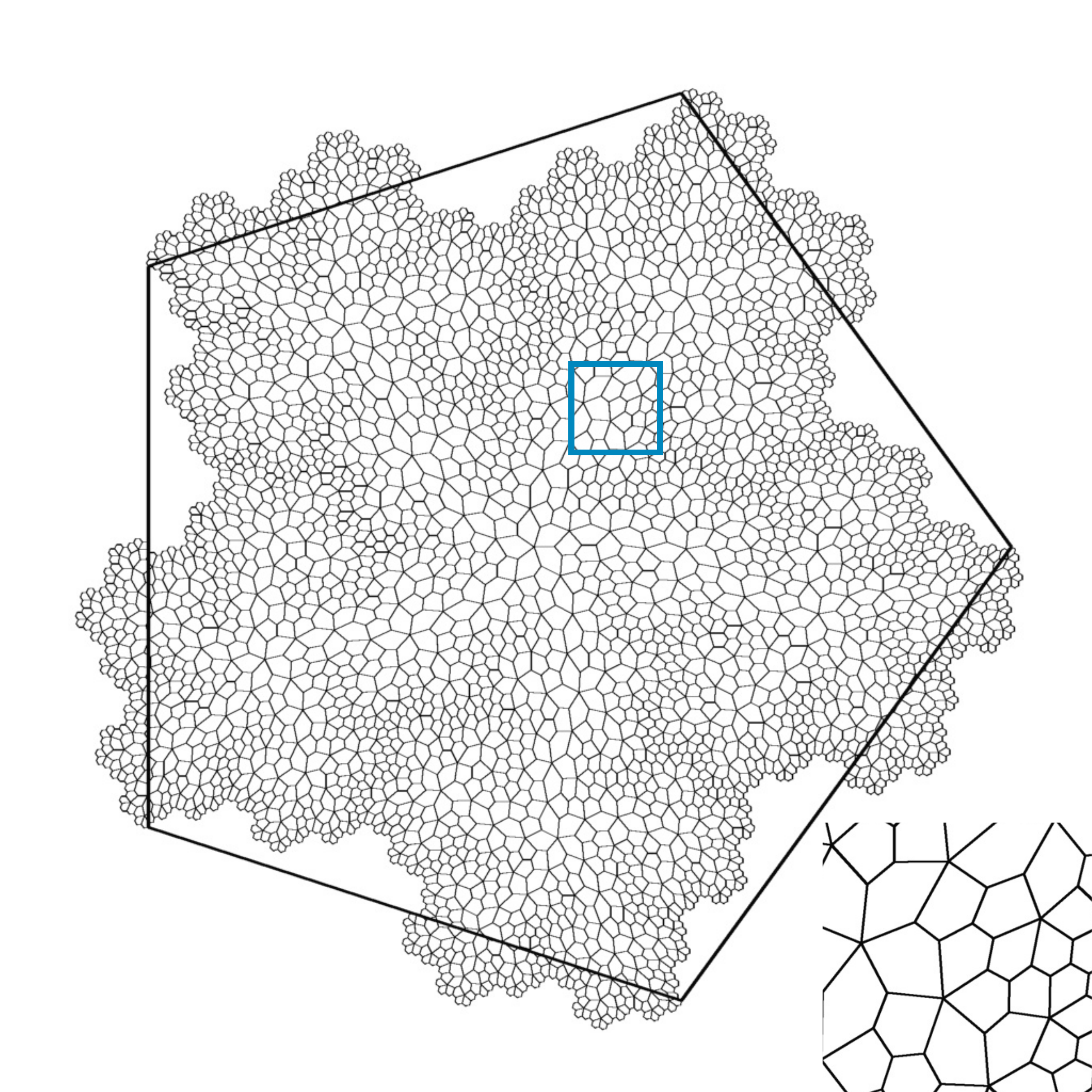}
	\end{subfigure}
	\hfill
	\begin{subfigure}[b]{0.48\textwidth}
		\centering 
		\includegraphics[width=1\textwidth]{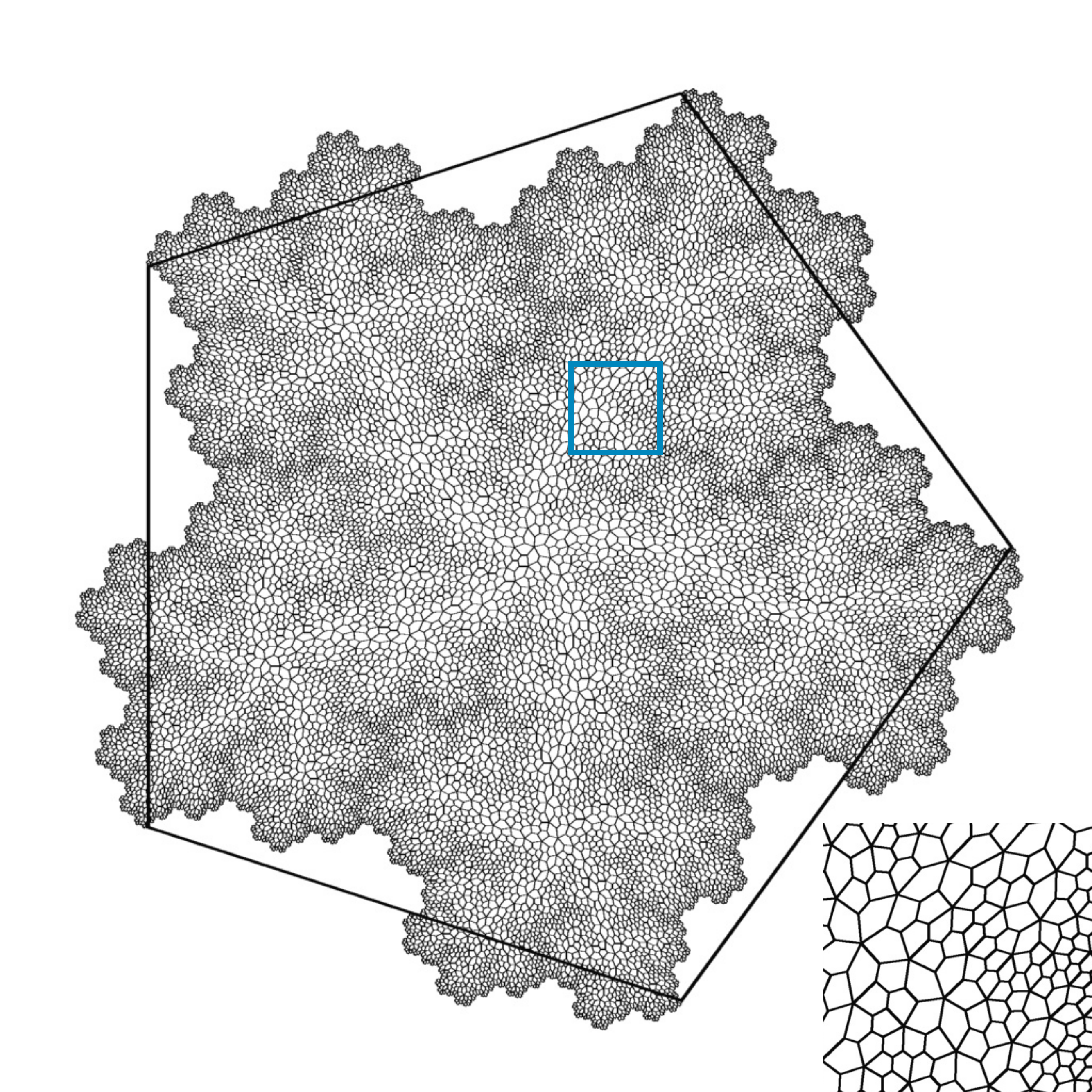}
	\end{subfigure}
	\begin{subfigure}[b]{0.48\textwidth} 
		\includegraphics[width=1\textwidth]{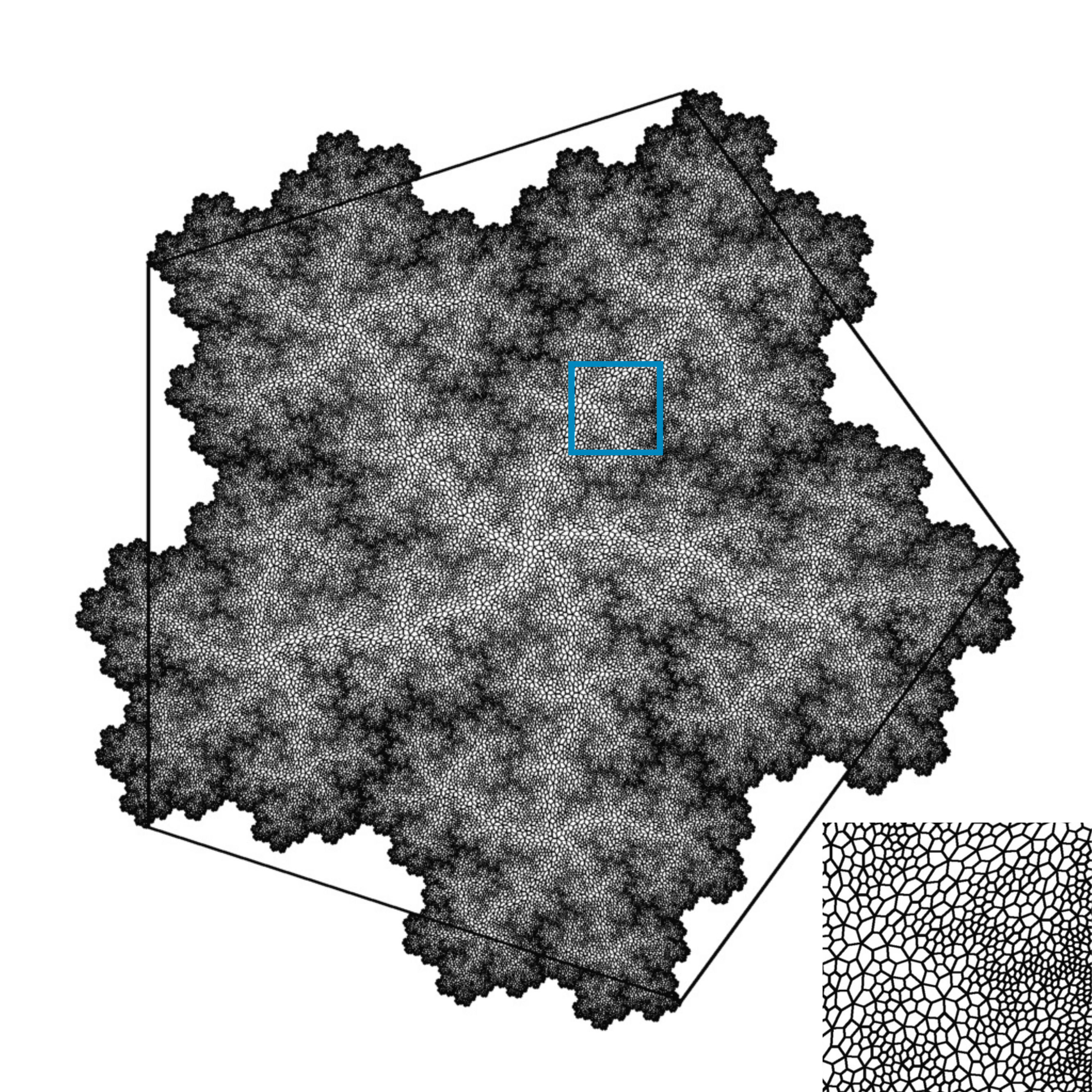}
	\end{subfigure}
	\hfill
	\begin{subfigure}[b]{0.48\textwidth} 
		\includegraphics[width=1\textwidth]{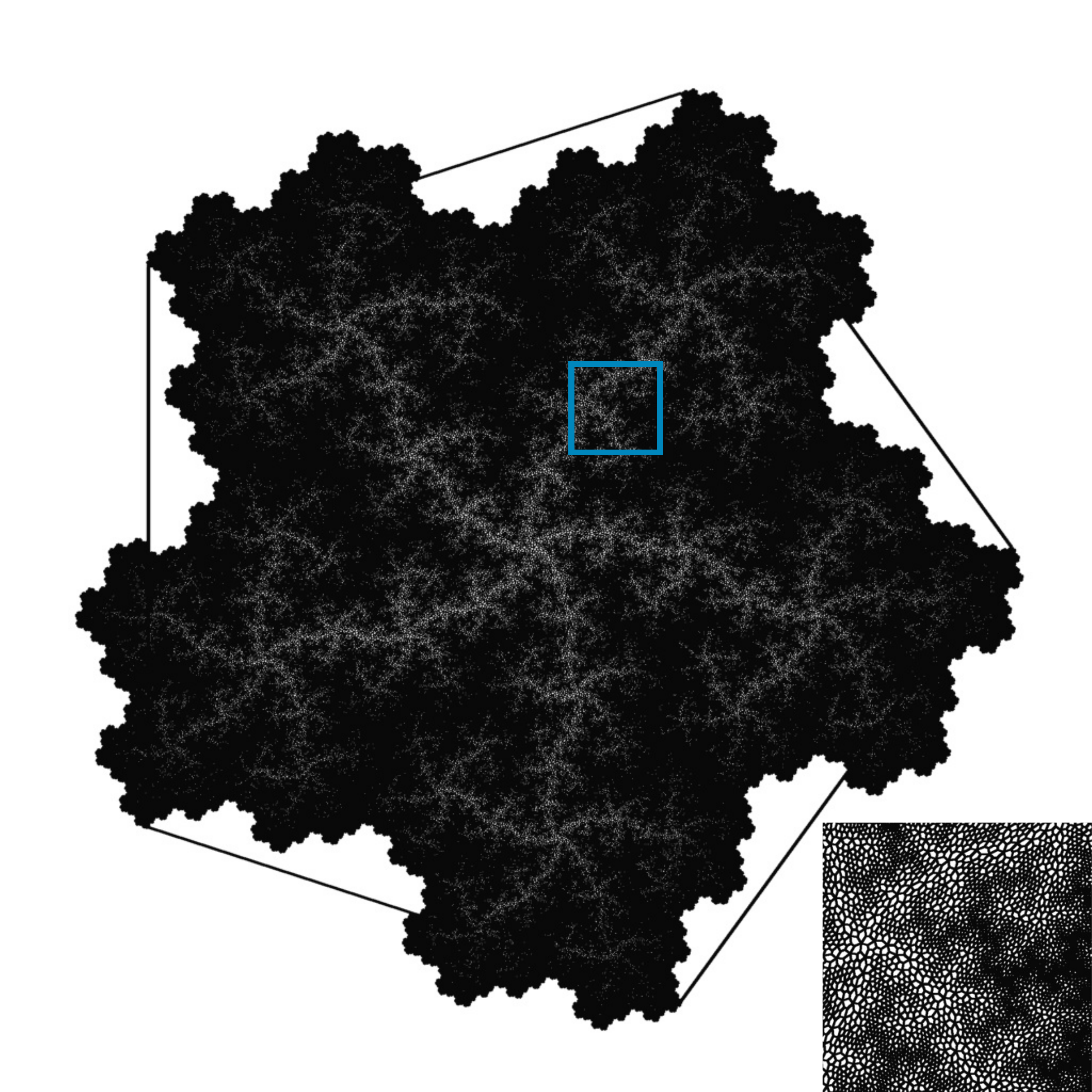}
	\end{subfigure}
	\caption{Repeated pentagon snub subdivision refinement applied to a regular pentagon. From upper left to lower right, figures show 5th, 6th, 7th, and 8th refinement step. Shown are the edges resulting from the corresponding refinement step.}
	\label{fig:SubdivisionEdges}
\end{figure}

To explore the self-similarity of the presented images further, we explore two rendering options for an iteratively refined pentagon.
The first is based on edges, while the second is based on vertices.
The images in Figure~\ref{fig:SubdivisionEdges} continue the series of images from Figure~\ref{fig:FirstSnubRefinementSteps}, i.e., they show the fifth to eighth refinement step of the underlying pentagon.
As discussed in the previous two paragraphs, the edge lengths of the newly inserted edges shrinks with each refinement step.
Thereby, structures arise that vaguely remind of lightning during a thunderstorm.
In the current resolution, the next---ninth---refinement step would show a completely black picture.
Note the gradually arising self-similarity as provided by the zoom-in boxes.

\begin{figure}
	\centering
	\includegraphics[width=0.7\textwidth]{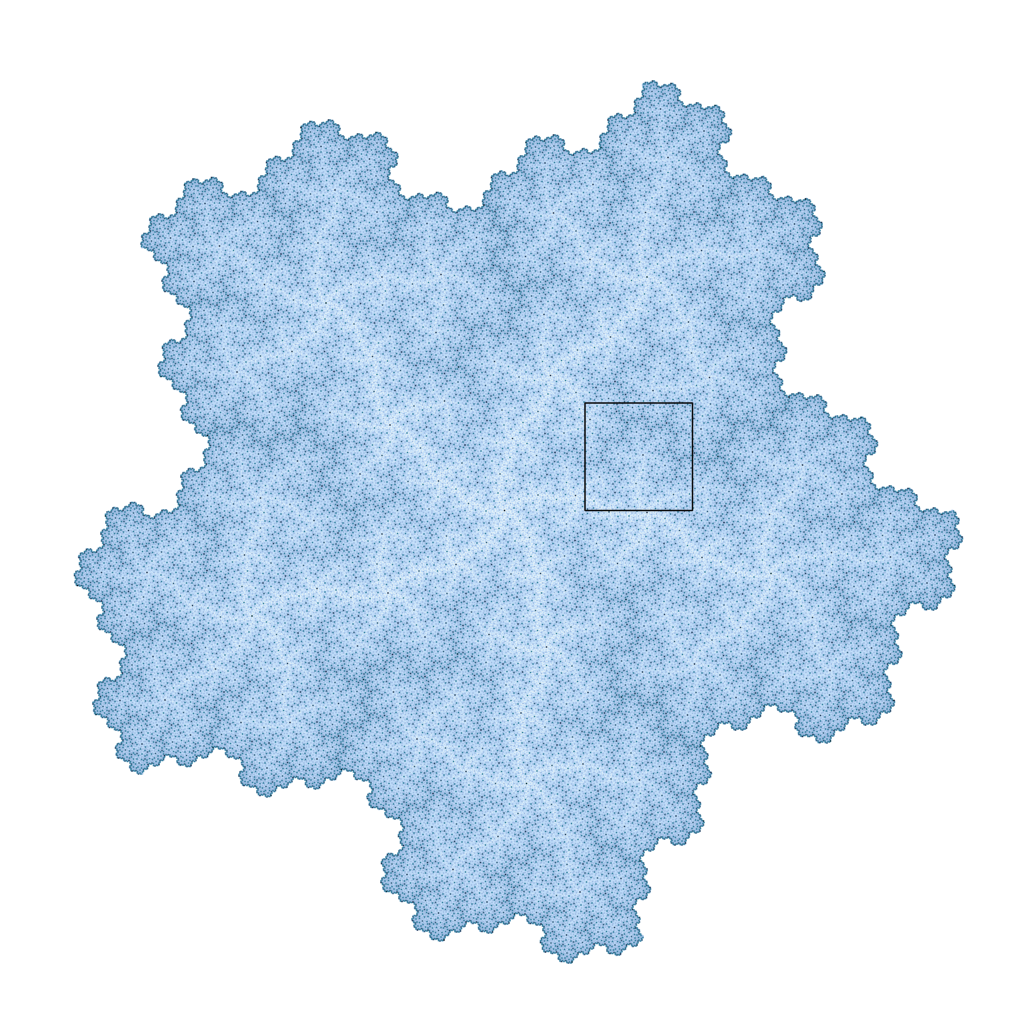}
	\begin{minipage}[b]{0.29\textwidth}
		\includegraphics[width=1.\textwidth]{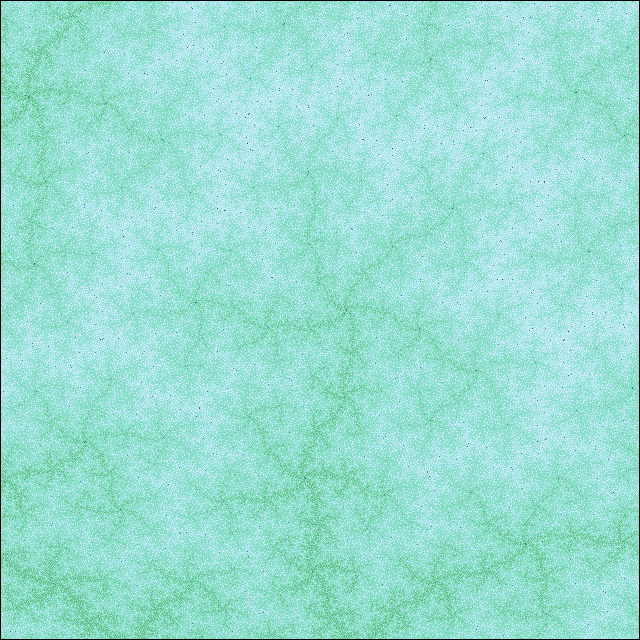}\\
		\vspace{0.5cm}
		\includegraphics[width=1.\textwidth]{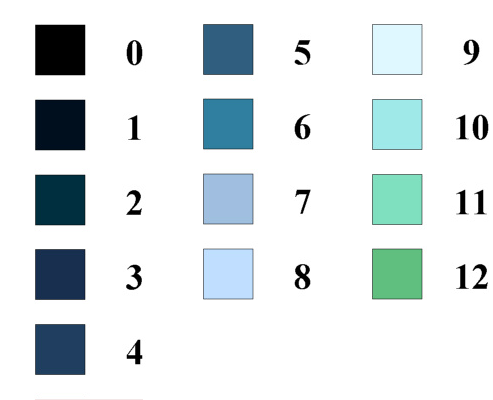}\\
	\end{minipage}
	\caption{
		Pixels colored according to the refinement step in which the first vertex is created within this pixel.
		The five vertices of the input pentagon are colored in step~0.
		Throughout the refinement steps~1--9, colors get gradually brighter, while for the refinement steps~10--12, they gradually become green.
		The left image shows the subdivided pentagon after refinement step~9, while the right image shows the highlighted version after refinement step~12.}
	\label{fig:colorLevels}
\end{figure}

In contrast to the illustrations of Figure~\ref{fig:SubdivisionEdges}, in the image in Figure~\ref{fig:colorLevels}, each pixel is colored according to the first refinement step, in which it contains a vertex of the iteratively subdivided regular pentagon.
The later this pixel is hit by a vertex, the brighter it is colored.
This is most notable on the boundary curve of the subdivided pentagon, whose pixels are created fairly early on in the process.
In comparison, the way along one of the wave-like curves from one of the corners of the original pentagon to the center point takes longer to be refined.
Thus, in the center, around one dark pixel created in the very beginning, there is a collection of comparably bright pixels that are filled late in the process.

Even though the refinement process can be repeated infinitely often, this image does not change after a finite number of refinement steps, because each inner pixel is painted after nine refinements. 
Similar to the well-known images of the Mandelbrot set~\cite[Part~IV]{pickover1998chaos}, one can zoom into the image and compute a new image for a smaller area of the pentagon to a finer level, producing self-similar images, see the corresponding zoom-in in Figure~\ref{fig:colorLevels}.

\begin{figure}
	\hspace{0.066\textwidth}
	\begin{subfigure}[t]{0.4\textwidth}
		\includegraphics[width=\textwidth]{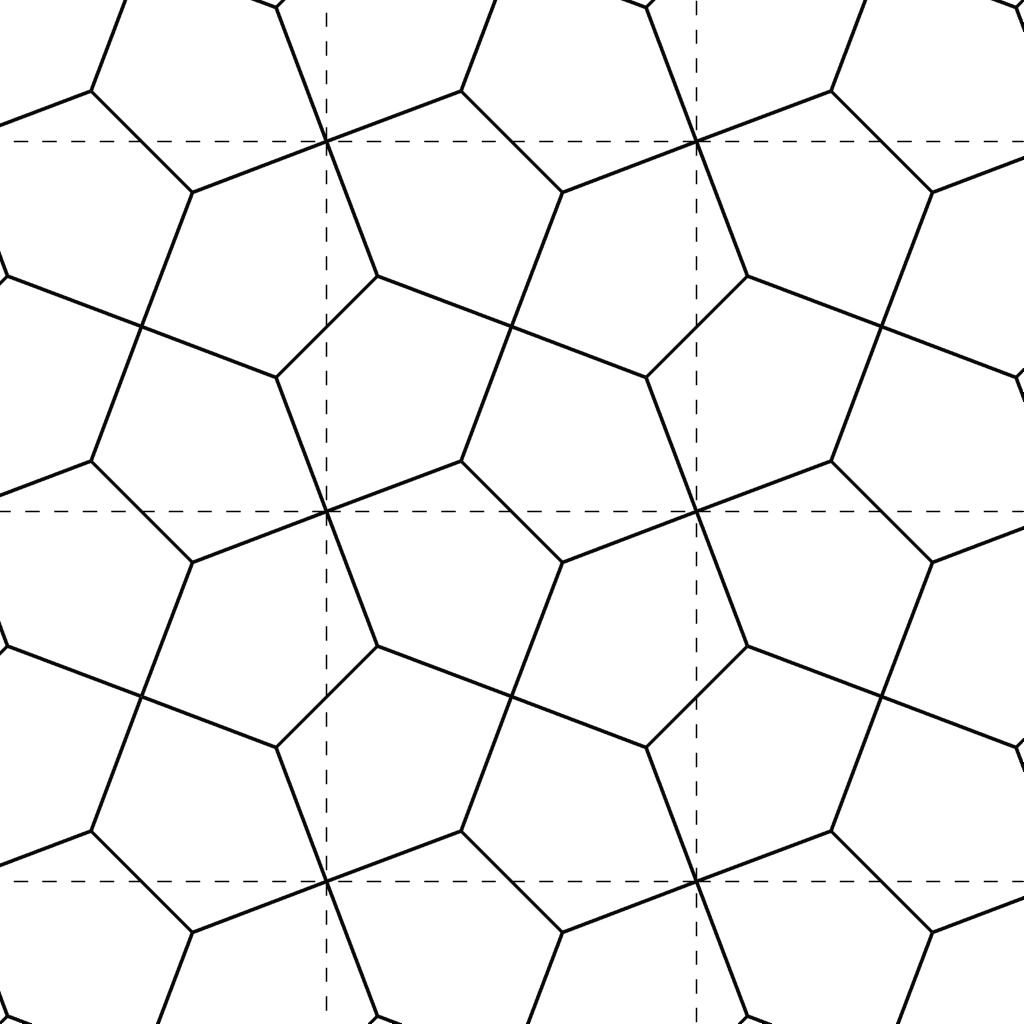}
		\caption{1st refinement step on a regular square grid.}
		\label{fig:WeavingSquare}
	\end{subfigure}
	\hspace{0.066\textwidth}
	\begin{subfigure}[t]{0.4\textwidth}
		\includegraphics[width=\textwidth]{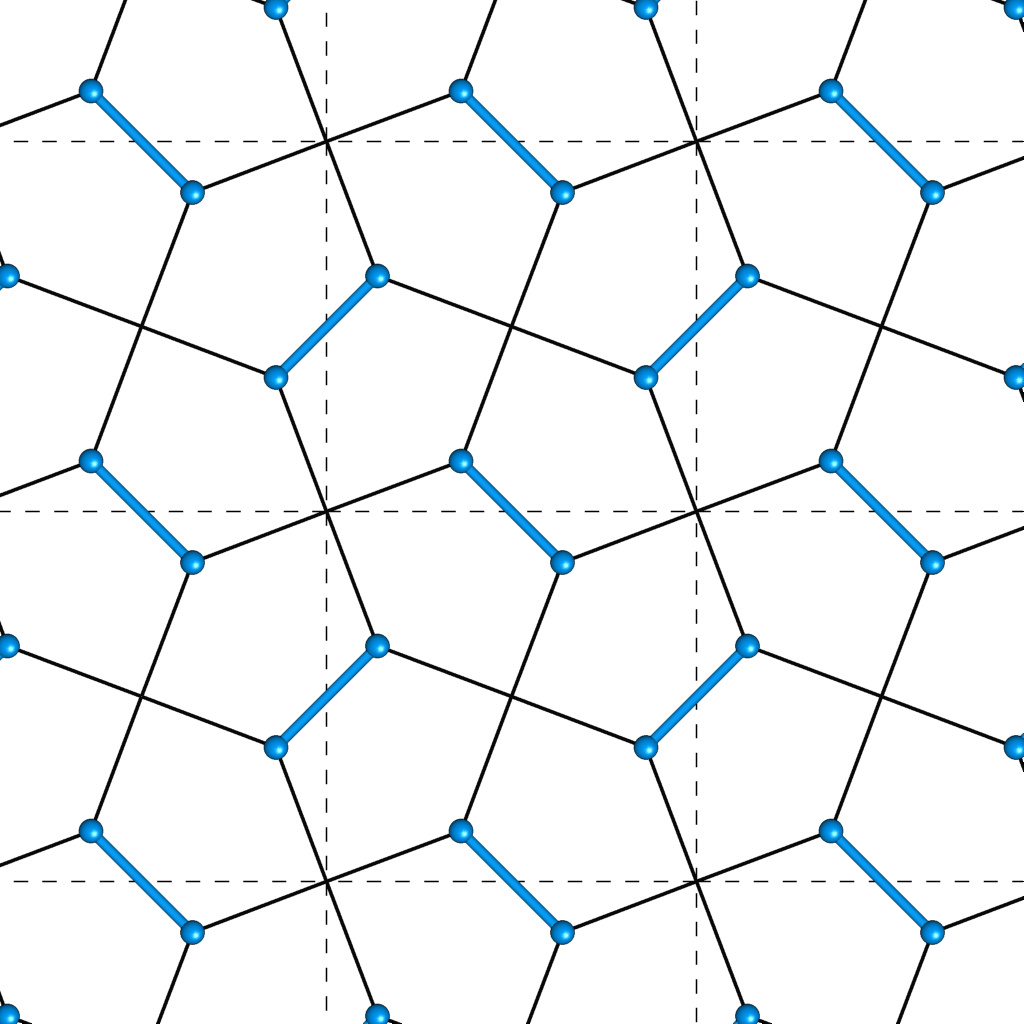}
		\caption{Highlighting the middle edges of all Z-triplets.}
		\label{fig:WeavingSquareMiddleEdges}
	\end{subfigure}
	\hspace{0.066\textwidth}\\
	
	\vspace{0.5cm}
	
	\hspace{0.066\textwidth}
	\begin{subfigure}[t]{0.4\textwidth}
		\includegraphics[width=\textwidth]{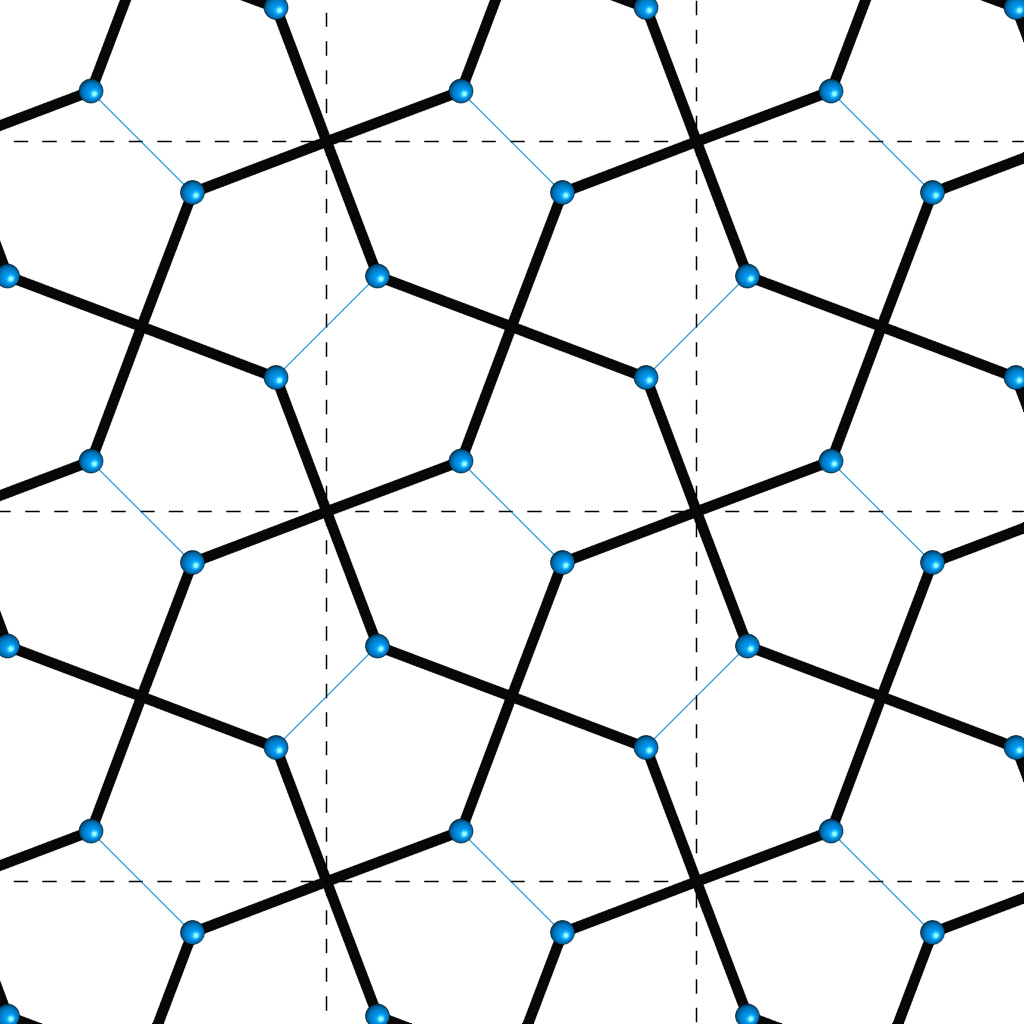}
		\caption{Glue pentagons along the middle Z-triplet edges.}
		\label{fig:WeavingSquareGluing}
	\end{subfigure}
	\hspace{0.066\textwidth}
	\begin{subfigure}[t]{0.4\textwidth}
		\includegraphics[width=\textwidth]{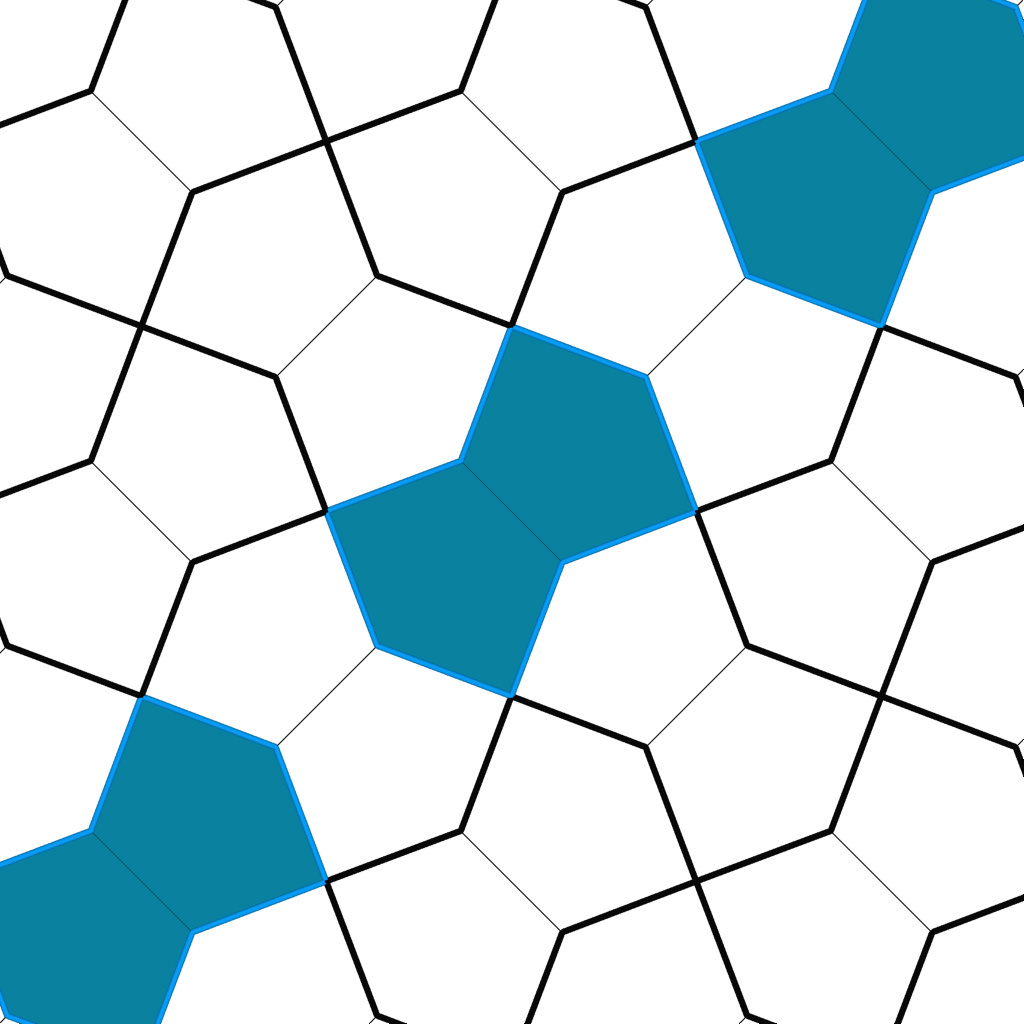}
		\caption{Color a strand, following central Z-triplet edges.}
		\label{fig:WeavingSquareStrip}
	\end{subfigure}
	\hspace{0.066\textwidth}\\
	
	\vspace{0.5cm}
	
	\hspace{0.066\textwidth}
	\begin{subfigure}[t]{0.4\textwidth}
		\includegraphics[width=\textwidth]{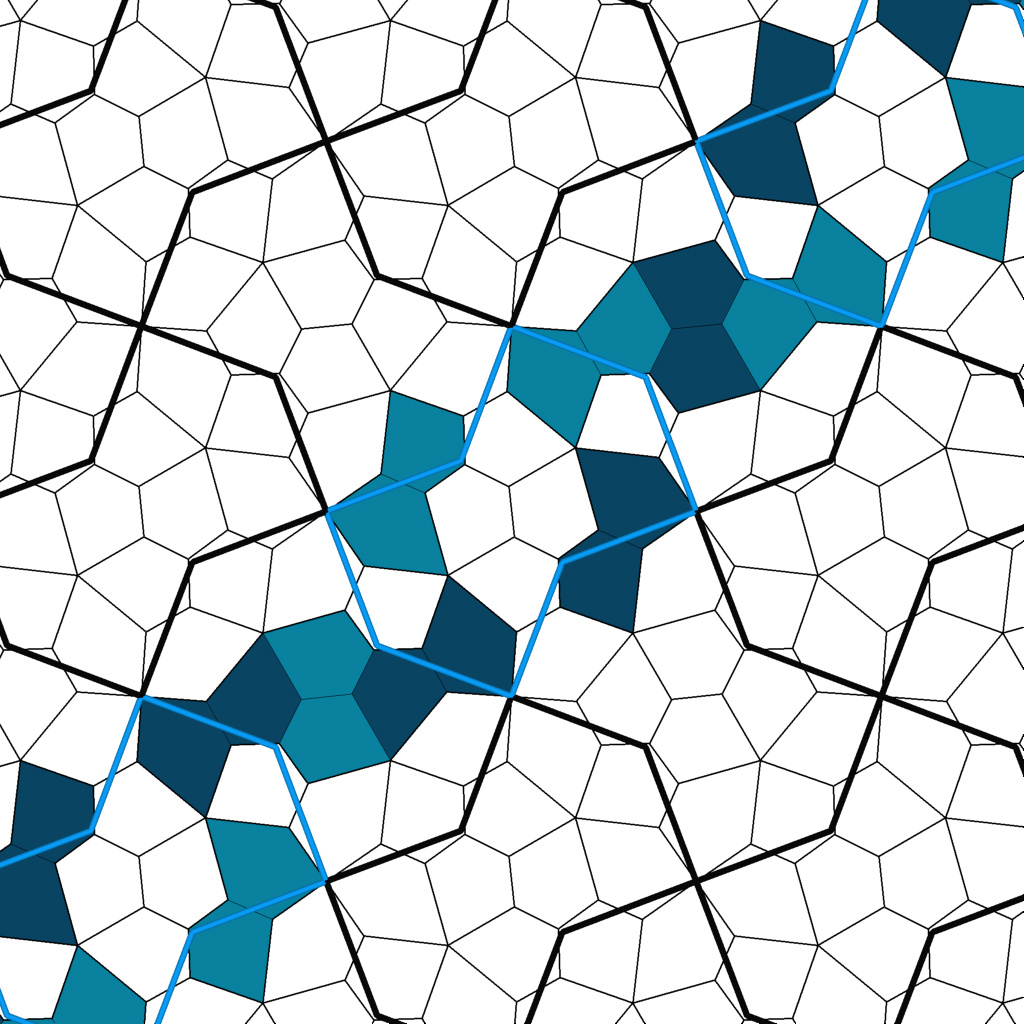}
		\caption{2nd refinement: a strand splits into two twisted strands.}
		\label{fig:WeavingSquare2ndRefinement}
	\end{subfigure}
	\hspace{0.066\textwidth}
	\begin{subfigure}[t]{0.4\textwidth}
		\includegraphics[width=\textwidth]{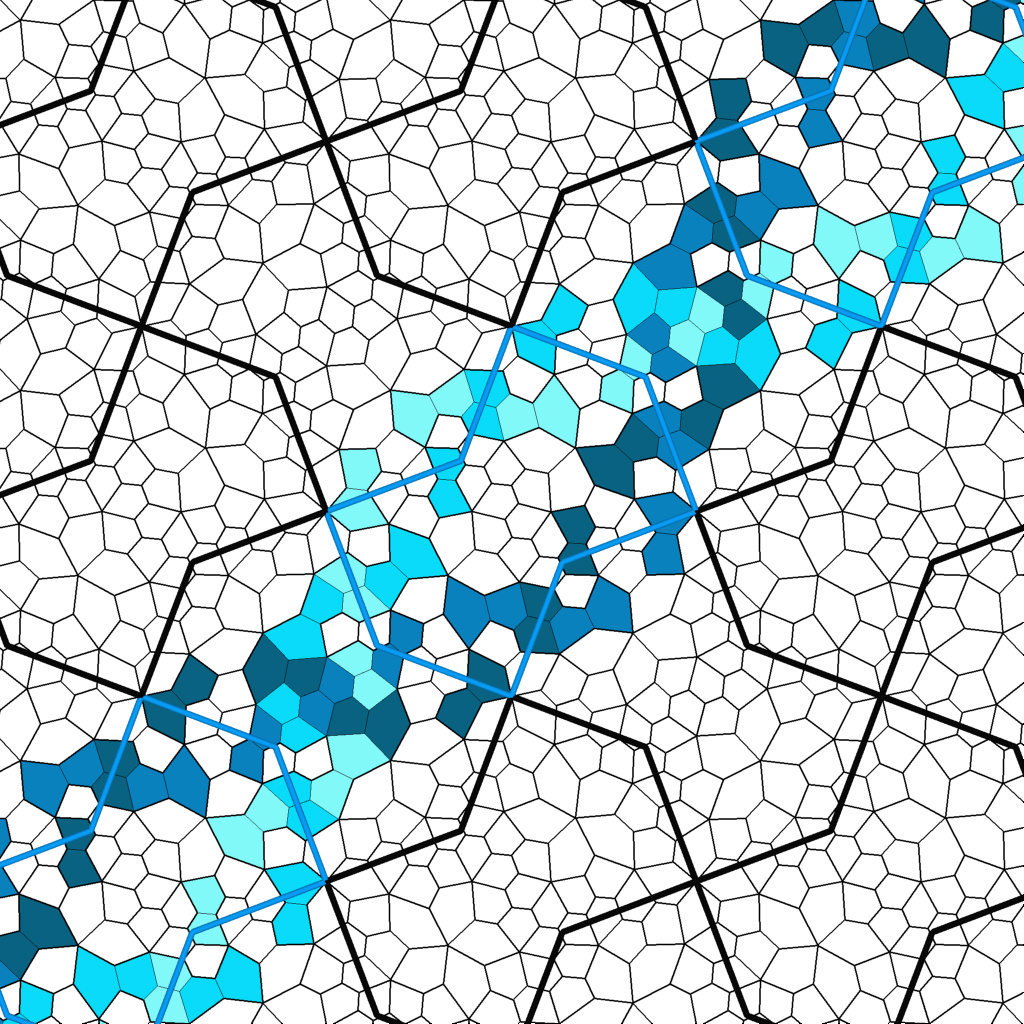}
		\caption{3rd refinement: a twisted pair of twisted strands.}
		\label{fig:WeavingSquare3rddRefinement}
	\end{subfigure}
	\hspace{0.066\textwidth}\\
	\caption{Creating a weaving pattern on a refined square grid by gluing tiles and coloring them systematically.}
	\label{fig:SquareGridWeaving}
\end{figure}

\section{Weaving patterns resulting from the pentagon snub subdivision scheme}
\label{sec:WeavingPatterns}

Aside from fractal properties, the subdivided object exhibits another interesting behavior, when coloring \rev{pairs of} its pentagons appropriately.
To illustrate the coloring, consider a regular square grid to which one refinement step of the pentagon snub subdivision scheme is applied, see Figure~\ref{fig:WeavingSquare}.

Note that by construction, each of the pentagons shown has exactly one edge that arises as the middle edge of a Z-triplet.
Furthermore, when fixing a pentagon, the vertex opposite to said edge is incident to the middle edge of another Z-triplet, confer the highlighted middle edges in Figure~\ref{fig:WeavingSquareMiddleEdges}.
By pairwise gluing all pentagons that share a middle edge of a Z-triplet, the mesh is reduced to octagons, see Figure~\ref{fig:WeavingSquareGluing}.

The coloring is applied as follows: An initial octagon~$\mathcal{O}$ and a color for~$\mathcal{O}$ are chosen.
By construction, if~$\mathcal{O}$ does not contain a pentagon on the boundary, exactly two vertices of~$\mathcal{O}$ are incident to middle Z-triplet edges outside of~$\mathcal{O}$.
These two edges connect~$\mathcal{O}$ to two other octagons~$\mathcal{O}'$ and~$\mathcal{O}''$.
Both are colored in the same color as~$\mathcal{O}$ and the coloring is propagated in this way across the pentagon mesh.
For a single color, this results in a strand, see Figure~\ref{fig:WeavingSquareStrip}.

When refining the mesh further, these colored strands are split into further pentagons and corresponding glued octagons.
These can be colored as strands again and it can be observed that each original strand splits into two smaller, interwoven ones.
This behavior is illustrated in Figure~\ref{fig:WeavingSquare2ndRefinement}, where the glued octagons from the first refinement step are still visible and the octagons from the second refinement step clearly weave along the former strand.
In fact, this observation generalizes: With each additional refinement step, each strand is split into two interwoven strands that follow their parental one.
Thus, after the third refinement step, each original strand from the first step is split into four twisted strands, see Figure~\ref{fig:WeavingSquare3rddRefinement}.

In fact, the observations made for the coloring of the refined square grid solely depend on properties of the subdivision scheme.
Therefore, the coloring procedure can be applied to any bounded or unbounded 2-manifold mesh subdivided by the pentagon snub subdivision scheme.

\rev{
	In the following, we will continue the investigation of the pattern induced by the snub pentagon subdivision scheme.
	First, we discuss our naming choice (calling it a ``weaving'') and discuss related textile analogies.
	Following this, we present a physical realization of the weaving pattern via paper models.
	The final part of this section is devoted to generalizing this illustrative approach of weaving patterns to other subdivision schemes aside from snub pentagon subdivision.
}

\subsection{\rev{Naming choice: Why call the pattern a ``weaving''?}}
\label{sec:NamingChoice}

\rev{
	The creation process of the pattern, as outlined above for the regular square grid, has two central components that provide an allusion to traditional textile processes.
	This is, on the one hand side, the creation of intertwined strands.
	These arise, as a refinement step splits each strand present before, compare Figures~\ref{fig:WeavingSquareStrip} and~\ref{fig:WeavingSquare2ndRefinement}.
	On the other hand side, in each refinement step, the pattern is created via strands that overlap each other.
	This can be seen in Figure~\ref{fig:WeavingSquareStrip}, where the colored strand passes below two strands running orthogonally.
}

\rev{
	Focusing on the first element, as shown in Figure~\ref{fig:WeavingSquareStrip}, the initial application of the subdivision scheme creates single strands.
	However, subsequent subdivision refinement steps split these strands, see Figures~\ref{fig:WeavingSquare2ndRefinement} and~\ref{fig:WeavingSquare3rddRefinement}.
	Note that these subdivided strands still follow the direction of the original one.
	After the second refinement step, the single initial strand is split into two strands that are crossing above and below each other while following said direction.
	These two strands are then split by the third refinement step into two even smaller strands respectively.
	Each pair is intertwined and crosses above and below, while the pairs themselves are also intertwined, crossing above and below each other, just as the parental strands did.
	This behavior of the strands, following a common direction while being intertwined, i.e., crossing above and below each other, is reminiscent of \emph{braiding} or \emph{plaiting} as the result is ``considerably greater in length than in width or diameter''~\cite[p.~38]{seiler1994textiles}.
}

\rev{
	However, we do not consider the braiding of the individual strands to be the visually dominant piece of the final pattern.
	This is rather given by the crossing of the strands with other strands that run into a different direction.
	That, in turn, is reminiscent of the process in weaving, where ``it is thus possible to obtain at least two distinct interlacings of warp which are separated by the weft thread''~\cite[71]{seiler1994textiles}.
	This perspective does, at least at a global level, describe the behavior of the different strands in the pattern described here.
	This motivates our naming choice of calling the resulting pattern a \emph{weaving}.
}

\rev{
	As a final remark, note that this naming choice has to be taken with a pinch of salt.
	In classical weaving processes, the thread that provides the weft is completely moved below or above the threads that form the warp.
	First, this means that the warp is fixed, while only the weft is moved actively.
	However, in the following (see Figure~\ref{fig:PentagonsWeavingPaperModel}) we will see that this perspective is not always correct for more complex scenarios than the regular square grid.
	In those, ``[b]oth thread groups are equally active''~\cite[p.~42]{seiler1994textiles}, which classifies active-active intertwining.
	Second, in weaving, all strands that form the thread, are moved above or below another thread together.
	Figure~\ref{fig:WeavingSquare3rddRefinement} reveals that the situation is more delicate for the pattern considered here.
	At points where two directions meet orthogonally, their individual strands do interweave in a complex pattern.
	Some smaller strands go below those from the orthogonal direction, while others move above.
	This technique is called \emph{ply-splitting}, see~\cite[p.~43]{seiler1994textiles}.
	More specifically, in the case of the regular square grid pattern, we encounter a \emph{right-angled} or\emph{diagonal} ply-splitting, see~\cite[pp.~44/45]{seiler1994textiles}.	
	In these two aspects, the pattern investigated here is not a weaving.
	However, as the general visual impression is still reminiscent of a weaving pattern, we stick with this terminology.
	This is to be kept in mind throughout the following discussion.
	Similarly, our reference to a \emph{weaving pattern} does not imply the explanatory or procedural component that is carried by, e.g., a knitting pattern.
	Our patterns are simply the induced patterns on the geometry.
}

\subsection{\rev{Physical realization of the pattern}}

\rev{
	Before we turn to considering the impact of this weaving pattern for other subdivision schemes than the pentagonal snub subdivision, we will briefly investigate the possibility of creating physical models from the above observations.
	This is to assert that the presented pattern does not only provide a coloring, but a pattern that can be physically realized via a weaving process.
}

\rev{
	Starting with the simple case of a subdivided square grid as presented in Figure~\ref{fig:SquareGridWeaving}, it helps to rotate the pictures by $\tfrac{\pi}{4}$, as the strands thereby obtain a horizontal or vertical alignment.
	We can now fabricate a pattern to be printed on paper and cut out to provide a set of strands, which can subsequently be intertwined to provide the final pattern.
}
	
\begin{figure}
	\begin{minipage}{0.6\textwidth}
		\begin{subfigure}{1.\textwidth}
			\includegraphics[width=1.0\textwidth]{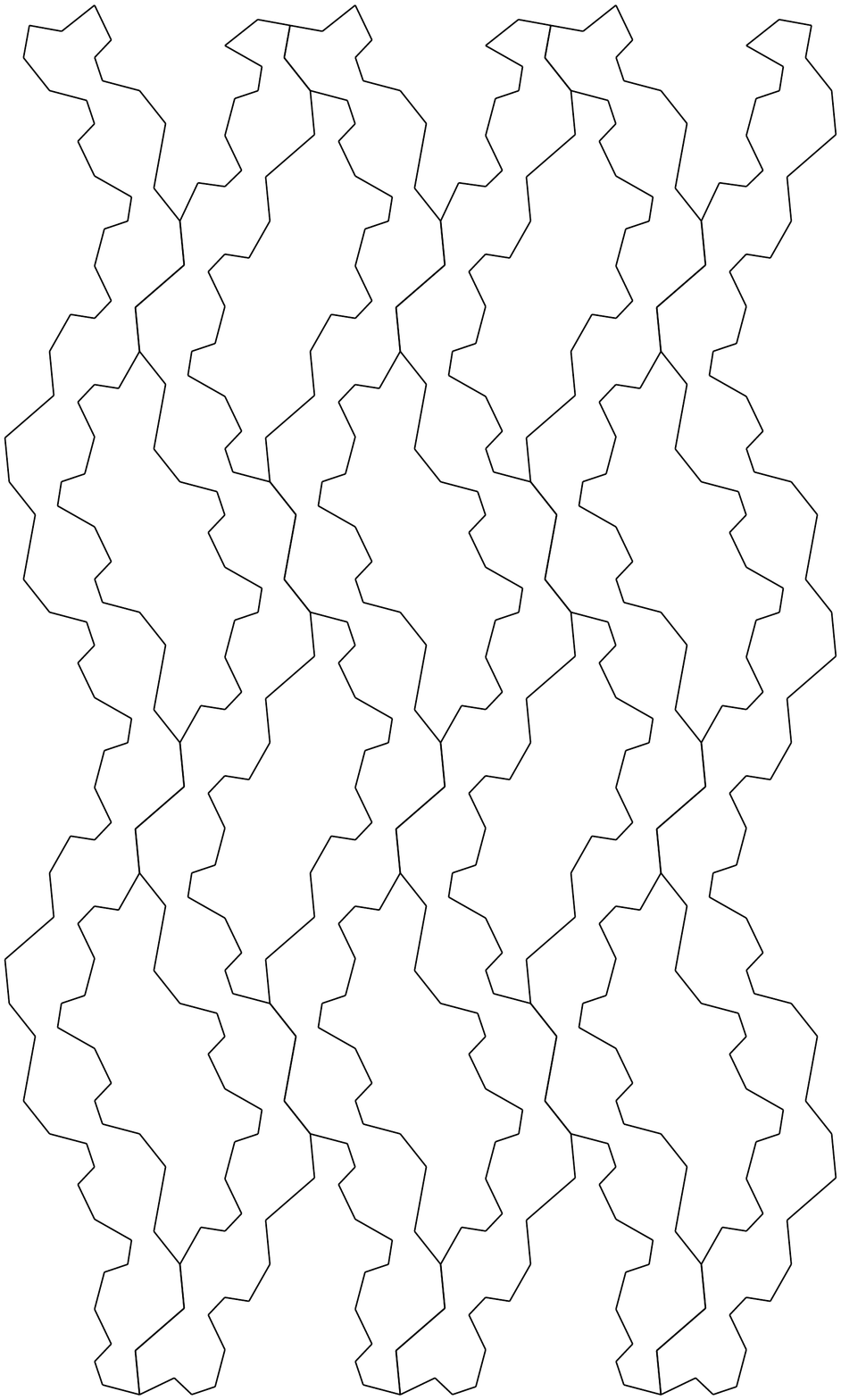}
			\caption{One set of paper modules.}
			\label{fig:SquareGridSinglePaperModule}
		\end{subfigure}
	\end{minipage}
	\begin{minipage}{0.4\textwidth}
		\begin{subfigure}{1.\textwidth}
			\includegraphics[width=1.0\textwidth]{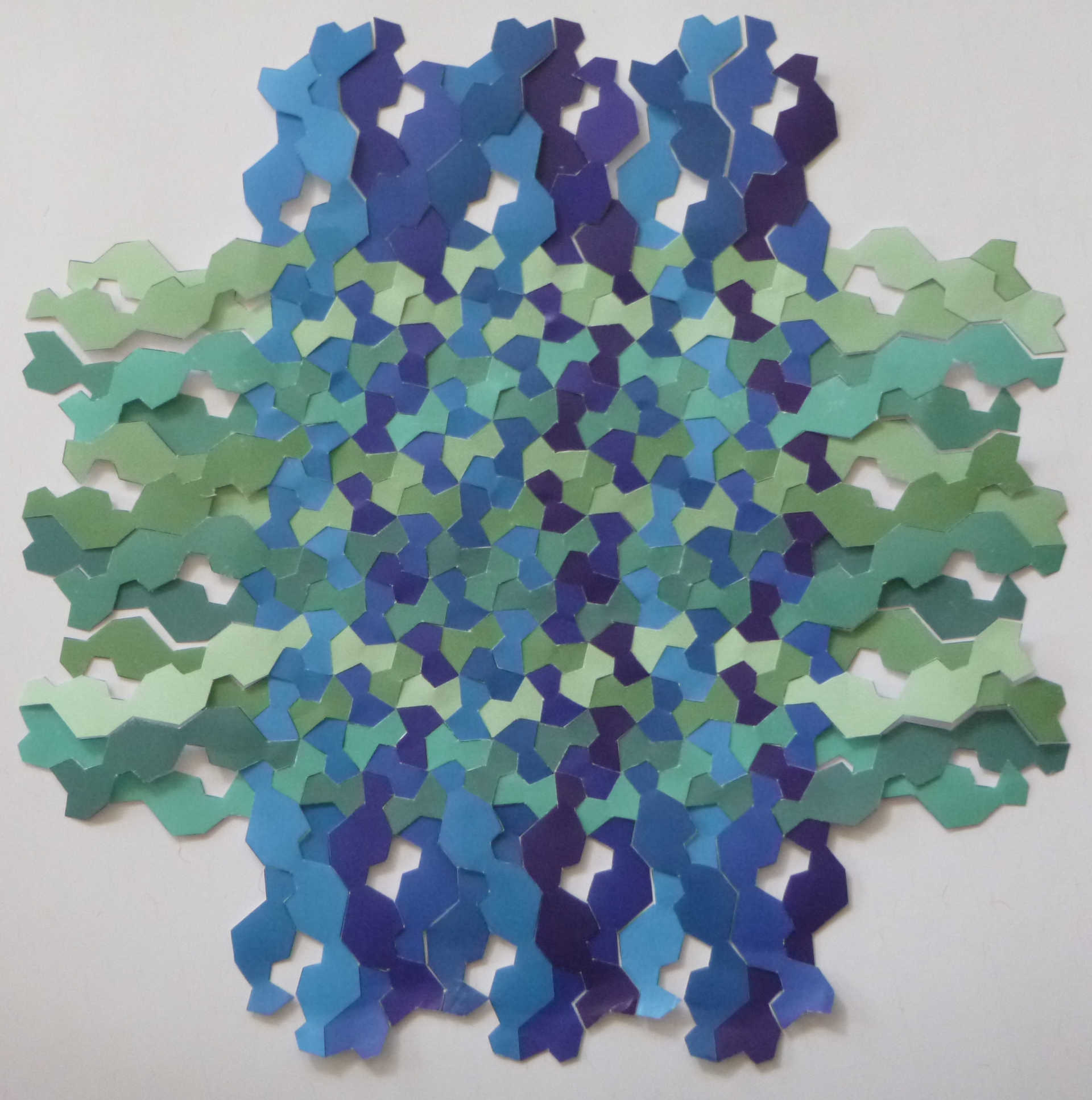}
			\caption{Physical realization with paper, highlighting intertwined strands. (Size: 27\,cm$\times$27\,cm)}
			\label{fig:SquareGridWeavingPhysical1}
		\end{subfigure}
		\begin{subfigure}{1.\textwidth}
			\includegraphics[width=1.\textwidth]{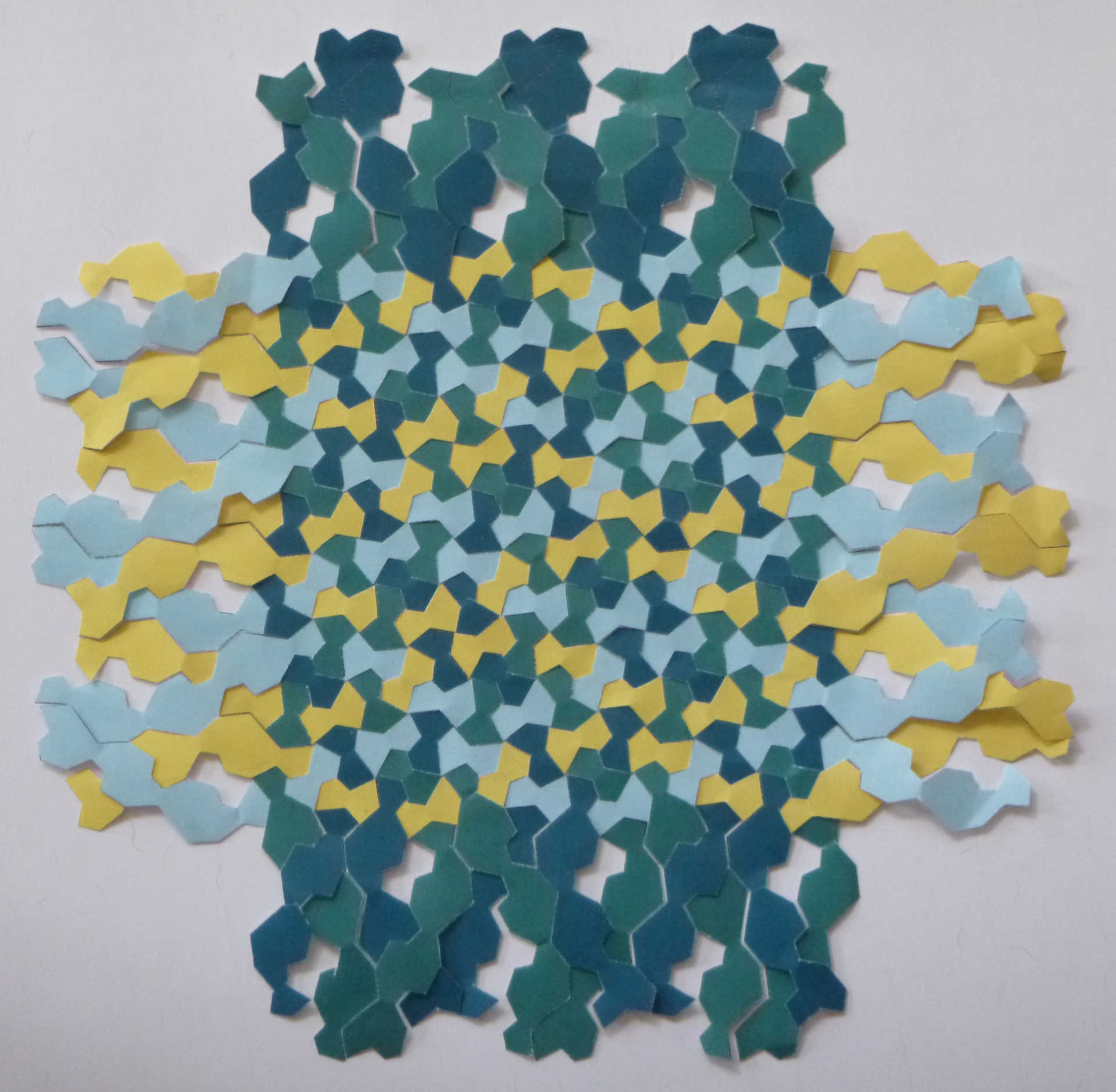}
			\caption{Paper model colored with minimal number of colors to separate the modules. (Size: 24.5\,cm$\times$24.5\,cm)}
			\label{fig:SquareGridWeavingPhysical2}
		\end{subfigure}
	\end{minipage}
	\caption{A single paper cut-out module, a paper model colored according to the description above, and a paper model colored such that every cut-out module is colored differently.
		Both physical models show weaving of a twice-subdivided square grid.
	}
	\label{fig:SquareGridPaper}
\end{figure}

\rev{
	A possible module\footnote{If you are interested in giving the process a try, find this module in the supplemental material to the article.} to be used is shown in Figure~\ref{fig:SquareGridSinglePaperModule}.
	This specific design is meant to be printed four times, providing a north-south, west-east, south-north, and east-west direction when assembling a weaving on the square grid as shown in Figure~\ref{fig:SquareGridWeaving}.
	That is, when assembling the pattern, the module as shown in Figure~\ref{fig:SquareGridSinglePaperModule} has to be printed four times, where three of the four copies have to be rotated by $\tfrac{\pi}{2}$, $\pi$, and $\tfrac{3 \pi}{2}$, respectively.
	Periodic repetition of this module or an appropriate reduction allows for the realization of larger or smaller number of squares in the grid, respectively.
	Figures~\ref{fig:SquareGridWeavingPhysical1} and~\ref{fig:SquareGridWeavingPhysical2} show corresponding results using different color schemes.
	The underlying regular square grid has been subdivided via the pentagon snub subdivision scheme two times. 
	The physical model covers several squares of the grid, which lies at a~$\tfrac{\pi}{4}$ angle to it.
}

\rev{
	The color schemes are chosen from a maritime palette to underline the wave-like, flowing behavior of the different strands.
	Furthermore, they illustrate the two different aspects of the pattern as discussed in Section~\ref{sec:NamingChoice}.
	The first scheme, utilizing twelve colors, highlights the intertwined pairs of strands that arise from subdividing single strands.
	It therefore alludes to the braiding component involved.
	The second scheme, using a reduced palette of four colors, provides a better overview of the general pattern created.
	Hence, the resulting model rather highlights the weaving part of the pattern.
}

\rev{
	The modules used for the presented paper realizations from Figure~\ref{fig:SquareGridPaper} were printed on A4-sized printer paper and cut using scissors and a pen knife. 
	In case of thicker material or a higher number of repetitions of the depicted block, the use of a laser cutter might be handy as the cutting part was by far the most time-consuming.
	We do not recommend to print the module in a smaller size since this makes the strands thinner and causes them to tear more easily. 
	While thicker material prevents this, it will likely be harder to assemble.
	While assembling the model, it is helpful to fix the top of those strands parallel to the warp direction.
	Then, with every weaving step, align two strands in weft direction.
	This procedure worked better in our experiments than completely twisting two parallel strands and trying to include the orthogonal ones afterwards, which was unsuccessful.
	For better orientation, we refer to a coloring such as shown in Figure~\ref{fig:SquareGridWeavingPhysical2}, which helps to align the modules.
}
	
\begin{figure}
	\begin{minipage}{0.5\textwidth}
		\begin{subfigure}{1.\textwidth}
			\includegraphics[width=1.0\textwidth]{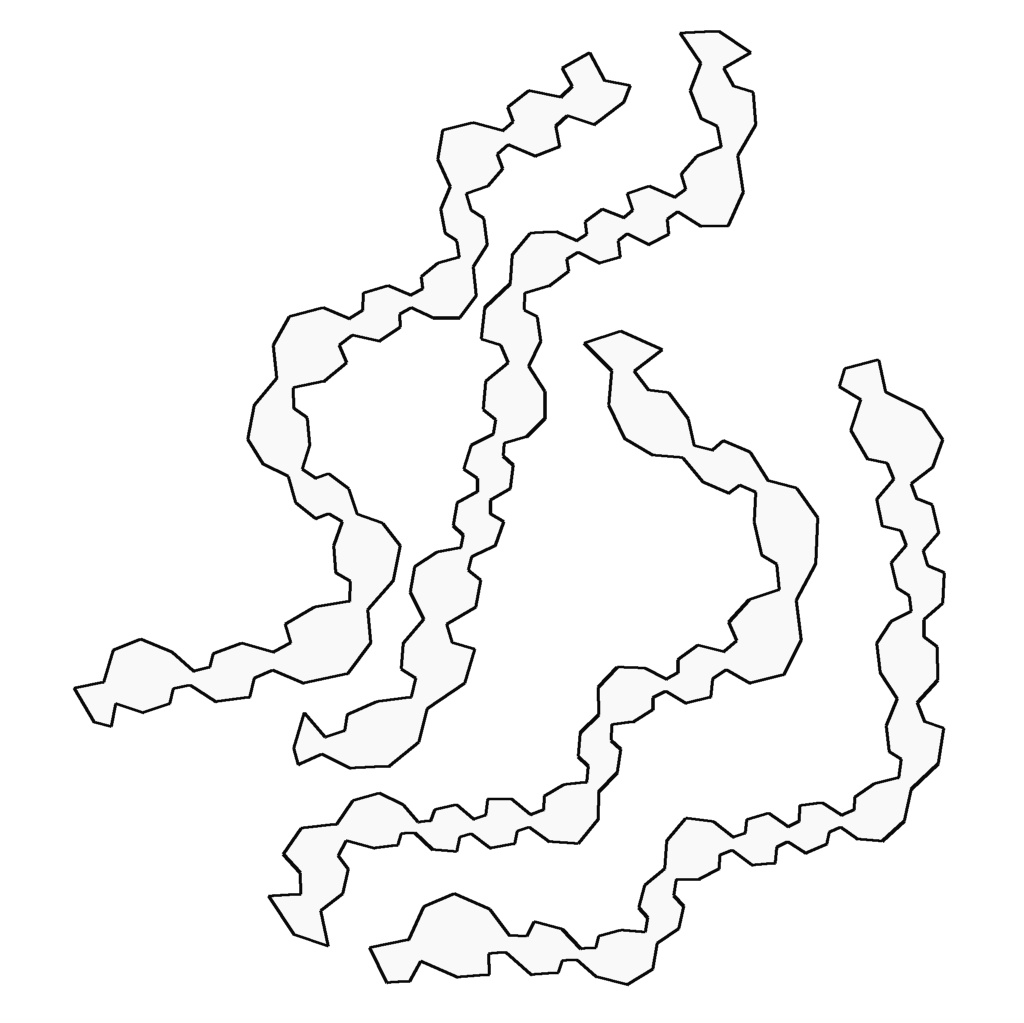}
			\caption{One set of paper modules.}
			\label{fig:SixPentagonsSinglePaperModule}
		\end{subfigure}
	\end{minipage}
	~
	\begin{minipage}{0.5\textwidth}
		\begin{subfigure}{1.\textwidth}
			\includegraphics[width=1.0\textwidth]{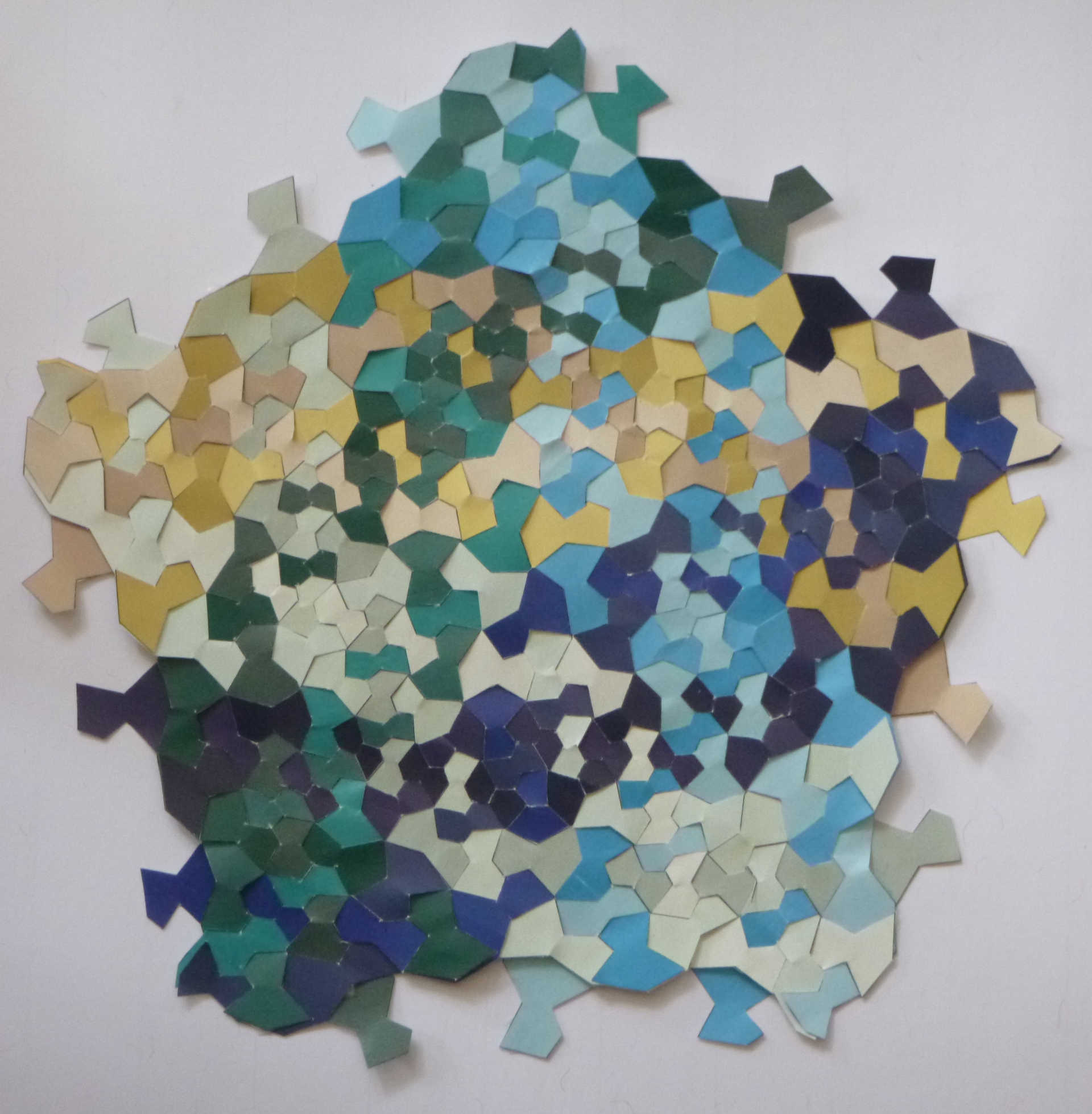}
			\caption{Physical realization of Figure~\ref{fig:FirstSnubRefinementSteps} as a paper model (Width/Height: 23\,cm$\times$22.5\,cm).}
			\label{fig:SixPentagonsWeavingPaperModel}
		\end{subfigure}
	\end{minipage}
	\caption{A set of paper models and a paper model colored according to the description above.
		The underlying input is a pentagon with five adjacent pentagons that have been subdivided three times via the pentagon snub subdivision scheme.
	}
	\label{fig:PentagonsWeavingPaperModel}
\end{figure}

\rev{
	The above discussed a paper model realization of the pentagon snub subdivision scheme when applied to a regular square grid, from which we derived the depicted weaving pattern.
	As a closing piece for this section, we offer a paper model realization of the subdivided pentagon as shown in Figure~\ref{fig:FirstSnubRefinementSteps}.
	Again, the pattern was printed on printer paper and cut using a pen knife.
	In contrast to the pattern derived from the square grid, here, five copies of the set of modules have to be used, representing the directions as induced by each side of the underlying pentagon.
	See Figure~\ref{fig:SixPentagonsSinglePaperModule} for a single module set and Figure~\ref{fig:SixPentagonsWeavingPaperModel} for the assembled paper model.
	As a color scheme, we chose to extend the maritime palette used before.
	Note that this model is substantially harder to assemble due to the lack of orthogonality in the meeting of strands.
	When assembling it, we found that it is best to work as a pair as some strands need to be connected to several other ones in order to prevent the entire structure from falling apart.
	The underlying input is a pentagon with five adjacent pentagons that have been subdivided three times via the pentagon snub subdivision scheme.
}

\subsection{\rev{Transferring the patterns to other subdivisions}}

\rev{
	The previous subsection was concerned with physical paper model realizations of the weaving pattern.
	Now, we are interested in generalizing the pattern to a broader range of subdivision schemes, aside from pentagon snub subdivision.
	We tackle this question by considering certain classes of meshes.
}

\subsubsection{Meshes from quadrilaterals}
\label{sec:MeshesFromQuadrilaterals}

\begin{figure}
	\centering
	\begin{subfigure}[t]{0.3\textwidth}
		\includegraphics[width=1.\textwidth]{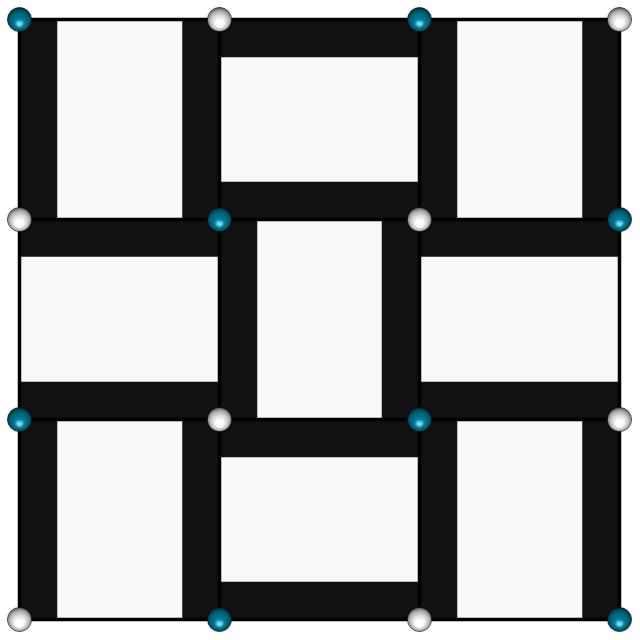}
		\caption{Relationship between a weaving and a two-coloring, where ${c_1=\text{blue}}$ and ${c_2=\text{white}}$.}
		\label{fig:SquareGridTwoColoring}
	\end{subfigure}
	\hfill
	\begin{subfigure}[t]{0.3\textwidth}
		\includegraphics[width=1.\textwidth]{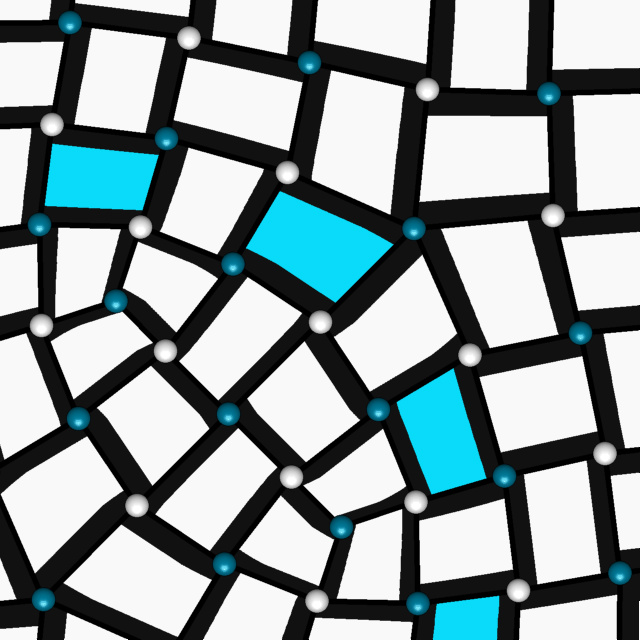}
		\caption{Two-coloring on the subdivision via Catmull-Clark~\cite{catmull1978recursively} in Figure~\ref{fig:CatmullClarkSubdivision}.}
		\label{fig:CatmullClarkTwoColoring}
	\end{subfigure}
	\hfill
	\begin{subfigure}[t]{0.3\textwidth}
		\includegraphics[width=1.\textwidth]{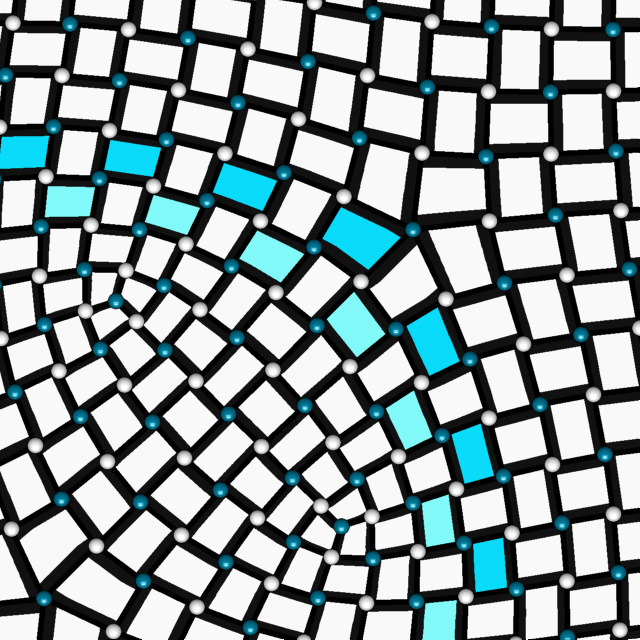}
		\caption{Two-coloring after another Catmull-Clark refinement step.}
		\label{fig:CatmullClarkTwoColoringSubdivided}
	\end{subfigure}
	\caption{Weaving from two-coloring: White-blue edges have the strand go on top, while blue-white edges have it go below. Also: Application of the weaving pattern to the subdivision scheme of Catmull-Clark~\cite{catmull1978recursively}.}
	\label{fig:Catmull-ClarkWeaving}
\end{figure}

\rev{
	Focusing on quadrilateral-based meshes first, we can make the following observation: Taken any quad-mesh, if the vertices are \emph{two-colorable} (i.e., if we can color all vertices of the mesh by either color~$c_1$ or color~$c_2$ such that no two vertices connected via an edge have the same color), the quad mesh can be interpreted as a weaving pattern.
	This is done by considering each quadrilateral to be the crossing of two strands, one above, one below.
	In order to determine, which strand goes where, we consider the edges of the quadrilateral.
	If a strand ``enters'' a quadrilateral via an edge between a $c_1$-vertex on the left and a $c_2$-vertex on the right, it will be on top, if the order is switched, it will be below.
	Consider Figure~\ref{fig:SquareGridTwoColoring} for an illustration of this, where a strand changes from above to below when passing an edge with a~$c_1$ vertex on the left and a~$c_2$ vertex on the right.
}
\rev{
	The condition of admitting to a two-coloring can be easily satisfied for any quadrilateral mesh that is a topological disc. 
	It is done by coloring the vertices via a disc growing approach: Color a first vertex without loss of generality by color~$c_1$, color all vertices that are connected to it with color~$c_2$, and continue to their neighbors, etc., while always alternating colors.
	Given this observation, let us revisit the quadrilateral-based subdivision schemes discussed in Section~\ref{sec:SubdivisionSchemes}.
}

\rev{
	By the above, a mesh subdivided by the scheme of Catmull-Clark~\cite{catmull1978recursively}, can be interpreted as a weaving.
	This holds in particular also for meshes of arbitrary topology by performing the following two-coloring after performing a refinement step: Assign color~$c_1$ to all new vertices inserted on the edges during the subdivision, while assigning color~$c_2$ to all old vertices as well as to those new vertices at the face centers (compare Figures~\ref{fig:CatmullClarkSubdivision} and~\ref{fig:CatmullClarkTwoColoring}, where edge-vertices are colored white and all other vertices have been colored blue).
	This will lead to a two-coloring, independent of the topology of the mesh.
}

\rev{
	If the input mesh already did admit to a weaving pattern via a corresponding two-coloring, a Catmull-Clark refinement step splits one strand into two parallel strands (as opposed to the pentagonal subdivision scheme, where the strands are intertwined), see Figure~\ref{fig:CatmullClarkTwoColoringSubdivided} for an illustration of this.
	Information of the weaving, as to what strand is above or below, is lost during a refinement step, see Figure~\ref{fig:Catmull-ClarkWeaving}.
}

\rev{
	By the above, for the subdivision scheme of Catmull-Clark~\cite{catmull1978recursively}, we can introduce a weaving pattern based on the property that after each refinement step, all resulting faces are quadrilaterals.
	In the case of the \emph{mid-edge} subdivision scheme~\cite{peters1997simplest}, this is only the case if the input mesh consists solely of quadrilateral faces and vertices of degree four, as every other face will simply be replicated and every vertex of degree different to four will give rise to a face of corresponding degree.
	This only leaves the regular square grid as valid input for which we can apply the technique given above.
	Turning to the subdivision scheme of Doo-Sabin~\cite{doo1978behaviour}, as illustrated in Figures~\ref{fig:MidEdgeSubdivision} and~\ref{fig:DooSabinSubdivision}, one refinement step via Doo-Sabin is equivalent to taking two \emph{mid-edge} refinement steps.
	Therefore, for both subdivision schemes, the only input which can be subdivided and associated to a weaving pattern via the strategy used above is the regular square grid, with a resulting weaving pattern as shown in Figure~\ref{fig:SquareGridTwoColoring}.
}

\subsubsection{Meshes from triangles}

\rev{
	Turning to triangle meshes, we can make another general observation.
	Let us assume that the considered mesh is purely composed of triangles and that the vertices of the mesh can be colored using two colors~$c_1$, $c_2$ such that every triangle has exactly one vertex of color~$c_1$ and two vertices of color~$c_2$.
	Given this, we can remove all edges from the mesh that span between two vertices colored by~$c_2$.
	Thereby, we obtain a mesh made entirely of quadrilaterals, which satisfies the two-coloring criterion described above.
	Hence, we can impose a weaving on corresponding triangle meshes by gluing pairs of triangles to form quadrilaterals, analogous to the pentagonal case, where we glued pentagons to octagons.
}

\begin{figure}
	\begin{subfigure}{0.3\textwidth}
		\includegraphics[width=1.\textwidth]{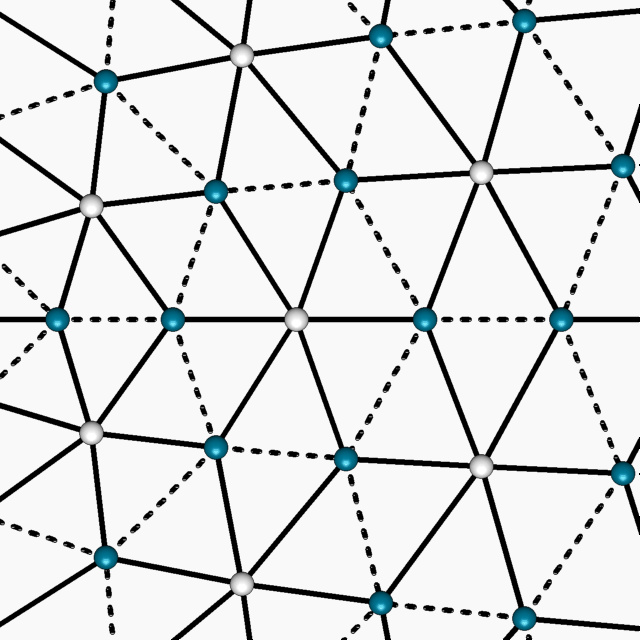}
	\end{subfigure}
	\hfil
	\begin{subfigure}{0.3\textwidth}
		\includegraphics[width=1.\textwidth]{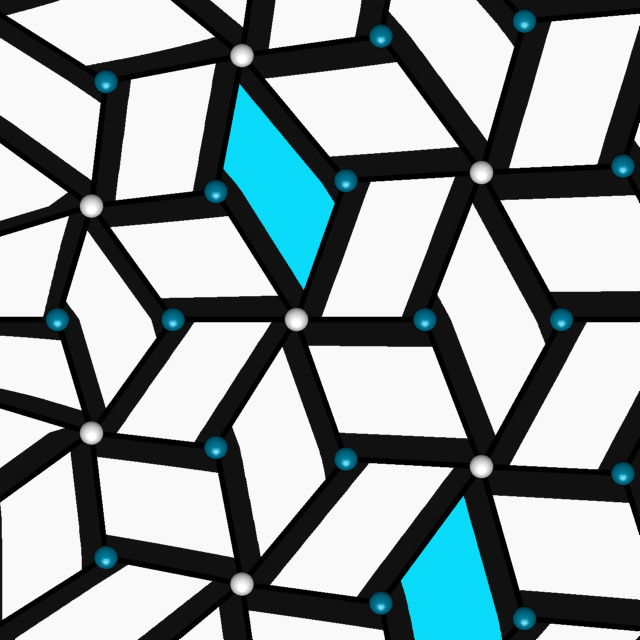}
	\end{subfigure}
	\hfil
	\begin{subfigure}{0.3\textwidth}
		\includegraphics[width=1.\textwidth]{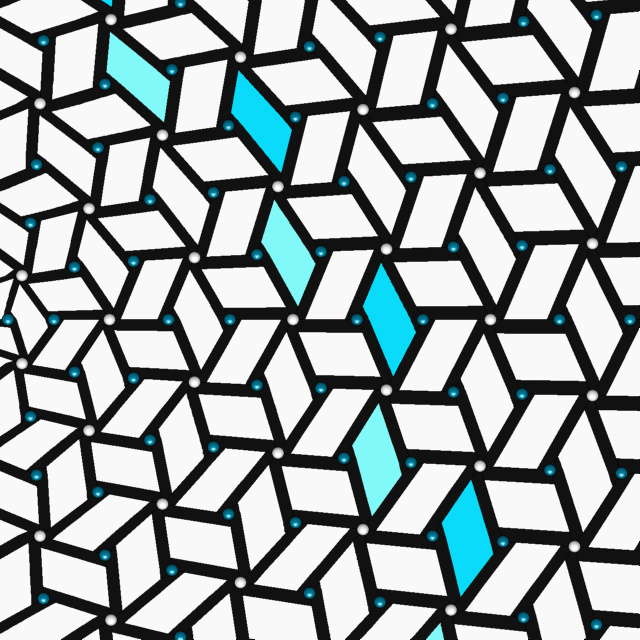}
	\end{subfigure}
	\caption{Weaving patterns from the subdivision scheme by Loop. Remove dashed edges between vertices both colored by the same color. Applying the subdivision splits a strand into two parallel ones.}
	\label{fig:LoopWeaving}
\end{figure}

\rev{
	In case of both the \emph{butterfly} subdivision scheme and the subdivision scheme by Loop, if we assume that the input mesh satisfies the coloring criterion sketched above, after applying a refinement step, we can obtain a valid new coloring as follows: keep colors for all old vertices, new vertices on an edge between a~$c_1$ and a~$c_2$ vertex are colored by~$c_2$, and new vertices on edges between two $c_2$-vertices are colored by~$c_1$.
	This yields a new triangle mesh that once more satisfies the coloring property.
	See Figure~\ref{fig:LoopWeaving} for an illustration of the triangle gluing as well as the behavior of the weaving under subsequent refinement steps.
}

\begin{figure}[b]
	\begin{subfigure}{0.3\textwidth}
		\includegraphics[width=1.\textwidth]{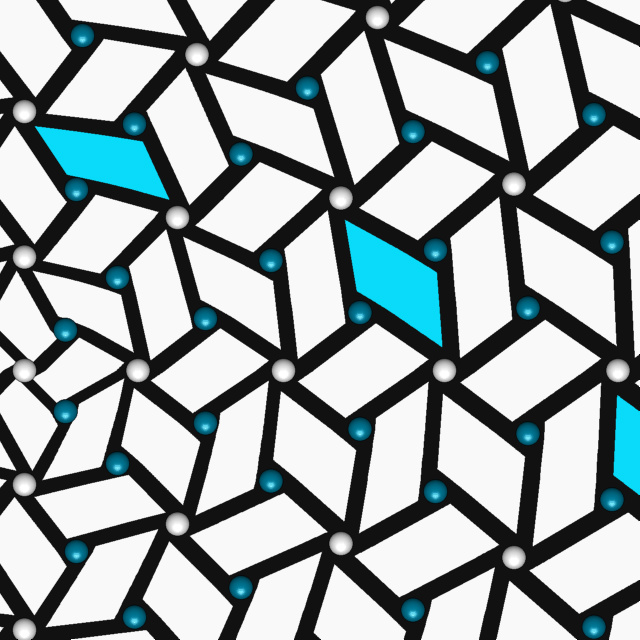}
	\end{subfigure}
	\hfil
	\begin{subfigure}{0.3\textwidth}
		\includegraphics[width=1.\textwidth]{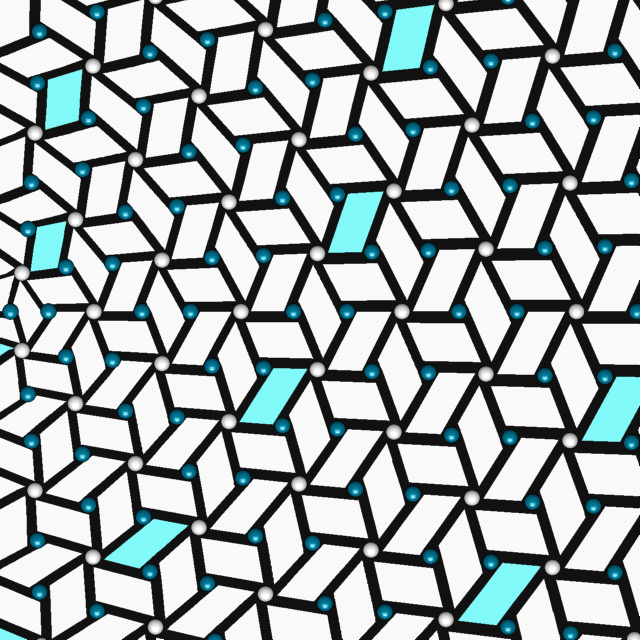}
	\end{subfigure}
	\hfil
	\begin{subfigure}{0.3\textwidth}
		\includegraphics[width=1.\textwidth]{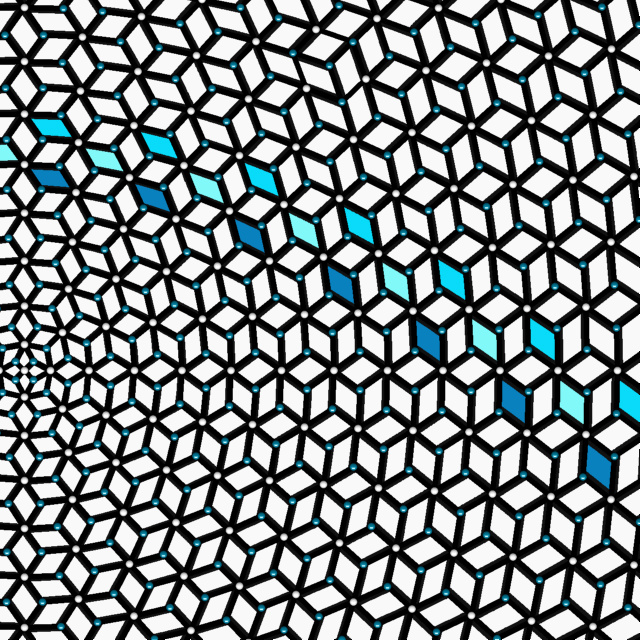}
	\end{subfigure}
	\caption{Weaving patterns from the $\sqrt{3}$ subdivision scheme. Applying the subdivision twice splits each original stand into three.}
	\label{fig:Sqrt3Weaving}
\end{figure}

\rev{
	Turning to the $\sqrt{3}$ subdivision scheme, we find that it always creates a weaving pattern, independent of the mesh it is applied to.
	This can be seen as follows: If we remove those edges that have just been flipped in the refinement step, we obtain a set of quadrilaterals, where each quadrilateral contains two vertices of the original mesh and two vertices created as face midpoints, pairwise diagonally in the quadrilateral.
	Thus, once more, we obtain a two-coloring on a set of quadrilaterals and hence a weaving by gluing sets of triangles.
	However, the behavior of the weaving pattern under subdivision is interesting.
	New strands are not parallel, but orthogonal to the strands of the previous step.
	Only after performing two refinement steps, each strand is separated into three, parallel strands, see Figure~\ref{fig:Sqrt3Weaving}.
}

\subsection{A general weaving pattern for meshes}

\rev{
	The previous sections discussed the weaving pattern for the pentagon snub subdivision scheme, suitable quadrilateral or triangle meshes, and their corresponding subdivision schemes.
	For some combinations, no topological or combinatorial restriction are necessary, while other need to satisfy certain properties.
	While the weaving patterns for meshes composed of quadrilateral faces directly work on these, the patterns for faces of uneven edge numbers, i.e., triangles and pentagons, work by gluing pairs of faces together and thereby introducing a weaving on a set of quadrilaterals or octagons.
	Here, we present a final weaving scheme that is independent of the underlying mesh and the subdivision scheme.
	Instead of gluing faces together, it splits the faces of the mesh.
}

\begin{figure}
	\begin{subfigure}[t]{0.3\textwidth}
		\includegraphics[width=1.\textwidth]{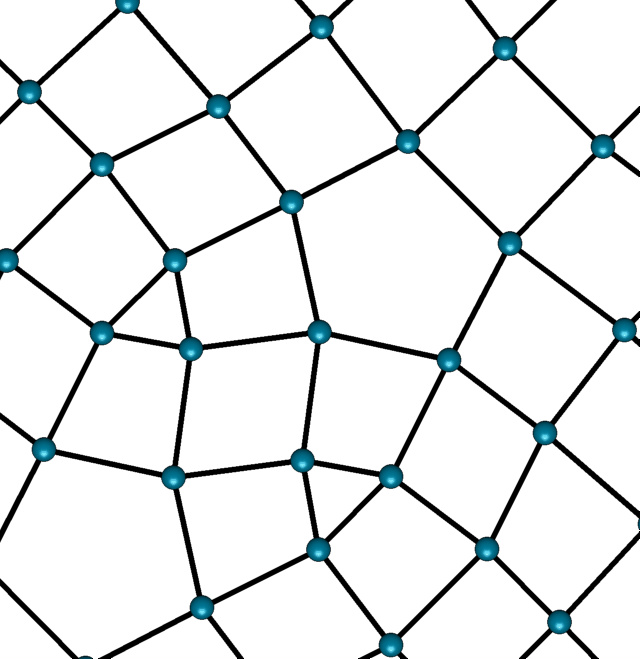}
		\caption{Arbitrary input mesh.}
	\end{subfigure}
	\hfil
	\begin{subfigure}[t]{0.3\textwidth}
		\includegraphics[width=1.\textwidth]{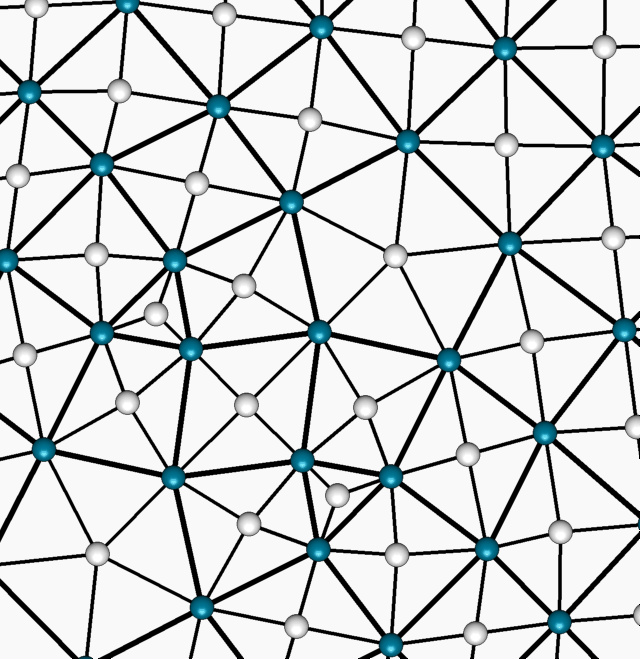}
		\caption{Adding face mid-points and edges to face-vertices.}
	\end{subfigure}
	\hfil
	\begin{subfigure}[t]{0.3\textwidth}
		\includegraphics[width=1.\textwidth]{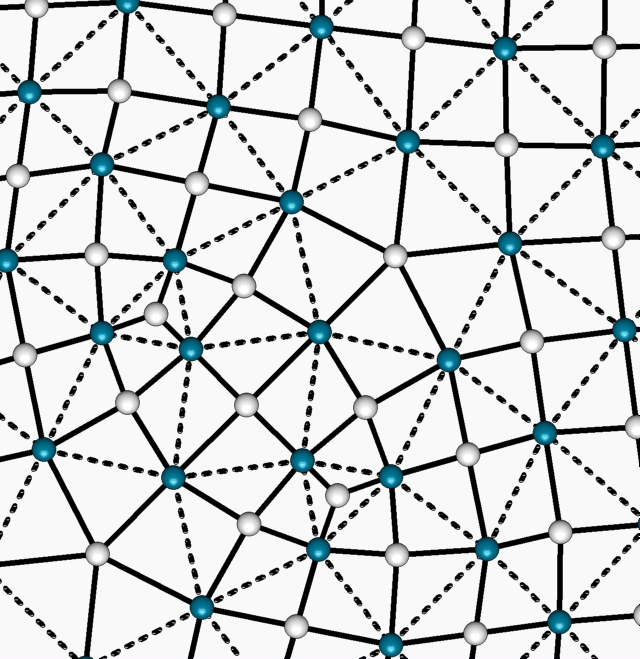}
		\caption{Removing old mesh edges.}
	\end{subfigure}\\
	\begin{subfigure}{0.3\textwidth}
		\includegraphics[width=1.\textwidth]{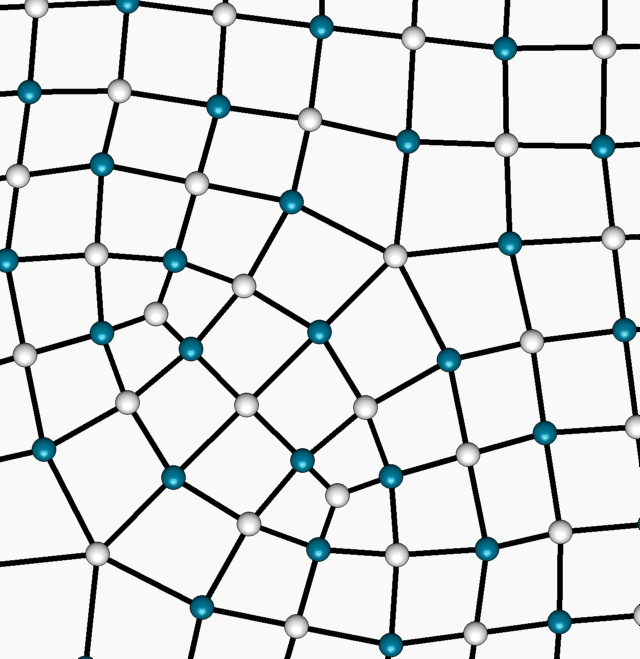}
		\caption{Obtained two-colored quad mesh.}
	\end{subfigure}
	\hfil
	\begin{subfigure}{0.3\textwidth}
		\includegraphics[width=1.\textwidth]{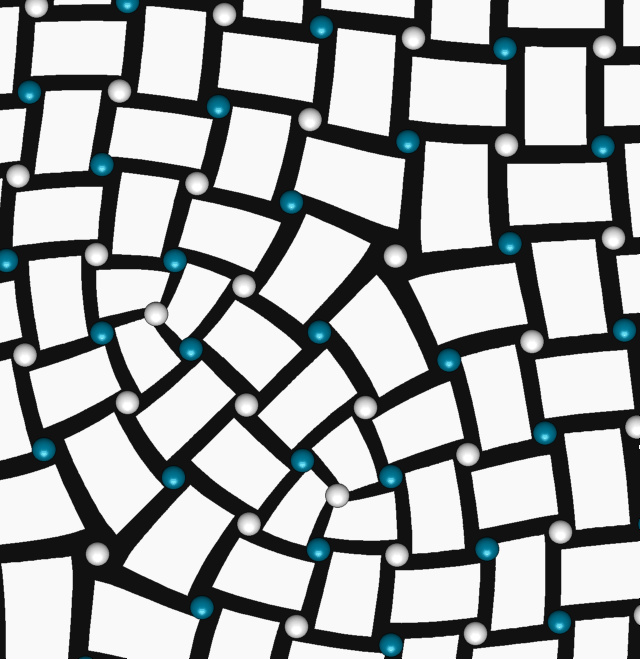}
		\caption{Induced weaving pattern from the quad mesh.}
	\end{subfigure}
	\hfil
	\begin{subfigure}{0.3\textwidth}
		\includegraphics[width=1.\textwidth]{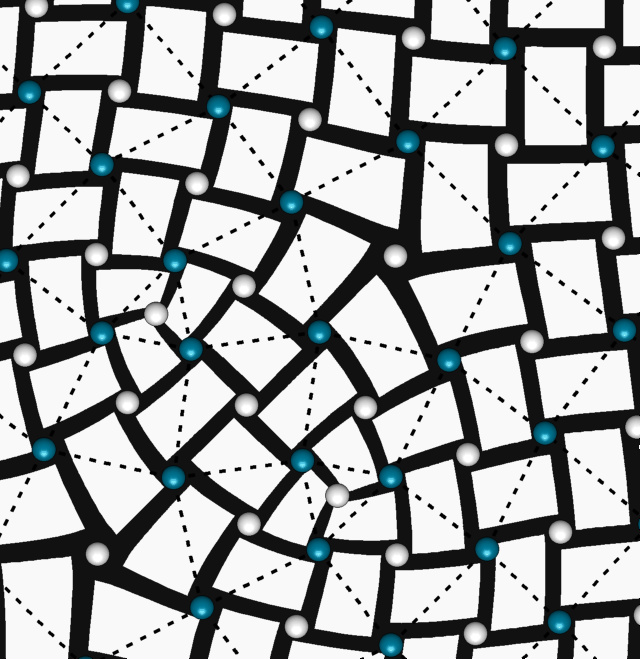}
		\caption{Weaving pattern overlapped with input mesh.}
		\label{fig:FaceSplitWeavingOverlayInput}
	\end{subfigure}
	\caption{Weaving pattern for an arbitrary input mesh.}
	\label{fig:FaceSplitWeaving}
\end{figure}

\rev{
	Given a mesh, possibly obtained from some subdivision scheme, we perform the following steps.
	First, for each face of the mesh, the face mid-point is added as new vertex to the mesh and connected via an edge to all vertices of the parental face.
	Then, the original edges of the mesh are deleted.
	Thereby, for each pair of faces that share an edge, their two midpoints, together with the two vertices of the edge, form a quadrilateral.
	This gives rise to a pure quadrilateral mesh on which a weaving pattern can be induced as described above, see Figure~\ref{fig:FaceSplitWeaving} for an illustration of these steps.
	Note that the process of Christian Mercat employs the same rationale to create Celtic links~\cite{mercat1996theorie}.
}

\rev{
	Just as for the weaving patterns induced by the triangle-based subdivision schemes and the pentagon snub subdivision scheme, in this case, the crossing of the weaving happens on an edge of the original mesh, see Figure~\ref{fig:FaceSplitWeavingOverlayInput}.
	This is in contrast to the weaving patterns as described in Section~\ref{sec:MeshesFromQuadrilaterals} build on quadrilateral meshes.
	There, the crossings of the weaving pattern take place within the faces of the original input mesh.
	For meshes with faces composed of an uneven number of sides, crossings on top of the original faces might only be possible by allowing, e.g., a different number of strands to enter a face than to leave it, or by including splits in strands on such faces.
	Aside from the weaving patterns described above, this motivates the following open question.
}

\begin{question}
	\rev{What additional, possible weaving patterns exist for specific of general meshes, where the crossing happens inside the input faces? In particular, what are corresponding patterns for the \emph{mid-edge} subdivision scheme~\cite{peters1997simplest} or to the subdivision scheme of Doo-Sabin~\cite{doo1978behaviour}?}
\end{question}

\section{Jigsaw puzzle artwork}

The weaving pattern described in the previous section completes the set of theoretical observations on the pentagon snub subdivision scheme.
This concluding section describes an artistic piece that illustrates the theory and makes it tangible \rev{beyond the presented paper models}.
The concept of subdivision schemes---with their property of splitting large domains into smaller, as-regular-as-possibly pieces---immediately suggests the creation of a jigsaw puzzle for their illustration.
This is a specifically good choice in this case, because it combines visual information via coloring the pieces as well as geometric information via the shape of the pieces.
In particular the latter adds a tactile experience to the illustration that is not present in images, animations, or computer simulations of subdivision.
Prominent examples of previous successful combinations of jigsaw puzzles and mathematics can be found at \emph{Kadon Enterprises, Inc.}~\cite{kadon1980gamepuzzles} or at \emph{nervous system}~\cite{rosenkrantz2007nervous}.

\begin{figure}
	\begin{subfigure}[t]{0.4225\textwidth}
		\includegraphics[width=1.\textwidth]{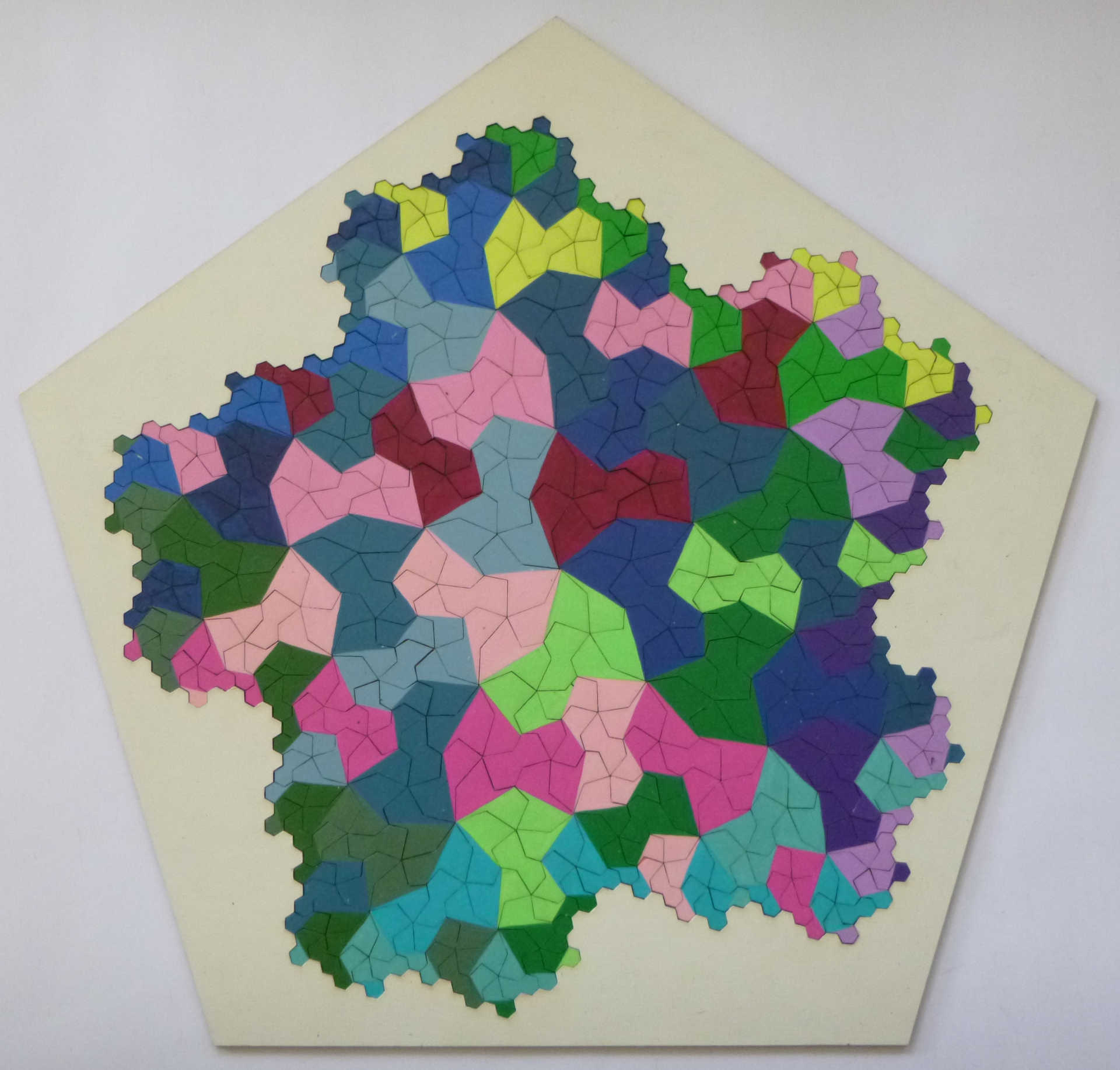}
		\caption{The jigsaw puzzle illustrating various theoretical aspects discussed in this paper. (50\,cm diagonal)}
		\label{fig:JigsawPuzzleComplete}
	\end{subfigure}
	\hfill
	\begin{subfigure}[t]{0.54\textwidth}
		\includegraphics[width=1.\textwidth]{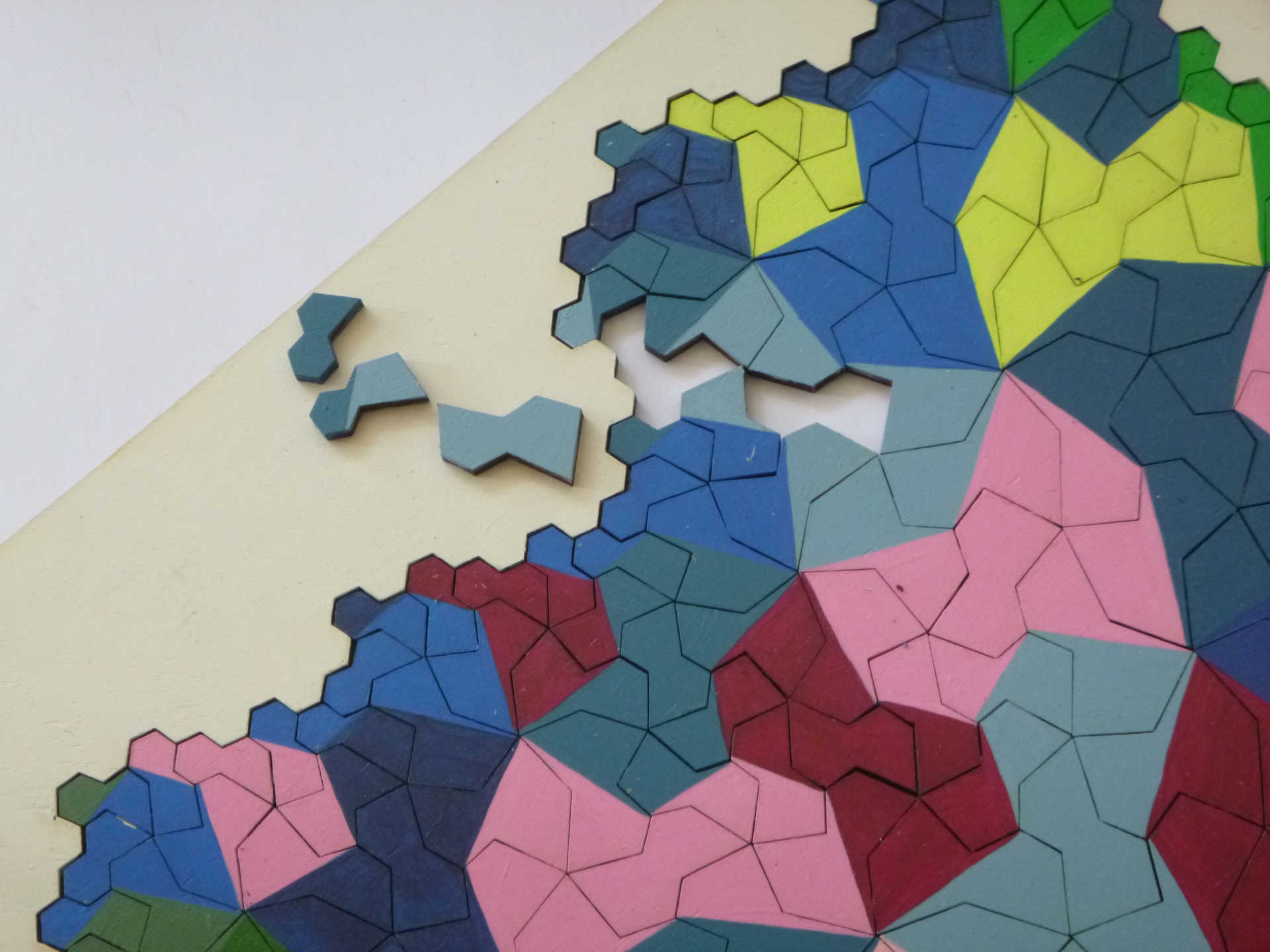}
		\caption{A border piece, a two-colored piece, and a uni-colored piece.}
		\label{fig:JigsawPuzzleHighlight}
	\end{subfigure}
	\caption{Final illustrating piece using the various possibilities of the discussed pentagon snub subdivision scheme.}
	\label{fig:JigsawPuzzle}
\end{figure}

The starting point for the specific jigsaw puzzle presented here in Figure~\ref{fig:JigsawPuzzle} is a regular pentagon that is subdivided four times via the pentagon snub subdivision scheme \rev{as discussed in Section~\ref{sec:PentagonSnubRefinement}}.
The resulting pentagons are then glued as described in Section~\ref{sec:WeavingPatterns} to create octagonal puzzle pieces.
However, not all pentagons have a ``partner'' to be glued to.
To overcome this, note that the smoothing operation is not applied to vertices along the boundary curve of the subdivided pentagon.
Thus, all outer angles along the boundary are~$\tfrac{2\pi}{3}$ or~$\tfrac{4\pi}{3}$.
This can be continued to an almost complete ring of perfect hexagons, which serve as gluing partners to those pieces that are left over.
Thereby, we create some enneagonal pieces along the boundary.
In particular, the hexagonal side of the boundary pieces helps with their placement as their hexagonal part fixes their inward/outward orientation.

The jigsaw puzzle is made from laser cut poplar plywood and is painted with acrylic paint. 
The diagonal of the surrounding pentagon measures 50\,cm. 
The edges made by the laser cut represent puzzle pieces obtained from the fourth refinement step, as described above.
However, the painted overlying tiling shows the weaving pattern associated to the third refinement step with a total of 20 twenty different colors.
Hence, most tiles are painted in two different colors.
Those tiles that only exhibit a single color are always at the center of the larger octagons from the coarser weaving pattern, see Figure~\ref{fig:JigsawPuzzleHighlight}.
The color scheme is reduced to three primary colors---green, blue, and pink. 
The different chosen shades made it possible to pair the interwoven strands within the same color class but still to distinguish them from the others.
For example, the light and dark green strands originate from one strand and are thus interwoven.

As touched upon in a blog post by the jigsaw puzzle designers from \emph{nervous system}~\cite{louis-rosenberg2020secret}, clean laser cutting is a non-trivial task.
The pieces and the frame of the jigsaw puzzle presented here have all been cut from a single piece of plywood.
Despite the laser cutter's high precision, the high number of cuts resulted in a certain leeway for the pieces.
However, after painting the pieces, these gaps closed by the wood swelling slightly.
When repeatedly creating the jigsaw puzzle, it could be beneficial to cut all pieces and the puzzle frame separately, taking into account the width of the introduced cuts, in order to create an even better fitting result.

The jigsaw puzzle is an illustrative collection of the aspects touched upon in this paper.
It does not only show the result of a refinement, but also the subdivision process itself as the coloring presents a coarser step than the individual puzzle pieces.
This is supported by the uni-colored pieces that always have to be placed at the center of one of the elements of the coarser step.
Handling the puzzle pieces provides a sense of how the different pentagons strive towards being regular.
However, the applied replacement by Z-triplets, operation~1, is also visible in the coloring: Those pieces that do exhibit two colors are not split into two convex pentagons.
Rather, they are split diagonally, which shows the replaced edge, while the two edges of the puzzle piece that slightly diverge from this edge indicate the two outer edges of the Z-triplet, see Figure~\ref{fig:JigsawPuzzleHighlight}.
Furthermore, assembling the jigsaw immediately introduces the uniform degree of all inner vertices present, as the convex corners of the pieces come together in multiples of three or five.
The fractal boundary curve is alluded to by the hexagonal pattern at the boundary of the jigsaw puzzle and its regular pattern does invoke images of the related Koch curve.
When completing the puzzle, the weaving patterns---that are present in any mesh resulting from the pentagon snub subdivision scheme---serve as a guiding motif.
The split of a strand into two interwoven strands that occurs with any further refinement step is encapsulated by the choice of colors and their shades.

\begin{figure}
	\begin{subfigure}[t]{0.4225\textwidth}
		\includegraphics[width=1.\textwidth]{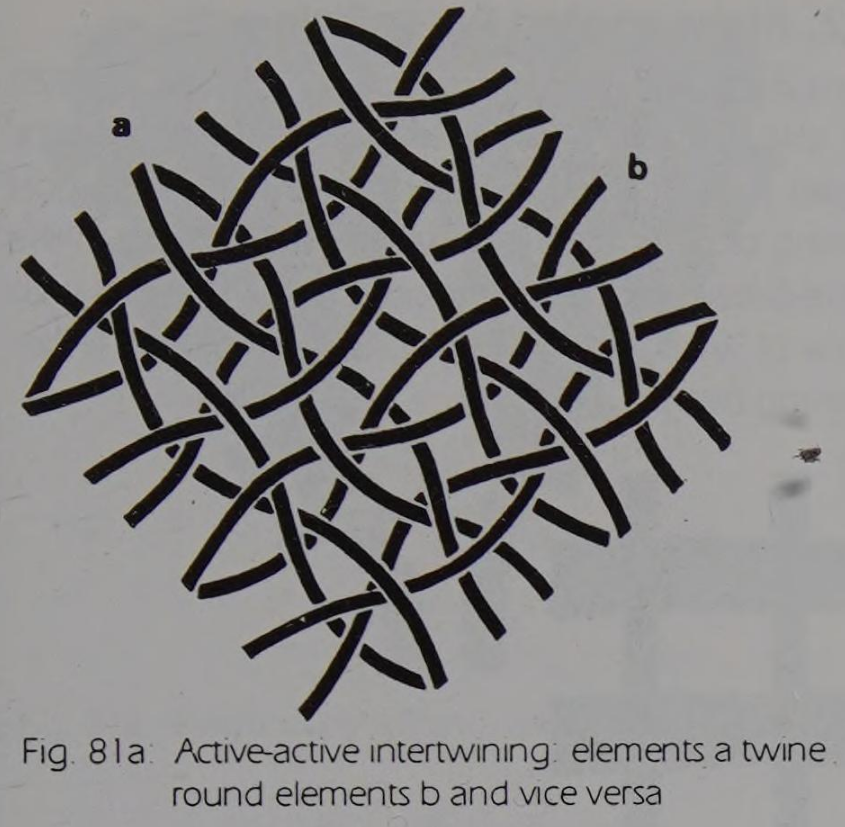}
		\caption{Figure 81a from~\cite{seiler1994textiles}, illustrating active-active intertwining.}
		\label{fig:IntertwinedSeiler}
	\end{subfigure}
	\hfill
	\begin{subfigure}[t]{0.54\textwidth}
		\includegraphics[width=1.\textwidth]{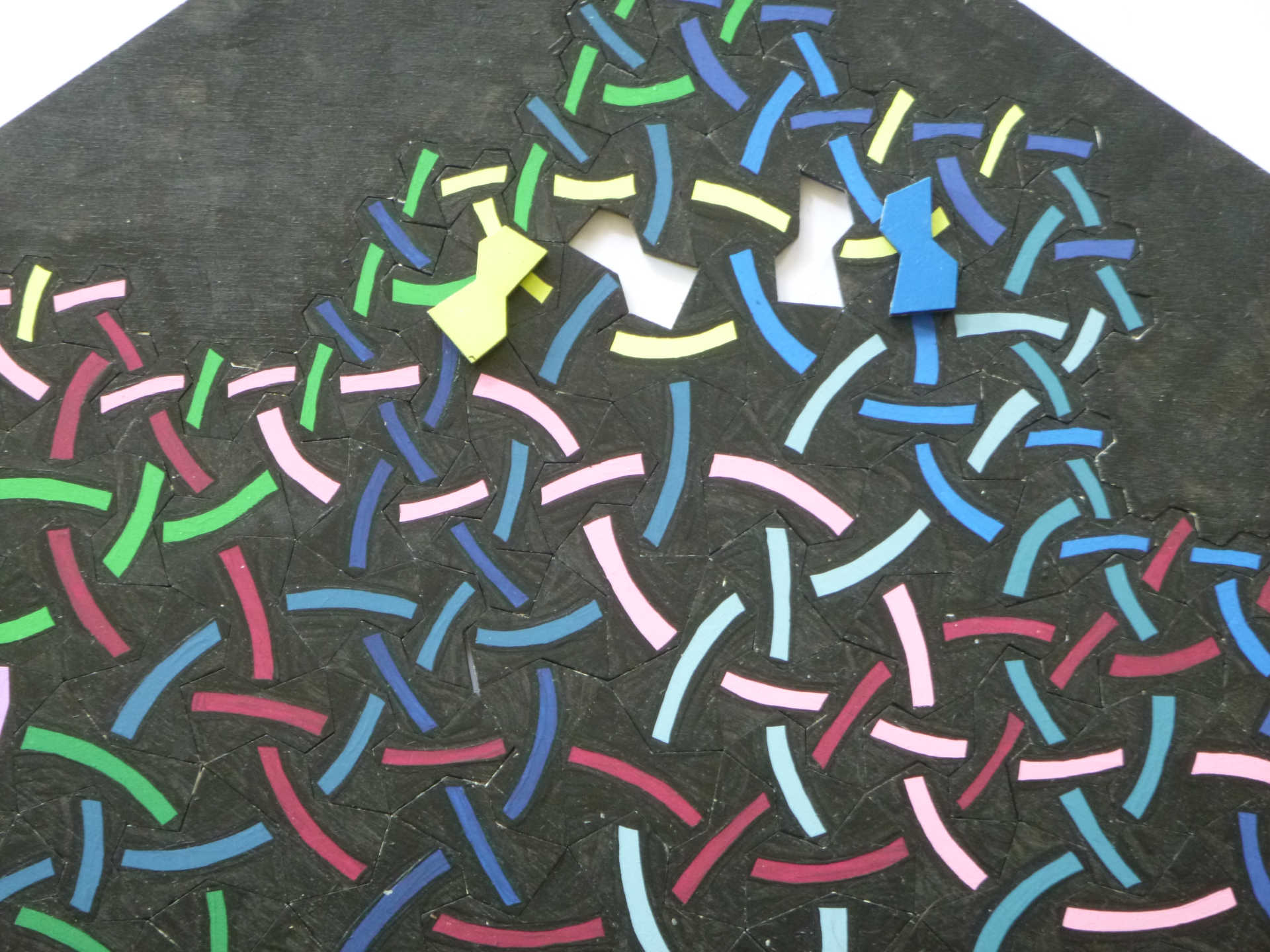}
		\caption{Two pieces flipped, their color matches that of the enclosed strands.}
		\label{fig:JigsawPuzzleBackHighlight}
	\end{subfigure}
	\begin{subfigure}{1.\textwidth}
		\includegraphics[width=1.\textwidth]{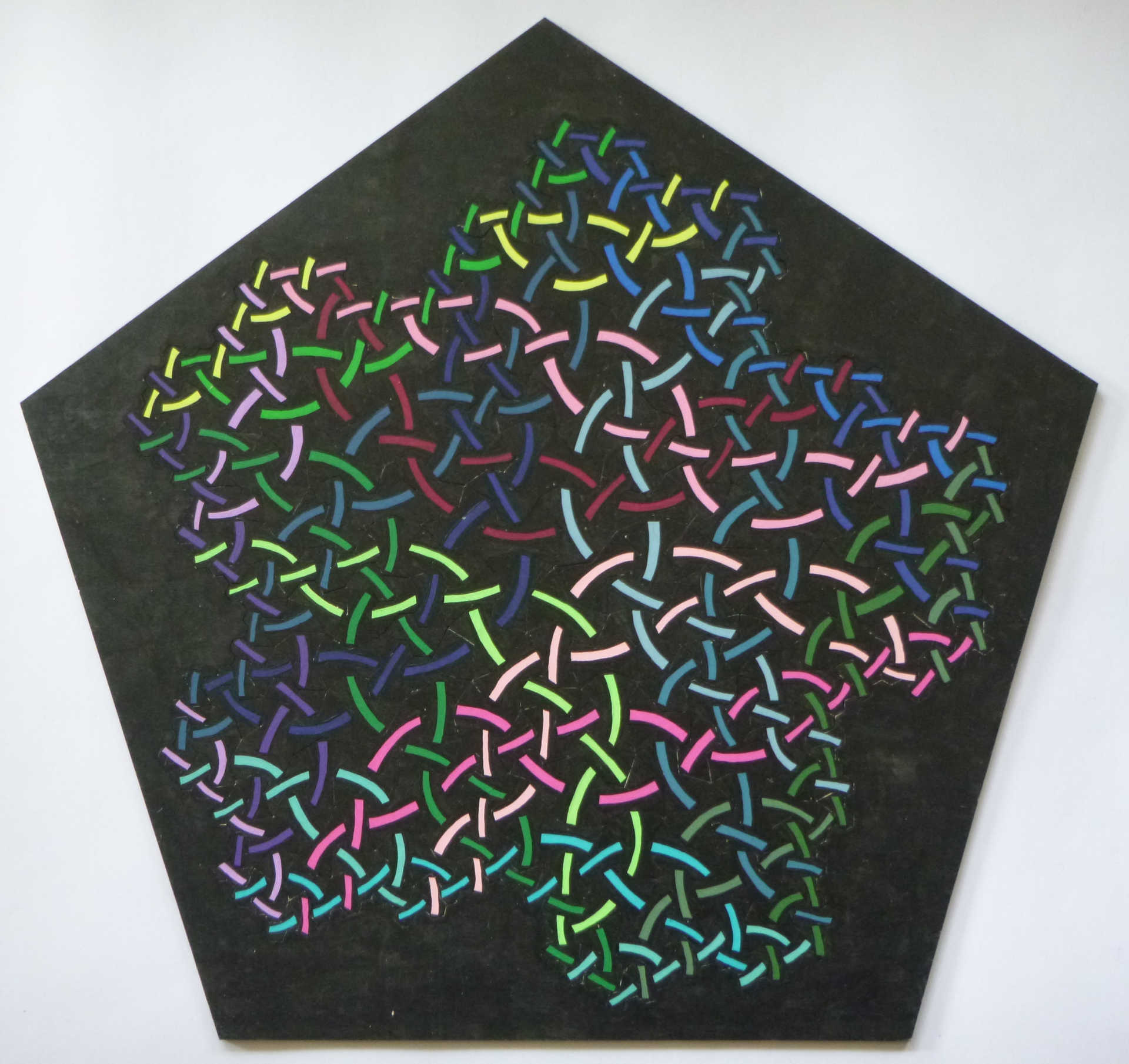}
		\label{fig:JigsawPuzzleBackComplete}
	\end{subfigure}
	\caption{Alternatively colored back-side of the jigsaw puzzle. (50\,cm diagonal)}
	\label{fig:JigsawPuzzleBack}
\end{figure}

\rev{
	Our jigsaw puzzle also offers a different coloring and thus a different assembly experience on its back-side, see Figure~\ref{fig:JigsawPuzzleBack}.
	While the tiles on the front-side are colored according to the weaving pattern of the previous refinement step, the back-side shows the weaving corresponding to the current refinement step.
	Thus, while the front side has two different kinds of pieces, those with one or those with two colors, the pieces on the back side only carry one color each.
	Therefore, this back coloring follows exactly the procedure as given in Section~\ref{sec:WeavingPatterns}.
	The colors of the strand pairs on this side of the jigsaw puzzle are chosen in a way such that they correspond to the front-side color of the enclosed puzzle-piece, see Figure~\ref{fig:JigsawPuzzleBackHighlight}.
	In particular, this side of the puzzle emphasizes the complex textile-like pattern behavior as discussed in Section~\ref{sec:NamingChoice}.
	This becomes apparent when comparing the drawn patterns on the jigsaw puzzle with the illustration of active-active intertwining, taken from~\cite{seiler1994textiles}, see Figure~\ref{fig:IntertwinedSeiler}.
	It is particularly interesting to note the differences between the patterns.
	For instance, the strands shown on our jigsaw puzzle always alternate between going above and below other strands, which is not true for the active-active intertwining pattern.
}

\rev{The paper model, with its strands of paper, illustrates the weaving component of the introduced coloring pattern.
The jigsaw puzzle adds to this by making the different refinement steps tangible, which offers manifold opportunities for exploration of the illustrated mathematics.
Hence, both models are good companions whenever the weather report does announce the next rainy afternoon.
They might well teach the explorers about some underlying principles of said report.}

\section*{Acknowledgments}

This research was partially funded by the Deutsche Forschungsgemeinschaft (DFG, German Research Foundation) -- 455095046.

\bibliography{literature}

\end{document}